\documentclass[a4paper,11pt]{amsart}
 \usepackage[margin=3cm]{geometry}
\usepackage{latexsym}
\usepackage{mathtools}
\usepackage{amsmath}
\usepackage{amsfonts}
\usepackage{mathrsfs}
\usepackage{tensor}
\usepackage{amscd}
\usepackage{amssymb}
\usepackage{amsfonts}
\usepackage{float}
\usepackage{esint}

\usepackage[usenames,dvipsnames]{xcolor}
\usepackage[colorlinks=true,linkcolor=Black,citecolor=NavyBlue]{hyperref}

\usepackage{caption}
\usepackage{graphicx}
\usepackage{color}
\allowdisplaybreaks 
\definecolor{Blue}{rgb}{0.,0.,1.}
\definecolor{Red}{rgb}{1.,0.,0.}
\definecolor{Green}{rgb}{0.,1.,0.}

\newcounter{smallarabics}
\newenvironment{arabicenumerate}
{\begin{list}{{\normalfont\textrm{(\arabic{smallarabics})}}}
  {\usecounter{smallarabics}\setlength{\itemindent}{0cm}
   \setlength{\leftmargin}{5ex}\setlength{\labelwidth}{4ex}
   \setlength{\topsep}{0.75\parsep}\setlength{\partopsep}{0ex}
   \setlength{\itemsep}{0ex}}}
{\end{list}}

\newcounter{smallroman}

\newcommand{\ben}{\begin{arabicenumerate}}  
\newcommand{\een}{\end{arabicenumerate}}

\def\init{\setcounter{equation}{0}}

\newtheorem{theoreme}{Theorem}[section]
\newtheorem{proposition}[theoreme]{Proposition}

\newtheorem{lemma}[theoreme]{Lemma}
\newtheorem{definition}[theoreme]{Definition}
\newtheorem{corollary}[theoreme]{Corollary}
\newtheorem{remark}[theoreme]{Remark}
\newtheorem{example}[theoreme]{Example}
\newcommand{\beq}{\begin{equation}}
\newcommand{\eeq}{\end{equation}}

\newcommand{\bex}{\begin{example}}
\newcommand{\eex}{\end{example}}
\def\bel{\begin{lemma}}
\def\eel{\end{lemma}}
\def\bet{\begin{theoreme}}
\def\eet{\end{theoreme}}
\def\bed{\begin{definition}}
\def\eed{\end{definition}}
\def\ber{\begin{remark}}
\def\eer{\end{remark}}

\newenvironment{refproof}[1]{\smallskip
\noindent{\bf Proof of {#1}.}\  }{\qeds}

\def\rr{{\mathbb R}}

\def\cc{{\mathbb C}}
\def\nn{{\mathbb N}}

\def\Im{{\rm Im}}
\def\Re{{\rm Re}}

\def\bar{\overline}

\def\cinf{C^\infty}
\def\proof{
\noindent{\bf Proof.}\ \ }

\def\cV{{\mathcal V}}

\def\cD{{\mathcal D}}
\def\cU{{\mathcal U}}

\def\cW{{\mathcal W}}

\def\i{{\rm i}}
\def\qed{$\Box$\medskip}
\def\qednoskip{$\Box$}
\def\p{ \partial}
\def\12{\frac{1}{2}}

\def\bbbone{{\mathchoice {\rm 1\mskip-4mu l} {\rm 1\mskip-4mu l}
{\rm 1\mskip-4.5mu l} {\rm 1\mskip-5mu l}}}
\def\one{\bbbone}

\def\coinf{C_0^\infty}

\def\cF{{\mathcal F}}

\def\cX{{\mathcal X}}

\def\tD{{\widetilde{D}}}
\def \p{ \partial}
\def\e{{\rm e}}

\def\s{{\rm s}}
\newcommand{\mat}[4]{\begin{pmatrix}#1 &#2  \\ #3 &#4 \end{pmatrix}}

\newcommand{\col}[2]{\begin{pmatrix}#1 \\#2\end{pmatrix}}

\newcommand{\traa}[1]{\mskip-6mu\upharpoonright_{#1}}

\def\cE{{\mathcal E}}

\def\WF{{\rm WF}}
\makeatletter
\newcommand*{\defeq}{\mathrel{\rlap{%
                     \raisebox{0.3ex}{$\m@th\cdot$}}%
                     \raisebox{-0.3ex}{$\m@th\cdot$}}%
                     =}
\makeatother
\makeatletter
\newcommand*{\eqdef}{=\mathrel{\rlap{%
                     \raisebox{0.3ex}{$\m@th\cdot$}}%
                     \raisebox{-0.3ex}{$\m@th\cdot$}}%
                     }
\makeatother

\def\rx{{\rm x}}

\def\bS{\mathbb{S}}

\def\vol{{\rm vol}}

\DeclareMathOperator{\Ker}{Ker}
\DeclareMathOperator{\Ran}{Ran}

\def\id{\one}

\newcommand{\Kerc}{{\rm Ker}_{\rm c}}
\newcommand{\Ranc}{{\rm Ran}_{\rm c}}
\newcommand{\Kersc}{{\rm Ker}_{\rm sc}}
\newcommand{\Ransc}{{\rm Ran}_{\rm sc}}

\def\dual{\!\cdot \!}
\def\CCR{{\rm CCR}}

\def\zero{{\mskip-4mu{\rm\textit{o}}}}

\def\cN{{\mathcal N}}

\def\tnab{\widetilde{\nabla}}

\def\tosim{\xrightarrow{\sim}}

\def\tf{\tilde{f}}

\def\zero{{\mskip-4mu{\rm\textit{o}}}}

\def\eucl{{\rm eucl}}
\def\dual{\!\cdot \!}
\def\calde{Calder\'{o}n }\def\tc{\tilde{c}}
\def\rR{{\rm R}}\def\Ric{{\rm Ric}}\def\rg{{\bf g}}
\def\rh{{\bf h}}
\def\scal{{\rm R}}\def\Riem{{\rm Riem}}
\def\tD{\widetilde{D}}
\def\nab{\nabla}
\def\trace{{\rm tr}}
\def\dvol{\mathop{}\!d{\rm vol}}
\def\Diff{{\rm Diff}}
\def\trg{\widetilde{\rg}}
  \def\tK{\widetilde{K}}\def\tV{\widetilde{V}}\def\trho{\widetilde{\varrho}}
\def\tM{\widetilde{M}}
\def\sc{\rm{sc}}\def\c{\rm{c}}
\def\tq{\widetilde{q}}
\def\Vect{{\rm Vect}}
\newcommand{\vD}[1]{\vec{D}_{#1, L}}
\def\vd{\vec{d}}\def\vdel{\vec{\delta}}

\def\tcE{\tilde{\cE}}
\def\tcF{\tilde{\cF}}
\def\TT{{\rm TT}}
\author{Christian G\'erard}
\address{Laboratoire de Math\'ematiques d'Orsay, Universit\'e Paris-Saclay, France}
\email{christian.gerard@math.u-psud.fr} 
\author{Micha{\l} Wrochna} 
\address{Mathematical Institute, Universiteit Utrecht, The Netherlands \vspace{-0.3cm}} \address{Mathematics \& Data Science, Vrije Universiteit Brussel, Belgium}
 \email{{m.wrochna@uu.nl}}

\title[IR-fixed Euclidean vacuum for linearized gravity on de Sitter]{{\large IR-fixed Euclidean vacuum}\\ { \large for linearized gravity on de Sitter space}}

\begin{document}

\maketitle

\vspace{-0.8cm}

\begin{abstract}We consider the Euclidean vacuum for linearized gravity on the global de Sitter space, obtained from the Euclidean Green's function on the $4$-sphere. We use the notion of Calder\'on projectors to recover a quantum state for the Lorentzian theory on de Sitter space. We show that while the state is gauge invariant and Hadamard, it is not positive on the whole of the phase space. We show however that a suitable modification at low energies yields a well-defined Hadamard state on  global de Sitter space.
\end{abstract}

\section{Introduction and main result}\label{sec0}\init

\subsection{Introduction}\label{ss:intro}

The quantization of linearized gravity poses significant mathematical challenges as compared to scalar fields and Dirac fermions. Building on earlier advances \cite{HS,FH,BDM,Gerard2025} (see e.g.~\cite{furlani,FP,hollands,DS,FS,GW1,Moretti2023,Iuliano2023} for results for Maxwell, Proca, Yang--Mills and Teukolsky fields), the existence of a well-defined state satisfying the Hadamard condition has only recently been shown on spacetimes with a compact Cauchy surface \cite{Gerard2024}.  Even so, requiring in addition that the state is invariant under spacetime symmetries can  lead to further  issues \cite{dS1,Miao2010,dS2,Glavan2023}.

\medskip

A particularly important case in which one expects various simplifications is gravity linearized around the global de Sitter solution. In coordinates, de Sitter space is $dS^{4}= \rr_{t}\times \bS^{3}$ equipped with the metric
$$
 \rg= - dt^{2}+ \cosh^{2}(t)\rh, 
 $$
 where $\rh$ is the canonical metric on $\Sigma=\bS^{3}$. Wick rotation $t\mapsto \i s$ yields the metric
$$
   \widetilde{\rg}= ds^{2}+ \cos^{2}(s)\rh,
$$
which is just the standard metric on the sphere $\bS^{4}$ away from the two poles. The de Donder gauge-fixed linearized Einstein  operator on $dS^{4}$ (denoted in the sequel by $D_2$) Wick rotates to an elliptic operator $\tD_2$  on $\bS^4$.  It is invertible, so the Euclidean Green's function obtained from $\tD_2^{-1}$ provides a natural candidate for the two-point function of a state after Wick rotating back to Lorentzian signature. The so-obtained ``state'', if well-defined,  is called \emph{Euclidean} or \emph{Bunch--Davies} \emph{vacuum}  in the physics literature.  An explicit expression for the Green's function was derived by Allen \cite{Allen1986}, and further properties, formulas in different gauges, symmetries  considerations and computations of physical quantities have been obtained  \cite{Allen1986,AFO,Faizal2012,Higuchi2001,Gazeau2023}, paralleled by results in the conformally flat chart \cite{dS1,Miao2010,Miao2011,dS2}.

\medskip

In the present paper we demonstrate that the Euclidean Green's function on $\mathbb{S}^4$ does actually \emph{not} yield a state on de Sitter space due to positivity issues. This observation is implicit  in the work of Faizal--Higuchi \cite{Faizal2012}, who {building on the work of Higuchi--Weeks \cite{higuchiweeks}} make a {formal} modification that kills off an infinite-dimensional space of pure gauge modes. We propose instead a mathematically rigorous finite-dimensional modification that resolves the problem. 
\medskip

To explain the main issue, let us first recall the relevant preliminaries on linearized gravity. Let $P$ be the  Einstein operator linearized around the de Sitter metric $\rg$, and let $K$ be the operator that generates linearized gauge transformations $u\mapsto u+Kw$. We will actually use a convention in which all operators are composed with a trace reversal $(Iu)_{ab}= u_{ab}- \frac{1}{2} \trace_{\rg}(u) \rg_{ab}$, see Sect.~\ref{sec1}, in particular $(Kw)_{ab}=\nabla_{(a} w_{b)}-\12 g_{ab}\nabla^c w_c$.  
The pre-symplectic space of smooth solutions modulo gauge is then the quotient space
$$
{\Ker P}/{\Ran K}, 
$$ 
equipped with the canonical pre-symplectic form (or Hermitian form, as we prefer to use the complex formalism). In quantization it is often more convenient to work on the dual level with an isomorphic space of compactly supported smooth tensors
\beq\label{eq:psd} 
{\Kerc  K^\star}/{\Ranc  P}, 
\eeq
where  $K^\star$ is the formal adjoint of $K$ for a non-positive Hermitian form  involving $I$, explicitly  $(K^\star u)_b=-2\nab^{a}u_{ab}$ in our convention. Then, in order to be interpreted as bi-solutions of $P$, two-point functions need to be defined as Hermitian forms on the quotient space \eqref{eq:psd}. As we are working with the complex formalism, to define a state one actually needs a pair of two-point functions ${\bf \Lambda}^\pm_2$ (called \emph{covariances} in the main part of the text to distinguish them from the associated operators and Schwartz kernels), see Subsect.~\ref{sec1.7}. 

The \emph{de Donder gauge} or \emph{harmonic gauge} consists in considering the  \emph{gauge-fixed}  operator  
 $$
 D_2= P+  K K^\star,
 $$ 
 which in contrast to $P$ is hyperbolic.
Then, solutions of $Pu=0$ are obtained by solving $D_2 u =0$ with the harmonic gauge condition $K^\star u=0$.  The original quotient space $\Ker P/ \Ran K$ is then isomorphic to
\[
(\Ker D_{2}\cap \Ker K^{\star})/K\Ker D_{1},
\] where $D_{1}= K^{\star}K$ is an hyperbolic operator acting now on $(0, 1)$-tensors, related to $D_{2}$ by the identity $D_{2}K= K D_{1}$.  The subspace $K\Ker D_{1}$ represents the residual gauge freedom.

Concerning quantization, let us stress that if we have candidates for two-point functions ${\bf \Lambda}^\pm_2$ which are bi-solutions of $D_2$, they do not necessarily induce  sesquilinear forms  on the physical space  \eqref{eq:psd}.   To ensure that they do, ${\bf \Lambda}^\pm_2$ need to satisfy the \emph{weak gauge-invariance} condition\footnote{In fact, $\bar{u}\dual  {\bf \Lambda}_{2}^{+}Pv=-\bar{u}\dual  {\bf \Lambda}_{2}^{\pm}K K^\star v=0$ for all $u\in \Kerc  K^\star$ and all test tensors $v$ iff \eqref{eq:gi} holds. If ${\bf \Lambda}_{2}^{+}$ is Hermitian then one gets automatically $\bar{Pv}\dual  {\bf \Lambda}_{2}^{+}u=0$ as well.  \label{footnote1}}
\beq\label{eq:gi}
 {\bf \Lambda}_{2}^{\pm}=0\hbox{ on }\Kerc  K^{\star}\times \Ranc K K^\star.
\eeq
Furthermore, candidates for two-point functions need to satisfy a \emph{positivity} condition to define a state. In order to ensure that the UV behaviour of fields is as good as possible, it is in addition highly desirable that ${\bf \Lambda}_2^\pm$ is defined by a pair of distributions which satisfies the \emph{Hadamard condition}.

As already mentioned, the Green's function of $\tD_2$ can be Wick rotated to yield a {candidate} for a two-point function in Lorentzian signature. One can attempt to show that the distributional kernel $\tD^{-1}_2(x, x')$ is analytic in $s,s'$ in a sufficiently large domain and Wick rotates to the Feynman propagator of a state,  but then it is not clear if the resulting expression has the correct positivity properties on the quotient space\footnote{We stress that  positivity condition on the physical quotient space underpins all other rigorous approaches to quantization of linear gauge theories, whether based on the Gupta--Bleuler, BRST or BV formalism.}.

Our approach  to these difficulties is to use the hypersurface $\Sigma=\mathbb{S}^3$ as an intermediary, seen both as the  Cauchy surface $\{t=0\}$ in de Sitter space and as the  boundary surface $\{s=0\}$ in $\mathbb{S}^4$ split  into the two hemispheres $\Omega^{\pm}= \{\pm s>0\}$.  In terms of Cauchy data at $\{t=0\}$, the Euclidean vacuum construction is  equivalent to the following.

\begin{definition} The \emph{\calde projectors} $\tilde c_2^\pm$ of $\tD_2$ are the projections to the subspace of Cauchy data of $L^2$ solutions of $\tD_2 u =0$  in $\Omega^{\pm}= \{\pm s>0\}\subset \bS^{4}$.
\end{definition}

Next, we define a new pair of operators $c_2^\pm$ by multiplying tensor components of $\tilde c_2^\pm$  by suitable powers of $\i$. The  operators $c_2^\pm$ play then the role of Cauchy data of two-point function candidates on the Lorentzian side.
In Appendix \ref{s:bunch-davies}  we show that the construction through \calde projectors is {formally} identical to the standard Bunch--Davies construction via a mode expansion to the extent the latter is well-defined. {Presently it is not known how to justify the convergent of mode expansions in this context, on the other hand, our construction is mathematically well-posed.}

\medskip

To carry out our analysis  we use the \emph{\TT-gauge}, i.e.~we use the residual gauge freedom to work with traceless tensors. Without going into details, this means in practice that the physical phase space is isomorphic to a quotient
$$
\cE_{\rm TT} / \cF_{\rm TT},
$$
equipped with a Hermitian form ${\bf q}_{I,2}$, 
where $\cE_{\rm TT}$ is the space of Cauchy data of smooth traceless solutions $D_2$ satisfying the harmonic gauge,  $\cF_{\rm TT}$ describes the remaining gauge freedom, and ${\bf q}_{I,2}$ can be computed explicitly. The list of sufficient conditions for $c_2^\pm$ to define a state becomes:
\beq\label{eq:cst}
\begin{array}{rl}
1)& c_{2}^{+}+ c_{2}^{-}= \one \mbox{ on } \cE_{\rm TT}, \\[2mm]
2)& c_{2}^{\pm*}{\bf q}_{I, 2}= {\bf q}_{I, 2}c_{2}^{\pm} \mbox{ on } \cE_{\rm TT},\\[2mm]
3)&  c_{2}^{\pm}K_{21\Sigma}f \in \Ran K_{21\Sigma}+ \Ran K_{20\Sigma}, \ \  f\in \Ran K^\dagger_{21 \Sigma},\\[2mm]
4)&\pm\bar{f}\dual {\bf q}_{I, 2}c_{2}^{\pm}f\geq 0, \ \  f\in \cE_{\rm TT},
\end{array}
\eeq
where $K_{21 \Sigma}$ is the Cauchy data version of the operator $K$ mapping $1$-tensors to $2$-tensors, $K_{21\Sigma}^\dagger$ is the Cauchy data version of $K^\star$, and $\Ran K_{20 \Sigma}$ is spanned by Cauchy data of scalar multiples of the de Sitter metric $\rg$.  With these notation,
\[
\cE_{\TT}=\Ker K_{21\Sigma}^{\dag}\cap \Ker K_{20\Sigma}^{\dag}, \ \cF_{\TT}=\Ran K_{21\Sigma}\cap \Ker K_{20\Sigma}^{\dag}.
\]

Condition 3) corresponds to \emph{weak gauge invariance}, and 4) is \emph{positivity}. The Hadamard condition can also be expressed in terms of $c_2^\pm$, see Prop.~\ref{prop:newtt}. Our first result can be formulated as follows.
 
\begin{theoreme}[cf.~Thm.~\ref{thm5.0}, Prop.~\ref{prop5.3}--\ref{prop-pos1}] \label{thm:main1} The Euclidean Green's function on $\mathbb{S}^4$ does  not define a state for linearized gravity on  $dS^4$ by Wick rotation. More precisely, $c_2^\pm$ satisfy properties \emph{1)--3)} and  the Hadamard condition, but do not satisfy \emph{4)} on the whole of $\cE_{\TT}$.
\end{theoreme}

Positivity on $\cE_{\TT}$ fails only on a finite dimensional subspace $\cE_{\TT, 4}$ and appears to be unrelated with the IR divergences in mode expansions in the conformally flat chart \cite{dS1,Miao2010,dS2}).

 The reason why 4) is a delicate condition is that the \calde projectors $c_{2}^{\pm}$ satisfy the analog of 4), but with ${\bf q}_{I, 2}$ replaced by its Euclidean version $\tilde{\bf q}_{I, 2}$ obtained from the Riemannian  scalar product of tensors on $\bS^{4}$.

 We first show that the physical space of Cauchy data decomposes as 
\beq\label{eq:decomp}
\cE_{\rm TT}= \cE_{\rm TT, \rm gauge}\oplus  \cF_{\rm TT},
\eeq
where $\cE_{\rm TT, \rm gauge }$ is a ``good'' subspace on which ${\bf q}_{I, 2}$ is non-degenerate and moreover ${\bf q}_{I, 2}$ equals $\tilde{\bf q}_{2}$. A first consequence of \eqref{eq:decomp} is that ${\bf q}_{I, 2}$ is non-degenerate on the physical quotient space $\cE_{\rm TT} / \cF_{\rm TT}$.

We can also decompose the pure gauge subspace $\cF_{\TT}$ as
\beq\label{eq:decomposition}
\cF_{\TT}= \cF_{\rm TT, \rm gauge}\oplus \cE_{\TT, 4}, 
\eeq
where $\cE_{\TT, 4}$ is an explicit $6$-dimensional subspace and  $c_{2}^{\pm}$ are invariant under $\cF_{\rm TT, \rm gauge}$ i.e.~
\[
{\bf q}_{I, 2}c_{2}^{\pm}f\in \Ran K_{21\Sigma}+ \Ran K_{20\Sigma}, \ \  f\in \cF_{\rm TT, \rm gauge}.
\]
It is easy to check that $\Ran K_{21\Sigma}^{\dag}\subset \cF_{\rm TT, \rm gauge}$ so the weak gauge invariance property 3) is indeed satisfied.

This non-trivial decomposition is related to the fact that the  \calde projectors for $D_{1}$ are {\em not defined} on the whole space of Cauchy data, see \ref{sec3.2.3}. This is a consequence of the fact that the Wick rotation $\tD_{1}$ of $D_{1}$ is {\em not invertible}, with a kernel spanned by the Killing $1$-forms on $\bS^{4}$.

\medskip

From these two decompositions, the positivity and gauge invariance of the (pseudo) state $\omega_{\eucl}$ obtained from $c_{2}^{\pm}$ can be completely analyzed.

First from the fact that ${\bf q}_{I, 2}$ equals $\tilde{\bf q}_{2}$ on $\cE_{\TT, \rm gauge}$ we obtain easily that 4) is satisfied on $\cE_{\TT, \rm gauge}$.  

Similarly one can check that 
${\bf q}_{I, 2}$ equals $-\tilde{\bf q}_{2}$ on $\cE_{\TT, 4}$ and hence $\pm {\bf q}_{I, 2}c_{2}^{\pm}$ are {\em negative} on $\cE_{\TT, 4}$, with ${\bf q}_{I, 2}(c_{2}^{+}- c_{2}^{-})$ negative definite.

Beside the non-positivity on the whole of $\cE_{\TT}$, this also shows that the {\emph strong gauge invariance}, i.e.
\beq\label{sgi}
{\bf q}_{I, 2}c_{2}^{\pm}f\in \Ran K_{21\Sigma}+ \Ran K_{20\Sigma}, \ \  f\in \cF_{\TT},
\eeq
is \emph{not} satisfied. 

At this point it is not difficult to modify  the Euclidean pseudo vacuum $\omega_{\eucl}$ to turn  it in to a well-defined, fully gauge invariant Hadamard state. In fact if $\pi$ is the projection on $\cE_{\TT, \rm gauge}\oplus \cF_{\TT, \rm gauge}$ along $\cE_{\rm TT, 4}$ then the following holds:

\begin{theoreme}[cf.~Thm.~\ref{thm5.1}]\label{thm:main2} The modified Euclidean vacuum $\omega_{\rm mod}$ defined by composition with $\pi$ is a well-defined Hadamard state for linearized gravity on global de Sitter space. It is invariant  under the action of $O(4)$. 
\end{theoreme}

Note that the modification is non-trivial, e.g.~$\omega_{\rm mod}\neq \omega_{\rm eucl}$; this is related to the failure of strong gauge invariance \eqref{sgi}. 

The modified two-point function is not invariant under the action of the full group of de Sitter isometries, though it is still invariant under de Sitter isometries which preserve $t=0$.

\medskip

We remark that in the literature \cite{moncrief,Higuchi1991,Higuchi1991a}, the space  of Killing $1$-forms for  de Sitter space has entered the discussion of perturbative quantum  gravity for an a priori different reason (in our problem it is linked to the decomposition \eqref{eq:decomposition}):   it coincides with the space of \emph{linearization instabilities}, which represents linear perturbations that are not relevant for non-linear Einstein equations.  In our construction, thanks to modifying the Euclidean vacuum state with the projection $\pi$, we do not have to remove any special subspace from the phase space of linearized gravity (for instance through imposing non-linear constraints).

\medskip

Finally, we show that the obstructions to positivity and the construction of the modified Euclidean vacuum $\omega_{\rm mod}$ are not specific to linearized gravity. In fact we carry out a similar analysis in the case of Maxwell fields, and obtain analogous conclusions. The role of  the space of problematic modes $ \cE_{\TT, 4}$  is played by as space denoted by  $\cE_{0}$ which consists of Cauchy data of solutions of the form $\alpha /\cosh^3(t) dt$, $\alpha\in \cc$.

\subsection{Plan of the paper} In Subsect.~\ref{sec0.2} we introduce notation used throughout the paper.

Sect.~\ref{sec1} explains our conventions for linearized gravity and introduces preliminaries on quantization. 

Sect.~\ref{sec2} briefly discusses Wick rotation of tensors. 

Sect.~\ref{sec3} introduces Calder\'on projectors for second order differential operators on compact manifolds with an interface. The main novelty are results in the case when the operator is not invertible (as in the case of the operator $\widetilde D_1$).

Sect.~\ref{sec4} focuses on Wick rotation of global de Sitter space. In particular it introduces key results on the spectral theory of the Wick-rotated operators $\widetilde D_2$, $\widetilde D_1$.
 
 Sect.~\ref{sec4c} studies at length the spaces $\cE_{\TT}$, $\cF_{\TT}$  and their decompositions.
 
Sect.~\ref{sec4b} shows crucial properties of the Euclidean vacuum pseudo state, including positivity on $\cE_{\rm TT, \rm gauge}$ and invariance under de Sitter symmetries, and proves Theorem \ref{thm:main1}.

Sect.~\ref{sec5} carries out the construction of the modified Euclidean vacuum state $\omega_{\rm mod}$ and proves Theorem \ref{thm:main2}.

\medskip Appendix \ref{s:appA} contains various auxiliary computations.

Appendix \ref{s:maxwell} repeats  the analysis in the case of Maxwell fields.

In Appendix \ref{s:bunch-davies} we describe the relation between  the formal construction of an Euclidean vacuum by mode expansion and the one we use based on \calde projectors. We also comment  on the formal construction of states for linearized gravity on de Sitter space by mode expansion that is found in the physics literature.
\subsection{Notation}\label{sec0.2}
We now collect various notation used throughout the paper.
We use the notation $A\defeq B$ or equivalenty $B\eqdef A$ to mean that $A$ is by definition equal to $B$.

\subsubsection{Isomorphisms of vector spaces}
If $E, F$ are vector spaces and $A\in L(E, F)$ we write $A: E\tosim F$ if $A$ is an isomorphism. If $E, F$ are topological vector spaces, we use the same notation if $A$  is a homeomorphism.
\subsubsection{Sesquilinear forms}
If $E$ is a complex vector space, its antidual is denoted by $E^{*}$. 
A sesquilinear form ${\bf A}$ on $E$ is an element of $L(E, E^{*})$ and its action on elements of $E$ is denoted by $\bar{u}\dual {\bf A}v$. 

We write ${\bf A}\in L_{\rm h/a}(E, E^{*})$ if ${\bf A}$ is Hermitian resp. anti-Hermitian.

Often $E$  is equipped with a reference Hilbertian scalar product $(\cdot| \cdot)$ and we associate to ${\bf A}$ the operator $A\in L(E)$ by $\bar{u}\dual {\bf A}v\eqdef(v| Au)$. We write $A\in L_{\rm h/a}(E)$ if $A= A^{*}$ resp. $A= - A^{*}$ where $A^{*}$ is the adjoint of $A$ for $(\cdot| \cdot)$.

\subsubsection{Operators on quotient spaces}\label{sec0.2.1}

Let $F_i\subset E_i$, $i=1,2$ be vector spaces and let $A\in L(E_1, E_2)$. Then the induced map
\[
[A]\in L( E_1/F_1, E_2/F_2),
\]
defined in the obvious way, is
\begin{equation}
\label{e0.01}
\begin{array}{rl}
1)&\hbox{well-defined if }A E_1\subset E_2\hbox{ and }A F_1\subset F_2,\\[2mm]
2)&\hbox{injective iff }A^{-1}F_2=F_1,\\[2mm]
3)&\hbox{surjective iff }E_2=A E_1+F_2. 
\end{array}
\end{equation}

\subsubsection{Sesquilinear forms on quotients}\label{sec0.2.2} 
Let now $F\subset E$ be vector spaces and let ${\bf A}\in L( E, E^*)$. We denote by $F^\circ\subset E^*$ the annihilator of $F$. Then the induced map
\[
[{\bf A}]\in L( E/F,(E/F)^*),
\]
defined as before, is
\begin{equation}
\label{e0.02}
\begin{array}{rl}
1)&\hbox{well-defined if }{\bf A}E\subset F^{\circ}, \, F\subset\Ker {\bf A},\\[2mm]
2)&\hbox{non-degenerate iff }F=\Ker{\bf A}. 
\end{array}
\end{equation}
If ${\bf A}$ is hermitian or anti-hermitian then the condition $F\subset\Ker\, {\bf A}$ implies the other one ${\bf A}E\subset F^{\circ}$ (and vice versa).

\subsubsection{Differential operators}\label{sec0.2.3}
if $V_{j}\xrightarrow{\pi}M$ are vector bundles over a manifold $M$, we denote by $\Diff^{m}(M; V_{i}, V_{j})$ the space of differential operators of order $m$ mapping smooth sections of $V_{i}$ to smooth sections of $V_{j}$.

Often $V_{j}$ is equipped with a Hermitian form on its fibers, denoted by $(\cdot| \cdot)_{V_{j}}$.

\subsubsection{A guide to the use of sub- and superscripts}\label{sec0.2.4}
- We denote by $M, V_{j}, D_{j}$ Lorentzian manifolds, vector bundles over $M$ and hyperbolic differential operators acting on sections of $V$. 

- The bundles $V_{j}$
 are usually equipped with a fiber Hermitian structure. 
 
 - Lorentzian metrics on $M$ are denoted by $\rg$. 
 
-  A differential operator in $\Diff^{m}(M; V_{i}, V_{j})$ is often denoted by $K_{ji}$.

- Sometimes  one can construct from $M$ by Wick rotation a Riemannian manifold  which will be denoted by $\tM$. The corresponding vector bundles and differential operators are denoted by $\tV_{j}, \tD_{j}, \tK_{ji}$ etc. 
 
 - The bundles $\tV_{j}$ are usually equipped with a fiber Hilbertian structure. 
 - The Riemannian metric on $\tM$ obtained from $\rg$ is denoted by $\trg$.

- If one fixes a Cauchy surface $\Sigma\subset M$, we denote by $\varrho_{j}: \cinf(M; V_{j})\to \coinf(\Sigma; V_{j}\otimes \cc^{2})$ the Cauchy data map for the hyperbolic operator $D_{j}$.

- We denote by  $U_{j}: \coinf(\Sigma; V_{j}\otimes \cc^{2})\mapsto \cinf(M; V_{j})$  the operator solving the homogeneous Cauchy problem for $D_{j}$ ie
\[
\begin{cases}
D_{j}U_{j}= 0, \\
\varrho_{j}U_{j}= \one.
\end{cases}
\]
- Often $\Sigma$ can also be considered as an hypersurface in the Wick rotated manifold $\tM$. The analogous trace map is denoted by $\tilde{\varrho}_{j}: \cinf(\tM; \tV_{j})\to \coinf(\Sigma; \tV_{j}\otimes \cc^{2})$.

- Differential operators $K_{ji}$ satisfying the identity $K_{ji}D_{i}= D_{j}K_{ji}$ have 'Cauchy surface' versions $K_{ji\Sigma}:  \coinf(\Sigma; V_{i}\otimes \cc^{2})\to  \coinf(\Sigma; V_{j}\otimes \cc^{2})$ defined by $K_{ji}= \varrho_{j}K_{ji}U_{i}$.

- As a rule objects related to the Cauchy surface $\Sigma$ will be decorated with a  subscript $_{\Sigma}$.

- usually $K_{ji\Sigma}$ have Riemannian  versions denoted by $\tilde{K}_{ji\Sigma}$.

- we extend this notation also to some subspaces of Cauchy data: if $\cE$ is some subspace of $\coinf(\Sigma; V_{j}\otimes \cc^{2})$ then $\tilde{\cE}$ will denote its Riemannian version.

\section{Linearized gravity}\label{sec1}\init
\subsection{Notation and background} \label{sec1.1}
 We start by fixing notation. Let $(M,\rg)$ be a $4$-dimensional  Lorentzian manifold.   
\subsubsection{Convention for the Riemann tensor}\label{sec1.1.1} 
We use the same convention as in e.g.~\cite{R,FH,BDM,Gerard2025} for the sign of the Riemann tensor $\rR\indices{_{abcd}}=\rR\indices{_{abc}^{e}}\rg_{ed}$, i.e.
\[
(\nab_{a}\nab_{b}- \nab_{b}\nab_{a}) u_{c}= \rR\indices{_{abc}^{d}}u_{d}
\]
on $(0,1)$-tensors in terms of the Levi-Civita connection $\nabla_a$ on $(M,\rg)$. The Ricci tensor  is the symmetric tensor
\[
\Ric_{ab}=\rR\indices{_{acb}^{c}}= \rR\indices{^{c}_{acb}},
\]
and the scalar curvature is $\scal= \rg^{ab}\Ric_{ab}$. If $\dim M=4$ then  Einstein equations with cosmological constant $\Lambda$, i.e.~
$
\Ric -\12\rg\scal +\Lambda \rg =0
$,
are equivalent to 
\beq\label{eq:einstein}
\Ric =\Lambda \rg,
\eeq 
and as a consequence of \eqref{eq:einstein} one gets $\scal=4\Lambda$.

\subsubsection{Hermitian forms on tensors}\label{sec1.1.2}
 We denote by 
\[
V_{k}\defeq\cc\otimes^{k}_{\rm s}T^{*}M
\]
 the complex bundle of symmetric $(0,k)$-tensors.  We will  only need  the cases $k=0, 1,2$.
$V_{k}$ is  equipped with the non-degenerate Hermitian form
\begin{equation}
\label{eq:uvk}
(u| u)_{V_{k}}\defeq\bar{u}\dual  k! (\rg^{\otimes k})^{-1}u.
\end{equation}
 In abstract index notation,
\[
 (u|u)_{V_{k}}= k! \,\rg^{a_{1}b_{1}}\cdots\rg^{a_{k}b_{k}}\bar{u}_{a_{1}\dots a_{k}} u_{b_{1}\dots b_{k}}.
\]
 For example for $k=2$ we have
\beq\label{e0.00}
(u|u)_{V_{2}}= 2\trace(u^{*}\rg^{-1}u\rg^{-1}).
\eeq
The $k!$ normalization differs from the most common convention, it has however the advantage that various expressions involving adjoints have a more symmetric appearance.

For $U\subset M$ open,  the Hermitian form \eqref{eq:uvk} on fibers induces a Hermitian form on compactly supported smooth sections,
\begin{equation}
\label{uvkU}
(u|v)_{V_{k}(U)}=  \int_{U}(u(x)| v(x))_{V_{k}}\dvol_{\rg}, \quad u, v\in \coinf(U;V_{k}).
\end{equation} 
The adjoint of  $A: \cinf_0(M; V_{k})\to \cinf(M; V_{l})$ for those Hermitian forms will be denoted by $A^{*}$.

If $\Sigma\subset M$ is a Cauchy surface, we set
\[
(u|v)_{V_{k}(\Sigma)}=  \int_{\Sigma}(u(x)| v(x))_{V_{k}}\dvol_{\rh}, \quad u, v\in \coinf(\Sigma;V_{k}),
\]
where $\dvol_{\rh}$ is the induced density on $\Sigma$.
\subsubsection{Decomposition of tensors}\label{sec1.1.3}
\def\Sig{\Sigma}
Let us assume that $M= I\times \Sigma$ where $I\subset \rr$ is an open interval,  $\Sigma$ a smooth manifold  with variables $(t, \rx)$ and
\[
\rg= - dt^{2}+ \rh(t, \rx)d\rx^{2},
\]
where $ \rh\in \cinf(M,  \otimes^{2}_{s}T^{*}\Sigma)$ is a smooth $t$-dependent Riemannian metric on $\Sigma$.  We set 
\[
V_{k\Sigma}= \cc\otimes_{\rm s}^{k}T^{*}\Sigma.
\]
\subsubsection{ Decomposition of \texorpdfstring{$(0,1)$-}{(0,1)-}tensors}\label{sec1.1.4}
 We identify  \beq\label{etiti.0}
 \begin{array}{l}
 \cinf(M; V_{1})\tosim\cinf(I; \cinf(\Sigma; V_{0\Sigma}))\oplus \cinf(I; \cinf(\Sigma; V_{1\Sigma}))\hbox{ by}\\[2mm]
  w\mapsto (w_{t}, w_{\Sig}), \\[2mm]
   w \eqdef w_{t}dt+ w_{\Sig}.
  \end{array}
 \eeq
  The scalar product $(\cdot| \cdot)_{V_{1}}$ reads then
 \[
 (w| w)_{V_{1}}= - | w_{t}|^{2}+ (w_{\Sig}| w_{\Sig})_{V_{1\Sigma}}=  - | w_{t}|^{2}+ (w_{\Sig}|\rh^{-1} w_{\Sig}).
 \]
\subsubsection{Decomposition of \texorpdfstring{$(0,2)$}{(0,2)}-tensors}\label{sec1.1.5}
Similarly we identify  
$$
 \cinf(M; V_{2})\tosim\cinf(I; \cinf(\Sigma; V_{0\Sigma}))\oplus \cinf(I; \cinf(\Sigma; V_{1\Sigma}))\oplus\cinf(I; \cinf(\Sigma; V_{2\Sigma}))
 $$
 by
 \beq\label{etiti.-1}
 \begin{array}{l}
  u\mapsto (u_{tt}, u_{t\Sig},u_{\Sig\Sig}), \\[2mm]
u\eqdef u_{tt}dt\otimes dt+ u_{t\Sig}\otimes dt+ dt\otimes u_{t\Sig}+ u_{\Sig\Sig}.
  \end{array}
 \eeq
The scalar product  $(\cdot| \cdot)_{V_{2}}$  reads:
\beq\label{e2.9b}
(u|u)_{V_{2}}= 2|u_{tt}|^{2}-  4(u_{t\Sig}| u_{t\Sig})_{V_{1\Sigma}} +  (u_{\Sig\Sig}| u_{\Sig\Sig})_{V_{2\Sigma}}.
\eeq

\subsection{The differential and its adjoint}\label{sec1.2}
Let  
\[
d: \begin{array}{l}
\cinf(M; V_{k})\to \cinf(M; V_{k+1})\\[1mm]
(d u)_{a_{1} \dots, a_{k+1}}= \nab_{(a_{1}}u_{a_{2}\dots, a_{k+1})},
\end{array}
\]
where $u_{(a_{1} \dots a_{k})}$ is the symmetrization of $u_{a_{1}\dots a_{k}}$,
and
\[
\delta: \begin{array}{l}
\cinf(M; V_{k})\to\cinf(M; V_{k-1})\\[1mm]
(\delta u)_{a_{1}, \dots, a_{k-1}}= -k\nab^{a}u_{aa_{1}\dots a_{k-1}}.
\end{array}
\]
With these (non-standard) conventions, we have $d^*=\delta$ w.r.t.~the Hermitian form \eqref{uvkU}.

\subsection{Operators on tensors}\label{sec1.3}
\subsubsection{Trace reversal}\label{sec1.3.1}
The operator of \emph{trace reversal} $I$
 is given by
 \[
I\defeq \one - \frac{1}{4}|\rg)(\rg|,
\]
where $\one$ is the identity and
\[
(\rg| : u_{2}\mapsto (\rg| u_{2})_{V_{2}},  \quad |\rg): u_{0}\mapsto u_{0}\rg,
\]
i.e.~$I$ is the orthogonal symmetry w.r.t.~the line $\cc \rg$. Equivalently,
\[
(Iu)_{ab}= u_{ab}- \frac{1}{2} \trace_{\rg}(u) \rg_{ab}, \quad \trace_{\rg}(u)\defeq \rg^{ab}u_{ab}= \12 (\rg| u)_{V_{2}}. 
\]
It satisfies 
\beq\label{idiot}
I^{2}= \one, \quad I= I^{*}\hbox{ on }\cinf(M; V_{2}).
\eeq
\subsubsection{Ricci operator}\label{sec1.3.2}
The \emph{Ricci operator} is
\[
\Riem(u)_{ab}\defeq \rR\indices{_{a}^{cd}_{b}}u_{cd}= \rR\indices{^{c}_{ab}^{d}}u_{cd}, \ \ u\in\cinf(M; V_{2}).
\]
The fact that $\Riem$ preserves symmetric $(0,2)$-tensors follows from the symmetries of the Riemann tensor. 
\begin{lemma}\label{lemma2.12} The Ricci operator satisfies:
 \beq\label{e2.5}
\begin{array}{rl}
i)&\Riem\, \rg=-\Ric,\\[2mm]
ii)&\Riem \circ I=  I\circ \Riem\hbox{ if }g\hbox{ is Einstein},\\[2mm]
iii)&\Riem= \Riem^{*}.
\end{array}
\eeq
\end{lemma}

\subsection{Lichnerowicz operators}\label{sec1.4}
Let  $-\square_{i}$  be the rough d'Alembertian acting on  sections of $V_{k}$, i.e.
\[
-\square_{i} u_{i}= - \rg^{ab}\nabla^{2}_{e^{a}, e^{b}}u_{i},
\]
where $(e_{a})_{0\leq a\leq d}$ is a local frame and $\nabla^{2}_{u, v}= \nabla_{u}\circ \nabla_{v}- \nabla_{\nabla_{u}v}$ for vector fields $u, v$  on $M$.

The {\em Lichnerowicz operators} acting on sections of $V_{k}$, see \cite{Lichnerowicz1961}, are defined by:
 \beq\label{e0.3}
 \begin{array}{l}
 D_{0,L}= -\square_{0},\\[2mm]
 D_{1, L}= - \square_{1} + \Ric\circ \rg^{-1} ,\\[2mm]
D_{2, L}= - \square_{2}+ \Ric\circ \rg^{-1} \circ\cdot+  \cdot\circ \rg^{-1}\circ \Ric + 2 \Riem .
 \end{array}
 \eeq
They satisfy
\[
D_{i, L}= D_{i, L}^{*}.
\]

 The proofs of the next facts can be found for instance in \cite{Boucetta1999}.  The Lichnerowicz operators satisfy
 \beq\label{e0.5}
\begin{aligned}
 D_{1, L}&=\delta_{\rm a}\circ d_{\rm a}+ d_{\rm a}\circ \delta_{\rm a}= \delta\circ d- d\circ \delta+ 2\,\Ric\circ \rg^{-1},\\
 D_{2, L}&= \delta\circ d- d\circ \delta+ 2(\Ric\circ \rg^{-1} \circ \cdot+  \cdot \circ \rg^{-1}\circ \Ric) + 4\,\Riem,
\end{aligned}
 \eeq
 where $d_{\rm a}, \delta_{\rm a}$ are the anti-symmetric differential and codifferential.
\begin{proposition}\label{prop0.-1}
 If $(M, \rg)$ is Einstein then:
 \[
 \begin{array}{l}
  D_{i+1, L}\circ d= d\circ D_{i, L}, \quad \delta \circ D_{i+1, L}= D_{i, L}\circ \delta, \ i= 0, 1\\[2mm]
  (\rg| \circ D_{2, L}= D_{0, L}\circ (\rg|,  \quad D_{2, L}\circ |\rg) = |\rg) \circ D_{0, L}.
  \end{array}
 \]
 \end{proposition}

\subsection{Linearized gravity as a gauge theory} \label{sec1.5}
Let $(M,\rg)$ be a globally hyperbolic spacetime of dimension $4$. We assume that $(M,\rg)$ is Einstein.
Let us introduce the differential operators
\beq\label{e2.4bb}
\begin{aligned}
P&\defeq - \square_{2}-  I \circ d \circ \delta + 2\,\Riem \in \Diff^2(M;V_2), \\
K&\defeq  I\circ d\in\Diff^1(M;V_1,V_2).
\end{aligned}
\eeq
$Pu=0$ is  the \emph{linearized Einstein equation} (in the trace-reversed form).  The condition $K^\star u=0$, where $K^{\star}$ is defined below,  is the linearized \emph{de Donder} or \emph{harmonic gauge}.

\subsubsection{Physical Hermitian form}\label{sec1.5.1}
We consider $V_{k}$, $k=0,1,2$ as Hermitian bundles, where the Hermitian forms on fibers is now
\beq\label{def-physical-scalar-product}
(u| v)_{I,V_{k}} \defeq (u|v)_{V_{k}},\ k= 0, 1, \quad  (u| v)_{I,V_{2}} \defeq (u| I v)_{V_{2}}. 
\eeq
The corresponding Hermitian form on smooth sections of $V_k$, $k=1,2$ is 
\begin{equation}\label{eq:hform}
(u|v)_{I,V_{k}(U)}=  \int_{U}(u(x)| v(x))_{I,V_{k}}\dvol_\rg, \quad u, v\in \coinf(U;V_k).
\end{equation}

We denote by $A^{\star}$ the corresponding formal adjoint of $A$ for $(\cdot| \cdot)_{I, V_{k}(M)}$ to distinguish it from the formal adjoint $A^*$ for   $(\cdot|\cdot)_{V_{k}(M)}$.  The two are related as follows:
\beq\label{e2.7}
\begin{array}{rl}
A^{\star}= I A^{*}I \hbox{ if }A: \cinf(M;V_{2})\to \cinf(M;V_{2}),\\[2mm]
A^{\star}= A^{*}I\hbox{ if }A: \cinf(M;V_{k})\to \cinf(M;V_{2}), \ k= 0, 1,\\[2mm]
A^{\star}= IA^{*}\hbox{ if }A: \cinf(M;V_{2})\to \cinf(M;V_{k}),\ k= 0, 1,\\[2mm]
A^{\star}= A^{*}\hbox{ if }A: \cinf(M;V_{i})\to \cinf(M;V_{j}), \ i, j\neq 2.
\end{array}
\eeq 
In particular,
\beq\label{eq:Kstar}
K^{\star}= K^{*}I=\delta \circ I \circ I=\delta.
\eeq 
\subsubsection{Operators in linearized gravity}\label{sec1.5.1b}
Let us set:
\[
D_{k}\defeq D_{k, L}- 2\Lambda, \ k=0, 1,2.
\]
Then 
\beq\label{e0.6}
\begin{array}{l}
 K^\star K= D_{1} =-\square_{1}-\Lambda= \delta\circ  d- d\circ \delta,\\[2mm]
P+KK^\star= D_2=-\square_{2} +2\Riem= \delta\circ  d- d\circ \delta+  4 \Riem + 2 \Lambda.
\end{array}
\eeq
The operator $D_{0}$ is useful in connection with the {\em traceless gauge}.

\subsubsection{Cauchy problem}\label{sec1.5.2}
Let $\Sigma\subset M$ a smooth space-like Cauchy surface.  
For $k=0, 1,2$  we set
\[
\varrho_{k}u= \col{u\traa{\Sigma}}{\i^{-1}\nabla_{\nu}u\traa{\Sigma}}= \col{f_{0}}{f_{1}}, \ u\in \cinf_{\sc}(M; V_{k}),
\]
where $\nu$ is the future directed unit normal to $\Sigma$. 

We denote by $U_{k}$ the operator which assigns to a set of Cauchy data  the corresponding solution in $\Kersc  D_{k}$, i.e.~$U_{k}$ is the inverse of $\varrho_{k} : \Kersc  D_{k}\to\coinf(\Sigma;V_{k}\otimes \cc^{2})$.  In other terms,
\beq\label{e00.3}
\begin{cases}
D_{k}U_{k}= 0, \\
\varrho_{k}U_{k}= \one.
\end{cases}
\eeq
\subsubsection{Conserved charges}\label{sec1.5.3}

Let $G_k$ be the causal propagator of $D_k$, i.e.~the difference of the retarded and advanced propagator. 
There exist  a unique Hermitian form ${\bf q}_{k}: \coinf(\Sigma; V_{k}\otimes \cc^{2})\to \coinf(\Sigma; V_{k}\otimes \cc^{2})^{*}$ called the {\em charge} of $D_{k}$,  such that 
\[
(\phi_{k}|\i G_{k}\phi_{k})_{V_{k}(M)}= \overline{\varrho_{k}u_{k}}\dual {\bf q}_{k}\varrho_{k}u_{k}
\]
for $\phi_{k}\in \coinf(M; V_{k})$ and $u_{k}= G_{k}\phi_{k}\in \Kersc D_{k}$.
By  \cite[Lemma 2.2]{Gerard2025}, one can compute ${\bf q}_{k}$ using the identity
$$
\begin{aligned}
&(u_{k}| D_{k}v_{k})_{V_{k}(J_{\pm}(\Sigma))}- (D_{k}u_{k}| v_{k})_{V_{k}(J_{\pm}(\Sigma))}\\
&= \pm \i^{-1}\overline{\varrho_{k}u_{k}}\dual {\bf q}_{k}\varrho_{k}u_{k}, \ \ u_{k}, v_{k}\in \coinf(M; V_{k}).
\end{aligned}
$$

\subsubsection{Operators on  Cauchy data and physical charge}\label{sec1.6.1}
We follow here \cite[Subsect.~2.4]{GW1}.

To the operator $K$ we associate an operator $K_{\Sigma}$ acting on Cauchy data by setting
\beq\label{eq:relc}
K_{\Sigma}\defeq \varrho_2 K U_1.
\eeq
Similarly, since $[I, D_{2}]=0$ we can define
\beq\label{eq:reld}
I_{\Sigma}\defeq \varrho_{2}I U_{2}= I\otimes \cc^{2}.
\eeq
We obtain that  ${\bf q}_{2}I_{\Sigma}= I_{\Sigma}^{*}{\bf q}_{2}$ and 
as in \ref{sec1.5.1} we define the Hermitian form
\[
 {\bf q}_{I, 2}\defeq {\bf q}_{2}\circ  I_{\Sigma},
\]
called the {\em physical charge} for $D_{2}$. 

We denote by  $K^{\dagger}_{\Sigma}$  the adjoint of $K_{\Sigma}$ for the Hermitian forms ${\bf q}_{1}, {\bf q}_{I, 2}$, i.e.
\beq\label{eq:defkdag}
\overline{K_{\Sigma}^{\dag}f}_{2}\dual  {\bf q}_{1}f_{1}= \overline{f}_{2}\dual {\bf q}_{I, 2} K_{\Sigma}f_{1}, \ \ f_{k}\in \coinf(\Sigma, V_{k}\otimes \cc^{2}).
\eeq
We have:
\[
K_{\Sigma}^{\dag}= \varrho_{1}K^{\star}U_{2}.
\]
\begin{lemma}\label{lem:cauchyrel}We have:
\ben
\item\label{cauchyrelit1} $K \circ U_1 = U_2 \circ  K_\Sigma$, $K^\star \circ  U_2=U_1 \circ  K^{\dagger}_{\Sigma}$;
\item\label{cauchyrelit0} $\varrho_2  \circ K =K_{\Sigma} \circ \varrho_1$ on $\Kersc  D_1$,  $\varrho_1 \circ  K^\star=K^{\dagger}_{\Sigma} \circ \varrho_2$ on $\Kersc  D_2$;
\item\label{cauchyrelit4} $K^{\dagger}_{\Sigma}  \circ K_{\Sigma}=0$.
\een
\end{lemma}

\subsection{Phase spaces}\label{sec1.6} We use the notation $\Ker_{\rm c}$ and $\Ran_{\rm c}$ (resp.~$\Ker_{\rm sc}$ and $\Ran_{\rm sc}$) to denote the kernel and range of differential operators acting on  compactly supported (resp.~space-compact)  smooth tensors.

\begin{proposition}\label{prop:Gpasses} The maps
\[
\begin{array}{l}
[G_2]:\dfrac{\Kerc  K^\star}{\Ranc  P}\longrightarrow \dfrac{\Kersc  P}{\Ransc  K},\\[3mm]
[\iota]:\dfrac{\Kersc  D_2\cap\Kersc  K^\star}{K\Kersc D_{1}}     \longrightarrow \dfrac{\Kersc  P}{\Ransc  K},
\end{array}
\]
where $\iota: \Kersc  D_2\cap\Kersc  K^\star\to \Kersc P$ is the canonical injection,  are well defined and bijective. 
\end{proposition}
Let us define the Hermitian forms ${\bf Q}_{k}$ on $\coinf(M; V_{k})$: 
\[
 \bar{u_{k}}\dual {\bf Q}_{k}u_{k}\defeq \i (u_{k}| G_{k}u)_{V_{k}(M)},
\]
and the {\em physical charge}
\[
 {\bf Q}_{I,2}\defeq {\bf Q}_{2}\circ I= I^{*}\circ {\bf Q}_{2}.
 \]
The trace reversal $I$ appears 

\begin{definition}
The \emph{physical phase space} is  the Hermitian space $(\cV_{P}, {\bf Q}_{P})$, where:
\[
\cV_{P}= \frac{\Kerc  K^\star}{\Ranc  P}, \quad [\bar{u}]\dual {\bf Q}_{P}[v]=\bar{u}\dual {\bf Q}_{I, 2}v, \ \ [u], [v]\in  \frac{\Kerc  K^\star}{\Ranc  P}.
\]
\end{definition}

${\bf Q}_{P}$ is a well-defined Hermitian form on $\cV_P$. 

%
%
%
%
\subsubsection{Phase space of Cauchy data}\label{sec1.6.2}
\begin{proposition}\label{prop:rhopasses} The induced map
\[
[\varrho_2]: \ \dfrac{\Kersc  D_{2}\cap\Kersc  K^{\star}}{K\Kersc D_{1} }\longrightarrow \dfrac{\Kerc K_{\Sigma}^{\dagger}}{\Ranc  K_\Sigma}
\]
is well defined and bijective.
\end{proposition}

\begin{proposition}\label{prop-added}
 The map 
 \[
 [\varrho_{2}G_{2}]: \Big(\dfrac{\Kerc  K^\star}{\Ranc  P}, {\bf Q}_{P}\Big)\longrightarrow\Big(\dfrac{\Kerc  K^{\dagger}_{\Sigma}}{\Ranc  K_\Sigma},{\bf q}_{I, 2}\Big)
 \]
 is an isomorphism of Hermitian spaces.
\end{proposition}
\subsection{Quantization}\label{sec1.7}

 The algebraic quantization of linear gauge theories is discussed in detail in \cite[Sect.~3]{GW1}. The algebraic framework reduces the quantization problem to showing the existence of physically relevant quantum states on the  CCR $*$-algebra $\CCR(\cV_{P}, {\bf Q}_{P})$ associated to the Hermitian space $(\cV_{P}, {\bf Q}_{P})$ defined in Subsect.~\ref{sec1.6}.   The notions of quasi-free states and covariances (or two-point functions) are explained in  \cite[Sect.~3]{GW1} and references therein.

\subsubsection{Covariances}\label{sec1.7.1}
In the complex formalism, a quasi-free state on the CCR $*$-algebra $\CCR(\cV_{P}, {\bf Q}_{P})$ is determined by a pair ${\bf \Lambda}_{P}^{\pm}$ of {\em covariances}, i.e.~of Hermitian forms on $\cV_{P}$  such that \[
\begin{cases}
{\bf \Lambda}_{P}^{\pm}\geq 0,\\
{\bf \Lambda}_{P}^{+}- {\bf \Lambda}_{P}^{-}= {\bf Q}_{P}.
\end{cases}
\]
We will consider quasi-free states $ \omega$ on  $\CCR(\cV_{P}, {\bf Q}_{P})$ with covariances obtained  from a  pair of continuous Hermitian forms ${\bf \Lambda}_{2}^{\pm}$ on $\coinf(M; V_{2})$ (called  the {\em spacetime covariances} of $\omega$) by:
 \beq\label{e00.1}
 [\bar{u}]\dual {\bf \Lambda}_{P}^{\pm}[u]= \bar{u}\dual  {\bf \Lambda}_{2}^{\pm}u, \ \ [u]\in \frac{\Kerc  K^\star}{\Ranc  P}.
 \eeq
\begin{lemma}\label{lemma.had}
 Suppose  that ${\bf \Lambda}_{2}^{\pm}\in L(  \coinf(M; V_{2}), \coinf(M; V_{2})^{*})$ are such that:
  \beq\label{defodefo}
\begin{aligned}
i)&\quad  D_{2}^{*} \circ {\bf \Lambda}^\pm_{2}={\bf \Lambda}^\pm_{2}  \circ D_{2} =0 , \\ 
ii)&\quad {\bf \Lambda}_2^+-{\bf \Lambda}^-_2 = {\bf Q}_{I, 2} \hbox{ on }\Kerc  K^{\star},\\
iii)&\quad {\bf \Lambda}_{2}^{\pm}=0\hbox{ on }\Kerc  K^{\star}\times \Ranc K K^\star,\\
v) & \quad {\bf \Lambda}_{2}^{\pm}= {\bf \Lambda}_{2}^{\pm*}, \ {\bf \Lambda}^\pm_{2}  \geq 0  \hbox{ on } \Kerc  K^{\star}.
\end{aligned}
\eeq
Then ${\bf \Lambda}_{2}^{\pm}$ are the covariances  of a quasi-free state on $\CCR(\cV_{P}, {\bf Q}_{P})$.\end{lemma}

Lemma \ref{lemma.had} (its proof is straightforward) has appeared in a very similar form in \cite{GW1,Gerard2025,Gerard2024}, but with $iii)$ replaced by the stronger condition 
$$
 {\bf \Lambda}_{2}^{\pm}=0\hbox{ on }\Kerc  K^{\star}\times \Ranc K.
$$
 The weak gauge-invariance condition $iii)$ is actually sufficient, see the discussion around \eqref{eq:gi} in Subsect.~\ref{ss:intro}.
 
\subsubsection{Hadamard condition} \label{sec1.7.2}
We use  the following definition of Hadamard states  \cite{SV},  cf.~\cite[Subsect.~3.4]{GW1} and references therein. The general consensus is that only states satisfying the  \emph{Hadamard condition} (the \emph{Hadamard states}) are  physical. 
We recall that
 $$
 \cN=\{ (x,\xi)\in T^*M\setminus\zero :  \xi\cdot \rg^{-1}(x)\xi = 0 \}
 $$
is  the  characteristic set of the wave operator on $(M,\rg)$, and 
$$
\cN^{\pm}=\cN\cap \{(x, \xi):\pm v\dual \xi>0 \,\ \forall v\in T_{x}M\hbox{ future-directed time-like}\}
$$
are its  two connected components.

To formulate the Hadamard condition, we need to identify  the Hermitian forms ${\bf \Lambda}_{2}^{\pm}$  with distributional kernels  $\Lambda_{2}^{\pm}(\cdot, \cdot)\in \cD'(M\times M; L(V_{2}))$, called  {\em two-point functions}.  

This identification is equivalent to embedding $\coinf(M; V_{2})$ in $\cD'(M; V_{2})$ and is defined by the identity
\beq\label{e00.2bb}
\bar{u}\dual {\bf \Lambda}^{\pm}_{2}v\eqdef\int_{M\times M}(u(x)| \Lambda_{2}^{\pm}(x, y)v(y))_{V_{2}}\dvol_{\rg}(x)\dvol_{\rg}(y)
\eeq
for all $u, v\in \coinf(M; V_{2})$. 
One can of course use other Hermitian forms on  the fibers of $V_{2}$ to do this identification, like for example $(\cdot| \cdot)_{I, V_{2}}$. This change amounts  to composing  $\Lambda_{2}^{\pm}(x, y)$ by smooth  linear operators acting on the fibers of $V_{2}$ over $x$ and $y$ and it does not change the Hadamard condition \eqref{defidefi} below.

Note that most of the literature on QFT on curved spacetimes uses the real formalism. The  commonly used \emph{real two-point functions} $\omega_2(x,y)$ are  obtained by the formula $\omega_2=\Re (\Lambda^\pm_2 \pm \12 Q_{I,2})$, where $Q_{I,2}$ is the distribution obtained from ${\bf Q}_{I,2}$ using the above identification, see e.g.~\cite[Sect.~4]{G}.

\begin{definition}\label{defhadama}
  A quasi-free state $\omega$ on $\CCR(\cV_{P}, {\bf Q}_{P})$ given by  covariances $\boldsymbol\Lambda_{2}^{\pm}$ as in Lemma \ref{lemma.had} is {\em Hadamard} if in addition to \eqref{defodefo} it satisfies:
\beq\label{defidefi}
 \WF(\Lambda_{2}^{\pm})'\subset \cN^{\pm}\times \cN^{\pm}.
 \eeq
\end{definition}

\subsubsection{Hadamard condition on a Cauchy surface}\label{sec1.7.3}
One can equivalently consider Hermitian forms  $\boldsymbol\lambda^{\pm}_{2\Sigma}$  on  the space of Cauchy data $\coinf(\Sigma; V_{2}\otimes \cc^{2})$ called {\em Cauchy surface covariances}.
Namely   if we have a pair of Hermitian forms 
\[
\boldsymbol\lambda_{2\Sigma}^{\pm}\in L(\coinf(\Sigma; V_{2}\otimes \cc^{2}), \coinf(\Sigma; V_{2}\otimes \cc^{2})^{*}),
\]
  then we  set
 \beq\label{e00.2b}
 {\bf \Lambda}_{2}^{\pm}= (\varrho_{2}G_{2})^{*}\boldsymbol\lambda_{2\Sigma}^{\pm}(\varrho_{2}G_{2})\in L(  \coinf(M; V_{2}), \coinf(M; V_{2})^{*}).
 \eeq
Note that $\boldsymbol\lambda_{2\Sigma}^{\pm}$ generate covariances $[\boldsymbol\lambda_{2\Sigma}^{\pm}]$ on   $({\Kerc K^{\dagger}_{\Sigma}}/{\Ranc K_\Sigma},{\bf q}_{I, 2})$ iff
  \beq
  \label{defodefo-cauchy}
  \begin{aligned}
 i)&\quad \boldsymbol\lambda_{2\Sigma}^{+}-\boldsymbol\lambda_{2\Sigma}^{-} = {\bf q}_{I, 2} \hbox{ on }\Kerc K_{\Sigma}^{\dag},\\
 ii)&\quad \boldsymbol\lambda_{2\Sigma}^{\pm}= 0\hbox{ on }\Kerc K_{\Sigma}^{\dag}\times \Ranc K_{\Sigma},\\
 iii) & \quad \boldsymbol\lambda_{2\Sigma}^{\pm}= \boldsymbol\lambda_{2\Sigma}^{\pm*}, \ \boldsymbol\lambda^\pm_{2\Sigma}  \geq 0 \mbox{ on } \Kerc K_{\Sigma}^{\dag}.
 \end{aligned}
  \eeq
We will call property  \eqref{defodefo-cauchy} {\it ii)} the {\em strong gauge invariance} property.
 
 However, in order for the spacetime covariances ${\bf \Lambda}_{2}^{\pm}$ given by  \eqref{e00.2b} to satisfy \eqref{defodefo} the following weaker conditions are 
 sufficient:
  \beq
  \label{defodefo-cauchy-weak}
  \begin{aligned}
 i)&\quad \boldsymbol\lambda_{2\Sigma}^{+}-\boldsymbol\lambda_{2\Sigma}^{-} = {\bf q}_{I, 2} \hbox{ on }\Kerc K_{\Sigma}^{\dag},\\
 ii)&\quad \boldsymbol\lambda_{2\Sigma}^{\pm}= 0\hbox{ on }\Kerc K_{\Sigma}^{\dag}\times \Ranc K_{\Sigma}K_{\Sigma}^{\dag},\\
 iii) & \quad \boldsymbol\lambda_{2\Sigma}^{\pm}= \boldsymbol\lambda_{2\Sigma}^{\pm*}, \ \boldsymbol\lambda^\pm_{2\Sigma}  \geq 0 \mbox{ on } \Kerc K_{\Sigma}^{\dag}.
 \end{aligned}
  \eeq
 We call property  \eqref{defodefo-cauchy-weak} {\it ii)} the {\em weak gauge invariance} property.

Since ${\bf q}_{I, 2}$ is non-degenerate, we can also define $c_{2}^{\pm}:\coinf(\Sigma;V_{2}\otimes \cc^{2})\to \cinf(\Sigma;V_{2}\otimes \cc^{2})$ by
\beq\label{e00.4}
\boldsymbol\lambda^{\pm}_{2\Sigma}\eqdef \pm {\bf q}_{I, 2}\circ  c_{2}^{\pm}.
\eeq
\begin{proposition}\label{prop:states1} Conversely, suppose $c_2^\pm: \coinf(\Sigma;V_{2}\otimes \cc^{2})\to \cinf(\Sigma;V_{2}\otimes \cc^{2})$ is a pair of operators such that:
\beq\label{condit-hada}
\begin{array}{rl}
i)&c_2^++c_2^-=\one,\\[2mm]
ii)&c_2^\pm : \Ranc K_{\Sigma} K_{\Sigma}^\dag\to \Ran K_{\Sigma}, \\[2mm]
iii)&{\bf q}_{I, 2}\circ  c_{2}^{\pm}= c_{2}^{\pm*}\circ  {\bf q}_{I, 2}, \quad \pm {\bf q}_{I, 2} \circ  c^\pm_2 \geq 0 \hbox{ on } \Kerc  K_{\Sigma}^\dag.\\[2mm]
\end{array}
\eeq
Then ${\bf \Lambda}_{2}^{\pm}$ given by \eqref{e00.2b} and \eqref{e00.4} are the covariances of a quasi-free state on $\CCR(\cV_{P}, {\bf Q}_{P})$. Furthermore if for some neighborhood $\cU$ of $\Sigma$ in $M$ we have: 
\beq\label{eq:Hadamard}
{\it iv)}\ \ \WF(U_{2}\circ c_{2}^{\pm})'\subset (\cN^{\pm}\cup\cF)\times T^{*}\Sigma
\eeq over $\cU\times \Sigma$,
where $\cF\subset T^{*}M$ is a conic set with $\cF\cap \cN= \emptyset$, 
then the associated state  is Hadamard. 
\end{proposition}
The proof of \eqref{condit-hada} is analogous to the one found in \cite[Sect.~3.4]{GW1}. The proof of the statement on the Hadamard condition can be found in \cite[Sect.~11.1]{G}.

\begin{remark}
 Condition \eqref{condit-hada} {\it iii)} on $c_{2}^{\pm}$ is equivalent  to the weak gauge invariance property  \eqref{defodefo-cauchy-weak} {\it iii)}. The corresponding condition for the strong gauge invariance property  \eqref{defodefo-cauchy} {\it iii)} is
 \[
 c_2^\pm : \Ranc K_{\Sigma}\to \Ran K_{\Sigma}.
 \]
 \end{remark}

\begin{remark}
If $[c_{2}^{\pm}, I_{\Sigma}]=0$ then we can replace  the first condition in \eqref{condit-hada} {\it iii)} by the simpler
\[
{\bf q}_{2}\circ  c_{2}^{\pm}= c_{2}^{\pm*}\circ  {\bf q}_{2}.
\]
Conversely 
 if $c_{2}^{\pm}$ satisfy the conditions in Prop.~\ref{prop:states1} then  setting
\[
\hat{c}_{2}^{\pm}= \12(c_{2}^{\pm}+ I_{\Sigma}\circ  c_{2}^{\pm}\circ  I_{\Sigma}),
\]
we obtain that $\hat{c}_{2}^{\pm}$ also satisfies  the conditions in Prop.~\ref{prop:states1} and $[\hat{c}_{2}^{\pm}, I_{\Sigma}]=0$. The only point deserving some attention is the microlocal condition {\it iv)}, which follows from the fact that $\WF(I)'$ and $\WF(I_{\Sigma})'$ are included in the diagonal of $T^{*}M\times T^{*}M$ and $T^{*}\Sigma\times T^{*}\Sigma$ respectively.
\end{remark}
\subsection{TT gauge condition}\label{sec1.8}
We now explain  how  to add the traceless condition $(\rg| u_{2})_{V_{2}}=0$ to the harmonic gauge condition $\delta u_{2}=0$, following \cite{FH}.
These two conditions together are called the {\em TT gauge condition}. 

It is  a  natural gauge fixing condition, as it does not single out a particular Cauchy surface and is preserved by all isometries of the spacetime $(M, \rg)$.

In the rest of the paper we will set
\[
\begin{array}{l}
K_{21}= K= I\circ d: \cinf(M; V_{1})\to \cinf(M; V_{2}),\\[2mm]
K_{20}= |\rg): \cinf(M; V_{0})\to \cinf(M; V_{2}),\\[2mm]
K_{10}= d: \cinf(M; V_{0})\to \cinf(M; V_{1}).
\end{array}
\]
The adjoints defined as in \eqref{e2.7}, are
\[
K_{21}^{\star}= \delta,\quad K_{20}^{\star}= -(\rg|, \quad K_{10}^{\star}= \delta,
\]
and we have
\begin{equation}
\label{e4.10}
D_{i}K_{ij}= K_{ij}D_{j}, \quad K_{20}^{\star}K_{21}= - 2 K_{10}^{\star}, 
\end{equation}
as well as the  adjoint identities, using that $D_{k}= D_{k}^{\star}$.
\subsubsection{Operators on Cauchy data}\label{sec1.8.1}
In addition to $K_{21\Sigma}$, $K_{21\Sigma}^{\dag}$ we can   define more generally $K_{ij\Sigma}, K_{ij\Sigma}^{\dag}$, by
\[
K_{ij\Sigma}\defeq\varrho_{i}K_{ij}U_{j}, \quad K_{ij\Sigma}^{\dag}\defeq \varrho_{j}K_{ij}^{\star}U_{i}.
\]
We then have:
\[
\begin{array}{l}
 \bar{f}_{2}\dual {\bf q}_{I, 2}K_{20\Sigma}f_{0}= \bar{K_{20\Sigma}^{\dag}f}_{2}\dual {\bf q}_{0}f_{0},\\[2mm]
 \bar{f}_{1}\dual {\bf q}_{1}K_{10\Sigma}f_{0}= \bar{K_{10\Sigma}^{\dag}f}_{1}\dual {\bf q}_{0}f_{0},
\end{array}
\]
  for $f_{i}\in \cinf_{\c}(\Sigma; V_{i}\otimes \cc^{2})$.
\subsubsection{Imposing the TT gauge}\label{sec1.8.2}

The TT gauge is imposed by intersecting the corresponding spaces  with $\Kersc K_{20}^{\star}$.

We assume in the rest of this subsection that $\Lambda\neq 0$ and we set $S_{0}\defeq -(4\Lambda)^{-1}K_{10}K_{20}^{\star}$. We denote by $\iota:\Kersc D_{2}\cap \Kersc K_{21}^{*}\cap \Kersc K_{20}^{\star}\to \Kersc D_{2}\cap \Kersc K_{21}^{*}$ the canonical injection.

\begin{proposition}\label{prop4.10}
 Assume that  $\Lambda\neq 0$. Then the quotient map $[\iota]$:
  \[
\bigg(\frac{\Kersc D_{2}\cap \Kersc K_{21}^{\star}\cap \Kersc K_{20}^{\star}}{K_{21}\Kersc D_{1}\cap \Kersc K_{20}^{\star}}, {\bf q}_{I, 2}\bigg)\longrightarrow \bigg(\frac{\Kersc D_{2}\cap \Kersc K_{21}^{\star}}{K_{21}\Kersc D_{1}}, {\bf q}_{I, 2}\bigg)
 \]
 is an isomorphism of Hermitian spaces with inverse
  \[
  [\one - K_{21}S_{0}]: \frac{\Kersc D_{2}\cap \Kersc K_{21}^{*}}{K_{21}\Kersc D_{1}}\longrightarrow\frac{\Kersc D_{2}\cap \Kersc K_{21}^{*}\cap \Kersc K_{20}^{\star}}{K_{21}\Ker_{\sc}D_{1}\cap \Ker_{\sc}K_{20}^{\star}}.
  \]
\end{proposition}
\begin{remark}\label{remark1}
 Since $K_{20}^{\star}K_{21}= -2 K_{10}^{\star}$ we have
 \[
 K_{21}\Kersc D_{1}\cap \Kersc K_{20}^{\star}= K_{21}(\Kersc D_{1}\cap \Kersc K_{10}^{\star}).
 \]
\end{remark}
\refproof{Prop.~\ref{prop4.10}}  
Note the identities 
\begin{equation}
\label{edd.e1}
D_{1}S_{0}= S_{0}D_{2}, \quad K_{20}^{\star}K_{21}S_{0}= K_{20}^{\star}+ (2\Lambda)^{-1}K_{20}^{\star}D_{2}, 
\end{equation}
obtained using \eqref{e4.10} and the fact that $D_{0}= K_{10}^{\star}K_{10}- 2\Lambda$. They imply that $[\iota]$ and $[\one - K_{21}S_{0}]$ are well defined. Clearly $(\one - K_{21}S_{0})\circ \iota= \one$. Moreover using \eqref{edd.e1} one obtains also that $\iota\circ (\one - K_{21}S_{0})= \one$ on $\Ker_{\sc}D_{2}$, modulo $K_{21}\Kersc D_{1}$. Therefore $[\iota]$ and $[\one- K_{21}S_{0}]$ are each other's inverses. \qed

\subsubsection{Phase space of Cauchy data}\label{sec1.8.3}
We have seen  in Prop.~\ref{prop:rhopasses} that 
\beq\label{e4.12b}
[\varrho_{2}]: \Big(\frac{\Kersc D_{2}\cap \Kersc K_{21}^{\dag}}{K_{21}\Kersc D_{1}}, {\bf q}_{I, 2}\Big)\to \Big(\frac{\Kerc K_{21\Sigma}^{\dag}}{\Ranc K_{21\Sigma}}, {\bf q}_{I, 2}\Big)
\eeq
is an isomorphism of Hermitian spaces. The same argument shows that
\beq\label{e4.12a}
\begin{aligned}
[\varrho_{2}]& : \Big(\frac{\Kersc D_{2}\cap \Kersc K_{21}^{\star}\cap \Kersc K_{20}^{\star}}{K_{21}\Kersc D_{1}\cap \Kersc K_{20}^{\star}}, {\bf q}_{I, 2}\Big) \\ & \quad \to \Big(\frac{\Kerc K_{21\Sigma}^{\dag}\cap \Kerc K_{20\Sigma}^{\dag}}{\Ranc K_{21\Sigma}\cap \Kerc K_{20\Sigma}^{\dag}}, {\bf q}_{I, 2}\Big)
\end{aligned}
\eeq
is an isomorphism of Hermitian spaces.

Note that $\rho_{2}S_{0}U_{2}\defeq S_{0\Sigma} =(-4\Lambda)^{-1}K_{10\Sigma}K_{20\Sigma}^{\dag}$. The Cauchy surface version of Prop.~\ref{prop4.10} is then the statement that 
 \[
 [\iota]:\Big(\frac{\Kerc K_{21\Sigma}^{\dag}\cap \Kerc K_{20\Sigma}^{\dag}}{\Ranc K_{21\Sigma}\cap \Kerc K_{20\Sigma}^{\dag}}, {\bf q}_{I, 2}\Big)\to \Big(\frac{\Kerc K_{21\Sigma}^{\dag}}{\Ranc K_{21\Sigma}}, {\bf q}_{I, 2}\Big)
 \]
is an isomorphism of Hermitian spaces with inverse
 \[
 [\one - K_{21\Sigma}S_{0\Sigma}]: \frac{\Kerc K_{21\Sigma}^{\dag}}{\Ran_{\c}K_{21\Sigma}}\to 
\frac{\Kerc K_{21\Sigma}^{\dag}\cap \Kerc K_{20\Sigma}^{\dag}}{\Ranc K_{21\Sigma}\cap \Kerc K_{20\Sigma}^{\dag}}.
\]

\begin{definition}\label{def-de-phase-spaces}
 We set
 \[
 \begin{array}{l}
 \cE_{\rm TT}\defeq \Kerc K_{21\Sigma}^{\dag}\cap \Kerc K_{20\Sigma}^{\dag},\\[2mm]
 \cF_{\rm TT}\defeq \Ranc K_{21\Sigma}\cap \Kerc K_{20\Sigma}^{\dag},\\[2mm]
  \cF_{\rm TT, w}\defeq \Ranc K_{21\Sigma}K_{21\Sigma}^{\dag}\cap \Kerc K_{20\Sigma}^{\dag},\\[2mm]
 \end{array}
 \]
 \end{definition}
 
By the previous discussion, the physical phase space is isomorphic to $(\cE_{\rm TT}/\cF_{\rm TT},{\bf q}_{I, 2})$.
Note that $\cF_{\rm TT}= K_{21\Sigma}\Kerc K_{10\Sigma}^{\dag}$ by \eqref{e4.10}, and ${\bf q}_{I, 2}= {\bf q}_{2}$ on $\cE_{\rm TT}$.  The subspace $\cF_{\rm TT, w}$ represents the restricted gauge transformations relevant for the weak gauge invariance condition expressed in different forms in \eqref{defodefo}, \eqref{defodefo-cauchy-weak} and \eqref{condit-hada}.

\subsubsection{Hadamard states in the TT gauge}\label{sec1.8.4}
Assume as in \ref{sec1.7.3} that we have a pair of Hermitian forms $\boldsymbol\lambda_{2\Sigma}^{\pm}\in L(\coinf(\Sigma; V_{2}\otimes \cc^{2}), \coinf(\Sigma; V_{2}\otimes\cc^{2})^{*})$. 

 They induce  covariances on $({\cE_{\rm TT}}/{\cF_{\rm TT}}, {\bf q}_{I, 2})$ if
 \beq
 \label{defodefo-cauchy-TT}
 \begin{aligned}
i)&\quad \boldsymbol\lambda_{2\Sigma}^{+}-\boldsymbol\lambda_{2\Sigma}^{-} = {\bf q}_{2} \hbox{ on }\cE_{\rm TT},\\
ii)&\quad \boldsymbol\lambda_{2\Sigma}^{\pm}= 0\hbox{ on }\cE_{\rm TT}\times \cF_{\rm TT},\\
iii) & \quad \boldsymbol\lambda_{2\Sigma}^{\pm}= \boldsymbol\lambda_{2\Sigma}^{\pm*}, \   \boldsymbol\lambda^\pm_{2\Sigma}  \geq 0 \mbox{ on } \cE_{\rm TT}.
\end{aligned}
 \eeq

 With  $c_{2}^{\pm}$ defined in \eqref{e00.4} the conditions in \eqref{defodefo-cauchy-TT} become:
\begin{equation}
\label{edd.e4}
 \begin{aligned}
i)&\quad c_{2}^{+}+ c_{2}^{-}= \one \hbox{ on }\cE_{\rm TT},\\
ii)&\quad c_{2}^{\pm}: \cF_{\rm TT}\to \Ran K_{21\Sigma}+ \Ran K_{20\Sigma},\\
iii) & \quad c_{2}^{\pm}{\bf q}_{I, 2}= {\bf q}_{I, 2}c_{2}^{\pm}, \  \pm \bar{f}\dual {\bf q}_{I, 2}c_{2}^{\pm}f\geq 0 \hbox{ on }\cE_{\rm TT}.
\end{aligned}
\end{equation}
If $[c_{2}^{\pm}, I_{\Sigma}]=0$ then  we can replace \eqref{edd.e4} {\it iii)} by
\beq\label{edd.e5}
\begin{aligned}
iii) & \quad c_{2}^{\pm}{\bf q}_{2}= {\bf q}_{2}c_{2}^{\pm}, \ \pm \bar{f}\dual {\bf q}_{2}c_{2}^{\pm}f\geq 0 \hbox{ on }\cE_{\rm TT},
\end{aligned}
\end{equation}
since $I_{\Sigma}= \one$ on $\Kerc K_{20\Sigma}^{\dag}$.

Similarly as before will call property  \eqref{defodefo-cauchy-TT} {\it ii)} or equivalently \eqref{edd.e4} {\it ii)}  the {\em strong  TT gauge invariance property}.

\subsubsection{Spacetime covariances} \label{sss:newTTc}
We now set
\beq
\label{edd.e3}
{\bf \Lambda}_{2,{\rm TT}}^{\pm}= (\varrho_{2}(\one - K_{21}S_{0})G_{2})^{*}\boldsymbol\lambda_{2\Sigma}^{\pm}(\varrho_{2}(\one - K_{21}S_{0})G_{2}).
\eeq

\begin{proposition}\label{prop:newtt}
Assume that  
\begin{equation}
\label{edd.e6}
 \begin{aligned}
i)&\quad c_{2}^{+}+ c_{2}^{-}= \one \hbox{ on }\cE_{\rm TT},\\
ii)&\quad c_{2}^{\pm}:\cF_{\rm TT,w}\to \Ran K_{21\Sigma}+ \Ran K_{20\Sigma},\\
iii) & \quad c_{2}^{\pm}{\bf q}_{I, 2}= {\bf q}_{I, 2}c_{2}^{\pm}, \ \pm \bar{f}\dual {\bf q}_{I, 2}c_{2}^{\pm}f\geq 0 \hbox{ on }\cE_{\rm TT}.
\end{aligned}
\end{equation}
Then ${\bf \Lambda}_{2, {\rm TT}}^{\pm}$ given by \eqref{edd.e3}  and \eqref{e00.4} are the covariances of a quasi-free state on $\CCR(\cV_{P}, {\bf Q}_{P})$. Furthermore, if for some neighborhood $\cU$ of $\Sigma$ in $M$ we have: 
\begin{equation}
\label{edd.e7}
 \begin{aligned}
iv)&  \ \WF(U_{2}\circ c_{2}^{\pm})'\subset (\cN^{\pm}\cup\cF)\times T^{*}\Sigma,
\end{aligned}
\end{equation}over $\cU\times \Sigma$,
where $U_{2}$ solves the Cauchy problem for $D_{2}$, see \ref{sec1.5.2}, and $\cF\subset T^{*}M$ is a conic set with $\cF\cap \cN= \emptyset$, 
then the associated state  is Hadamard. 
\end{proposition}

We call property \eqref{edd.e6} {\it ii)} the {\em weak TT gauge invariance} property.\medskip

\proof To prove the first statement of the proposition, we need to check that ${\bf \Lambda}_{2, {\rm TT}}^{\pm}$ vanish on $\Kerc K_{21}^{\star}\times \Ran_{\c}P$, or equivalently (since $G_{2}D_{2}=0$ and $P= D_{2}- K_{21}K_{21}^{\star}$) on $\Kerc K_{21}^{\star}\times \Ran_{\c}K_{21}K_{21}^{\star}$. Using property \eqref{edd.e6} {\it ii)}, this will hold if $(\one - K_{21}S_{0})$ maps $K_{21}K_{21}^{\star}\Ker_{\sc}D_{2}$ into itself.

We use \eqref{e4.10} and its adjoint identity to obtain:
\[
\begin{aligned}
 S_{0}K_{21}&= (2\Lambda)^{-1}K_{10}K_{10}^{\star}= (-4\Lambda)^{-1}K_{21}^{\star}K_{20}K_{10}^{\star},\\
 K_{21}S_{0}K_{21}&= (-4\Lambda)^{-1}K_{21}K_{21}^{\star}K_{20}K_{10}^{\star},
\end{aligned}
\]
which shows that $(\one - K_{21}S_{0})$ preserves $K_{21}K_{21}^{\star}\Ker_{\sc}D_{2}$. In fact if $u_{2}\in \Ker_{\sc}D_{2}$ then
\[
K_{21}S_{0}K_{21}K_{21}^{\star}u_{2}= - (4\Lambda)^{-1}K_{21}K_{21}^{*}v_{2}, 
\]
where $v_{2}= K_{20}K_{10}^{\star}K_{21}^{\star}u_{2}\in \Ker_{\sc}D_{2}$ using the first identity in \eqref{e4.10} and its adjoint identity.

To prove the second property we write 
\[
{\bf \Lambda}_{2,{\rm TT}}^{\pm}= (\varrho_{2}G_{2})^{*}\boldsymbol\lambda_{2\Sigma, {\rm TT}}^{\pm}(\varrho_{2}G_{2}),
\]
for
\[
\boldsymbol\lambda_{2\Sigma, {\rm TT}}^{\pm}= (\one - K_{21\Sigma}S_{0\Sigma})^{*}\boldsymbol\lambda_{2\Sigma}^{\pm}(\one - K_{21\Sigma}S_{0\Sigma}).
\]
Setting $\boldsymbol\lambda_{2\Sigma, {\rm TT}}^{\pm}\eqdef \pm {\bf q}_{I, 2}^{-1}c_{2, {\rm TT}}^{\pm}$, we have
\[
c_{2, {\rm TT}}^{\pm}= (\one - K_{21\Sigma}S_{0\Sigma})\circ c_{2}^{\pm}\circ (\one - K_{21\Sigma}S_{0\Sigma}).
\]
The operator $(\one - K_{21\Sigma}S_{0\Sigma})$ is a matrix of differential operators on $\Sigma$, so does not enlarge the wavefront set.
On the other hand we have
\[
U_{2}\circ (\one - K_{21\Sigma}S_{0\Sigma})= (\one - K_{21}S_{0})\circ U_{2}.
\]
Since $\one - K_{21}S_{0}$ is a differential operator on $M$, we have finally
\[
\WF(U_{2}\circ c_{2, {\rm TT}}^{\pm})'\subset \WF(U_{2}\circ c_{2}^{\pm})',
\]
so the Hadamard condition is satisfied by $c_{2, {\rm TT}}^{\pm}$. This completes the proof of the proposition. \qed

 If $[c_{2}^{\pm}, I_{\Sigma}]=0$ then  we can replace ${\bf q}_{ I,2}$ by  ${\bf q}_{ 2}$  in the previous proposition.
 
 \section{Wick rotation}\init\label{sec2}
 \subsection{Wick rotation}\label{sec2.1}
Let us assume that $M= I\times \Sigma$ where $I\subset \rr$ is some open interval with $0\in I$, $\Sigma$ a smooth manifold, and denote the points of $M$ by $(t, \rx)$. We assume  that $\rg$ is real analytic in $t$ and extends holomorphically to a strip $I\times \i \widetilde{I}$, where $\widetilde{I}$ is another open interval with $0\in \widetilde{I}$. We set $\tilde{M}= \tilde{I}\times \Sigma$, with variables $(s, \rx)$.

 \subsubsection{Wick rotation of tensors}\label{sec2.1.1}
 If $a\in \{0, \dots, d\}$ we set $\epsilon(a)= \delta_{a}^{0}$. If $A= (a_{1}, \dots, a_{n})$ we set $\epsilon(A)= \sum_{i}\epsilon(a_{i})$.
 
 If $T= T\indices{^{A}_{B}}(t)$ is a tensor, real analytic  in $t$, we  define the {\em Wick rotation} of $T$,
 \[
 \mathcal{W}T= \widetilde{T},
 \]
 by
 \[
 \widetilde{T}\indices{^{A}_{B}}(s)\defeq \i^{\epsilon(B)- \epsilon(A)}T\indices{^{A}_{B}}(\i s).
 \]
 Wick rotation consists in replacing $t$ by $\i s$, $dt$ by $\i ds$, $\p_{t}$ by $- \i \p_{s}$, keeping the $\rx$ variables and the associated tensor coordinates  fixed.

\subsubsection{Wick rotation of metric}\label{sec2.1.2}
 We set $\widetilde{\rg}= \mathcal{W}\rg$. In general $\trg$ is only a symmetric complex metric. Taking $\tilde{I}$ small enough we can assume that $\trg$ is non-degenerate. We  denote by $\widetilde{\nabla}$ the associated covariant derivative. Then, computing the Christoffel symbols, one easily checks that
 \[
 \widetilde{\nabla}_{\widetilde{X}}\widetilde{T}= \widetilde{\nabla_{X} T}.
 \]
  Similarly we have
  \[
  \rR(\trg)= \widetilde{\rR(\rg)},
  \]
where $\rR(\rg), \rR(\trg)$ denotes the Riemann tensor for $\rg$, $\trg$.

  The Wick rotation $\mathcal{W}$ commutes with tensor multiplication, action of tensors on tensors, raising and lowering indices, summation over repeated indices, etc.
  
  We set for $k=0, 1,2$:
 \[
 \tV_{k}= \cc\otimes \bigotimes_{\rm s}^{k}T^{*}\widetilde{M}, 
 \]
 and use the sesquilinear forms $(\cdot| \cdot)_{\tV_{k}}$ on the fibers of $\tV_{k}$ defined with $\trg$ instead of $\rg$. In general $(\cdot| \cdot)_{\tV_{k}}$ are not Hermitian. 
 
 We have 
 \[
 \cW (\rg| u)_{V_{2}}= (\trg| \cW u)_{\tV_{2}},
 \]
 hence
 \[
 \cW \circ I= \tilde{I}\circ \cW, \hbox{ for }\tilde{I}\tilde{u}= \tilde{u}- \frac{1}{4}(\trg| \tilde{u})_{\tV_{2}}\trg.
 \]
 
  \subsubsection{Wick rotation of differential operators}\label{sec2.1.3}
  If $D$ is a differential operator acting on tensors, with coefficients analytic in $t$, we define  its Wick rotation $\widetilde{D}$ by
\[
\widetilde{D}\widetilde{T}\defeq \widetilde{DT}.
\]

  \subsection{Traces \texorpdfstring{on $\Sigma$}{}}\label{sec2.2}
Let us assume additionally that $\rg= - dt^{2}+ \rh(t, \rx)d\rx^{2}$ where $t\mapsto \rh(t, \cdot)d\rx^{2}$ is analytic with values in Riemannian metrics on $\Sigma$,   so $\trg= ds^{2}+ \rh(\i s, \rx)d\rx^{2}$.  We identify $\Sigma$ with $\{0\}\times \Sigma$.

We will use 
  the decomposition of $(0, k)$-tensors on $M$ explained in \ref{sec1.1.3}. Using this decomposition the trace operators $\varrho_{k}$ become 
\[
\varrho_{k}u_{k}= \col{u_{k}\traa{\Sigma}}{\i^{-1}\p_{t}u_{k}\traa{\Sigma}}, \ \ u_{k}\in \cinf(M; V_{k}).
\]
We use the same decomposition for $(0, k)$-tensors on $\tM$ and define for $\widetilde{u}\in \cinf(\widetilde{M}; \tV_{i})$:
\[
\trho_{k}\widetilde{u}_{k}= \col{\widetilde{u}_{k}\traa{\Sigma}}{-\p_{s}\widetilde{u}_{k}\traa{\Sigma}}, \ \ \widetilde{u}_{k}\in \cinf(\tM; \tV_{k}).
\]
If $u_{k}$ is real analytic in $t$ then 
\[
\trho_{k}\cW u_{k}= \cF_{k\Sigma}\varrho_{k}u_{k},
\]
where
\begin{equation}\label{def-de-F_k}
\begin{array}{l}
\cF_{k\Sigma}\defeq \mat{\cF_{k}}{0}{0}{\cF_{k}},\\[2mm]
\cF_{0}= 1, \quad \cF_{1}= \mat{\i}{0}{0}{\one}, \quad \cF_{2}= \left(\begin{array}{ccc}
-1&0&0\\0&\i&0\\0&0&\one
\end{array}\right).
\end{array}
\end{equation}
If $A_{ij\Sigma}$ is a linear operator from $\cinf(\Sigma; V_{j}\otimes \cc^{2})$ to $\cinf(\Sigma; V_{i}\otimes \cc^{2})$ we set
\beq\label{ew.2}
\tilde{A}_{ij\Sigma}\defeq \cF_{i\Sigma}\circ A_{ij\Sigma}\circ\cF_{j\Sigma}^{-1}.
\eeq
 If we define $\tilde{I}_{\Sigma}$ by $\tilde{I}_{\Sigma}\trho_{2}\tilde{u}_{2}\defeq \trho_{2}\tilde{I}\tilde{u}_{2}$, then this notation is  coherent, i.e.
 \[
 \tilde{I}_{\Sigma}=  \cF_{2\Sigma}\circ I_{\Sigma}\circ\cF_{2\Sigma}^{-1}.
 \]
 
    \section{\calde projectors}\init\label{sec3}
    In this section we study \calde projectors for elliptic operators on a compact manifold. 
\subsection{Background}\label{sec3.1}

\subsubsection{Extendible distributions}\label{sec3.1.1}
Let $\tM$ be a compact manifold, and $\tV\xrightarrow{\pi}\tM$ a finite rank Hilbertian vector bundle. We equip $\tM$ with a smooth density $dx$. If $\tM$ is equipped with a Riemannian metric $\tilde{\rg}$, one chooses the induced density on $\tM$. We denote by   $\cinf(\tM; \tV)$ the smooth sections  of $\tV$ and we set:
\[
(u| u)_{\tM}= \int_{\tM}(u(x)| u(x))_{\tV}dx, \ \ u\in \cinf(\tM; \tV).
\]
If $\Omega\subset \tM$ is an open set with smooth boundary, we denote by $(\cdot| \cdot)_{\Omega}$ the analog scalar products on $\coinf(\Omega; \tV)$. 

 If $\cF(\tM; \tV)\subset \cD'(\tM; \tV)$  we denote by $\bar{\cF}(\Omega; \tV)\subset \cD'(\Omega; \tV)$ the image of $\cF(\tM; \tV)$ by the restriction map $\cD'(\tM; \tV)\to \cD'(\Omega; \tV)$. For example $\bar{\cD'}(\Omega; \tV)$ is the space of extendible distributions on $\Omega$, while $\bar{\cinf}(\Omega; \tV)$ is the space of smooth sections of $\tV$ over $\Omega$ which are bounded with all derivatives. 
 
 It is known that any element of $\bar{\cD'}(\Omega; \tV)$  can be extended by $0$ in $\tM\setminus\Omega^{\rm cl}$, i.e.~is the restriction to $\Omega$ of a distribution in $\cD'(\tM; \tV)$  supported in $\Omega^{\rm cl}$.
\subsubsection{Hypersurfaces}\label{sec3.1.2}
Let $\Sigma\subset \tM$ be a smooth compact hypersurface  such that $\tM\setminus \Sigma$ is the union of two disjoint open sets $\Omega^{+}$ and $\Omega^{-}$.    We equip   $\Sigma$ with a smooth density $d\rx$, equal to the induced Riemannian density if $\tM$ is equipped with a Riemannian metric $\trg$.

We  can consider the Hilbertian vector bundle $\tV\otimes \cc^{2}\xrightarrow{\pi}\Sigma$ and denote by $(\cdot| \cdot)_{\Sigma}$ the associated scalar product on $\cinf(\Sigma; \tV\otimes \cc^{2})$.

We embed $\cinf(\tM; \tV)$ into $\cD'(\tM; \tV)$ and $\cinf(\Sigma; \tV\otimes \cc^{2})$ into $\cD'(\Sigma; \tV\otimes \cc^{2})$ using the above scalar products.

We denote by $H^{s}(\tM; \tV), H^{s}(\Sigma; \tV\otimes \cc^{2})$  the canonical Sobolev spaces.

\subsubsection{Trace operators}\label{sec3.1.3}
  We fix a vector field $\nu$ transverse to $\Sigma$ pointing out of $\Omega^{+}$ and set for $u\in \cinf(\tM; \tV)$:
\[
\trho u\defeq  \begin{pmatrix}{u\traa{\Sigma}} \\ {\nu u\traa{\Sigma}} \end{pmatrix},
\]
and for $u\in \bar{\cinf}(\Omega^{\pm}; \tV)$:
\[^{}
\trho^{\pm} u\defeq \begin{pmatrix}{u\traa{\Sigma}} \\ {\nu u\traa{\Sigma}} \end{pmatrix}.
\]
If $(\tM, \tilde{\rg})$ is Riemannian, we take $\nu$ equal to the exterior unit normal vector field to $\Omega^{+}$.

We denote by $\trho^{*}: \cD'(\Sigma; \tV\otimes \cc^{2})\to \cD'(\tM; \tV)$ the adjoint of the operator $\trho: \cinf(\tM; \tV)\to \cinf(\Sigma; \tV\otimes \cc^{2})$ for the above scalar products.
\subsubsection{Green's formulas}\label{sec3.1.4}
 Let $\tD= \tD(x, \p_{x})$ be a second order differential operator acting on $\cinf(\tM; \tV)$. For simplicity we assume that its principal symbol $\tilde{p}(x, \xi)$ is scalar.  We denote by $\tD^{*}$ its formal adjoint on $\cinf(\tM)$ for the Hilbertian scalar product $(\cdot| \cdot)_{\tM}$.
 
 We can associate to $\tD$ a sesquilinear form ${\bf Q}_{\tM}$ on $\cinf(\tM; \tV)$, bounded on $H^{1}(\tM; \tV)$ such that  $(u| \tD v)_{\tM}= Q(u, v)_{\tM}$ for $u, v\in \cinf(\tM; \tV)$. In local coordinates we have
 \beq\label{e2.1}
 Q(u, v)_{\tM}= \int_{\tM}\sum_{|\alpha|, |\beta|\leq 1}(a_{\alpha, \beta}(x)\p_{x}^{\alpha}u| \p_{x}^{\beta}v)_{\tV} dx,
 \eeq
 for $a_{\alpha, \beta}\in \cinf(\tM; L(\tV))$.  For $u, v\in \bar{\cinf}(\Omega^{\pm}; \tV)$ we denote by $Q(u, v)_{\Omega^{\pm}}$ the analogous sesquilinear forms obtained by integrating only on $\Omega^{\pm}$.

In Lemma \ref{lemma1.1} below we denote by $L_{\rm h/a}(\tV\otimes\cc^{2})$ the linear maps on the fibers of $\tV\otimes \cc^{2}$ which are Hermitian/anti-Hermitian for the Hilbertian scalar product $(\cdot| \cdot)_{\tV\otimes \cc^{2}}$.
\begin{lemma}\label{lemma1.1}
\ben
\item There exists $\widetilde{\sigma}\in \cinf(\Sigma; L(\tV\otimes \cc^{2})))$ such that
\[
(u|\tD v)_{\Omega^{\pm}}- (\tD^{*}u| v)_{\Omega^{\pm}}= \pm (\trho^{\pm}u|\widetilde{\sigma} \trho^{\pm}v)_{\Sigma}, \ \ u, v\in \bar{\cinf}(\Omega^{\pm}; \tV).
\]
\item There exists $\widetilde{q}\in \cinf(\Sigma; L_{\rm h}(\tV\otimes \cc^{2}))$ such that for all $u, v\in \bar{\cinf}(\Omega^{\pm}; \tV)$,
\[
(u| \tD v)_{\Omega^{\pm}}+ (\tD u| v)_{\Omega^{\pm}}= Q(u, v)_{\Omega^{\pm}}+ \bar{Q}(v, u)_{\Omega^{\pm}}\mp (\trho^{\pm}u| \widetilde{q} \trho^{\pm}v)_{\Sigma}.
\]
\item If $[ \one_{\Omega^{+}},\tD-\tD^{*}]=0$,  then $\widetilde{\sigma}\in \cinf(\Sigma; L_{\rm a}(\tV\otimes \cc^{2})))$.
\item If $\Sigma$ is non characteristic for $\tD$,  then $\widetilde{\sigma}$ is non-degenerate.
\een
\end{lemma}
\proof Let us  fix two extension maps $e_{\Omega^{\pm}}: \bar{\cinf}(\Omega^{\pm}; \tV)\to \cinf(\tM; \tV)$. Let us first prove (1). 
Since $\tD$ is a local operator, we have:
\beq\label{e2.2}
\begin{aligned}
(u| \tD v)_{\Omega^{\pm}}- (\tD^{*}u| v)_{\Omega^{\pm}}&= (e_{\Omega^{\pm}}u|\one_{\Omega^{\pm}} \tD e_{\Omega^{\pm}}v)_{\tM}-(\tD^{*}e_{\Omega^{\pm}}u|\one_{\Omega^{\pm}} e_{\Omega^{\pm}}v)_{\tM}\\
&=(e_{\Omega^{\pm}}u|[\one_{\Omega^{\pm}}, \tD]e_{\Omega^{\pm}}v)_{\tM}.
\end{aligned}
\eeq
Computing in local coordinates  near $\Sigma$  we obtain that 
\beq\label{e2.2b}
(e_{\Omega^{\pm}}u|[\one_{\Omega^{\pm}}, \tD]e_{\Omega^{\pm}}v)_{\tM}= (\trho^{\pm}u|\widetilde{\sigma}^{\pm} \trho^{\pm}v)_{\Sigma},
\eeq
for some $\widetilde{\sigma}^{\pm}\in  L(\cinf(\Sigma; \tV\otimes \cc^{2}))$. If $f, g\in \cinf(\Sigma; \tV\otimes \cc^{2})$ we pick $u, v\in \cinf(\tM; \tV)$ such that  $\trho u= f$ and $\trho v= g$. Since 
\[
\begin{array}{rl}
&(u| \tD v)_{\Omega^{+}}- (\tD^{*}u| v)_{\Omega^{+}}+(u| \tD v)_{\Omega^{-}}- (\tD^{*}u| v)_{\Omega^{-}}\\[2mm]
&= (u| \tD v)_{\tM}- (\tD^{*}u| v)_{\tM}=0,
\end{array}
\]
we obtain that $(f| (\widetilde{\sigma}^{+}+ \widetilde{\sigma}^{-})g)_{\Sigma}= 0$, which proves (1). The fact that $\widetilde{\sigma}\defeq \widetilde{\sigma}^{+}$ is anti-selfadjoint if $[\one_{\Omega^{+}}, \tD- \tD^{*}]=0$ is immediate, using \eqref{e2.2}. This proves (3).

Let us now prove (2). We have
\[
\begin{aligned}
(u| \tD v)_{\Omega^{\pm}}+ (\tD u| v)_{\Omega^{\pm}}
&= (\one_{\Omega^{\pm}}e_{\Omega^{\pm}}u| \tD e_{\Omega^{\pm}}v)_{\tM}+ (\tD e_{\Omega^{\pm}}u| \one_{\Omega^{+}}e_{\Omega^{\pm}})_{\tM}\\
&=Q(\one_{\Omega^{\pm}}u, v)_{\tM}+ \bar{Q}(\one_{\Omega^{\pm}}v, u)_{\tM}.
\end{aligned}
\]
Computing in local coordinates using \eqref{e2.1} we obtain that
\[
Q(\one_{\Omega^{\pm}}u, v)_{\tM}+ \bar{Q}(\one_{\Omega^{\pm}}v, u)_{\tM}= Q(u, v)_{\Omega^{\pm}}+ \bar{Q}(v, u)_{\Omega^{\pm}}+  (\trho^{\pm}u| \widetilde{q}^{\pm}\trho^{\pm}v)_{\Sigma}
\]
for some $\widetilde{q}^{\pm}\in \cinf(\Sigma; L(\tV\otimes \cc^{2}))$. For  $v=u$  all the terms above are real hence $\widetilde{q}^{\pm}$ are Hermitian. To show that $\widetilde{q}^{-}= - \widetilde{q}^{+}$ we argue as in the proof of (1).

To prove (4) it suffices to work in local coordinates $(x^{0}, x')$ in which  $\Omega^{+}= \{x^{0}>0\}$. One can choose $\nu= - \p_{x^{0}}$ and check that, modulo a non-vanishing factor coming from the densities on $\tM$ and $\Sigma$, $\widetilde{\sigma}$ is invertible if $p(0, x', dx^{0})\neq 0$, where $p(x, \xi)\one_{\tV}$ is the (scalar) principal symbol of $\tD$.  \qednoskip
\subsection{\calde projectors}\label{sec3.2} The definition of \calde projectors for $\tD$ is standard if $\tD$ is invertible, we are however interested in generalizations when $\tD$ is merely Fredholm. We first start with the former case, and discuss the more general case subsequently.

\subsubsection{\calde projectors}\label{sec3.2.1} 

Let us assume from now on that $\tD$ is elliptic. This is in particular the case if $p(x, \xi)= \xi\dual \tilde{\rg}^{-1}(x)\xi$. Then   $\tD + \lambda$ is invertible for $\lambda\gg 1$ hence $\tD$ is Fredholm of index $0$.
\begin{definition}\label{def-de-caldero}
 Assume that $\tD:  H^{2}(\tM; \tV)\to L^{2}(\tM; \tV)$ is invertible.  By ellipticity $\tD: H^{s}(\tM; \tV)\to H^{s-2}(\tM; \tV)$ is invertible for any $s\in \rr$. One defines the {\em \calde projectors} $\tc^{\pm}$ by
 \[
 \tc^{\pm}f\defeq \mp \trho^{\pm}\tD^{-1}\trho^{*}\widetilde{\sigma} f, \ \  f\in \cinf(\Sigma; \tV\otimes \cc^{2}).
 \]
 \end{definition}
 
\begin{remark}
I $f\in \cinf(\Sigma; \tV\otimes \cc^{2})$ then $\trho^{*}\widetilde{\sigma} f$ belongs to $H^{-3/2-\epsilon}(\tM; \tV)$ for any $\epsilon>0$. Considering   $\tD^{-1}$ as a bounded map from $H^{-3/2-\epsilon}(\tM; \tV)$ to $H^{\12-\epsilon}(\tM; \tV)$ shows that  $u= \tD^{-1}\trho^{*}\widetilde{\sigma} f$ belongs to $H^{\12}(\tM; \tV)$. A standard argument of partial hypoellipticity using that $u=0$ in $\Omega^{\pm}$  shows that  $u$ belongs actually to $\bar{\cinf}(\Omega^{\pm}; \tV)$, so $\tc^{\pm}$ maps $\cinf(\Sigma; \tV\otimes \cc^{2})$ into itself.
\end{remark}
 
 \begin{proposition}\label{prop2.1}
The \calde projectors $\tc^{\pm}$ satisfy:
 \[
 \begin{array}{rl}
 i)& \tc^{+}+ \tc^{-}= \one, \\[2mm]
 ii)& (\tc^{\pm})^{2}= \tc^{\pm},\\[2mm]
 iii)&\hbox{ for }u\in \cD'(\tM; \tV), \ \tD u= 0\hbox{ in }\Omega^{\pm}\Rightarrow \trho^{\pm}u= \tc^{\pm}\trho^{\pm}u.
 \end{array}
 \]
 \end{proposition}
 We include the proof for reference.
 
 \medskip
 
\proof Let $f, g\in \cinf(\Sigma; \tV\otimes \cc^{2})$. We fix $v\in \cinf(\tM; \tV)$ such that $\trho v= g$ and set $w= \tD^{-1} \trho
^{*}\widetilde{\sigma} f$, $u^{\pm}= \mp w\traa{\Omega^{\pm}}$. Then by Lemma \ref{lemma1.1}:
\[
\begin{aligned}
&(g| \widetilde{\sigma}(\tc^{+}+ \tc^{-})f)_{\Sigma}= (\trho^{+}v| \widetilde{\sigma} \trho^{+}u^{+})_{\Sigma}+ (\trho^{-}v| \widetilde{\sigma} \trho^{-}u^{-})_{\Sigma}\\
&=(v| \tD u^{+})_{\Omega^{+}}- (\tD^{*}v| u^{+})_{\Omega^{+}}- (v| \tD u^{-})_{\Omega^{-}}+ (\tD^{*}v| u^{-})_{\Omega^{-}}\\
&=(\tD^{*}v| w)_{\tM}= (v| \trho^{*}\widetilde{\sigma} f)_{\tM}= (\trho v| \widetilde{\sigma} f)_{\Sigma}= (g | \widetilde{\sigma} f)_{\Sigma}.
\end{aligned}
\]
Since $\widetilde{\sigma}$ is non-degenerate, this proves {\it i)}.  Let us now prove {\it ii)}. For $f\in \cinf(\Sigma; \tV\otimes \cc^{2})$ we set $v= \tD^{-1}\trho^{*}\widetilde{\sigma} f$, so that $\mp \trho^{\pm}v= \tc^{\pm}f$. For $u\in \cinf(\tM; \tV)$ we have 
\[
(u| \tD\one_{\Omega^{\pm}}v)_{\tM}= (u|[\tD, \one_{\Omega^{\pm}}]v)_{\tM}= \mp (u| \trho^{*} \widetilde{\sigma} \trho^{\pm}v)_{\tM},
\]
using \eqref{e2.2b} and the fact that $\tD v= 0$ in $\Omega^{\pm}$. Therefore $\tD\one_{\Omega^{\pm}}v= \mp \trho^{*}\widetilde{\sigma}\trho^{\pm}v= \trho^{*}\widetilde{\sigma} \tc^{\pm}f$. Since $\tD$ is invertible this implies that $\one_{\Omega^{\pm}}v= \tD^{-1}\trho^{*}\widetilde{\sigma} \tc^{\pm}f$ hence $\tc^{\pm}f= \tc^{\pm}\circ \tc^{\pm}f$ by applying $\trho^{\pm}$ to the above identity. 

Finally $\tD\one_{\Omega^{\pm}}u= \mp \trho^{*}\widetilde{\sigma} \trho^{\pm}u$, hence $\one_{\Omega^{\pm}}u= \mp \tD^{-1}\trho^{*}\widetilde{\sigma} \trho^{\pm}u$, which implies that $\tc^{\pm}\trho^{\pm}u = \trho^{\pm}u$ and proves {\it iii)}. \qed
\subsubsection{Selfadjointness of \calde projectors}\label{sec3.2.2}
Let us now assume that there exist a smooth involution $\chi: \tM\to \tM$ with  
\[
\chi: \Omega^{\pm}\to \Omega^{\mp}, \  \chi= \id\hbox{ on }\Sigma,
\]
 Without loss of generality we can assume that the density $dx$ is invariant under $\chi$. 

We interpret $\chi$ as a reflection in $\Sigma$ and 
we assume that $\chi$ lifts to a bundle  involution $\kappa: \tV\to \tV$ preserving the fiber scalar product.   Since $\chi= \id$ over $\Sigma$, $\kappa$ preserves the fibers of $\tV$ over $\Sigma$; we denote by $\kappa_{\Sigma}$
its restriction to $\tV$over $\Sigma$. 
To define the trace operator $\trho$ we  can choose a vector field  $\nu$ such that $\chi^{*}\nu= - \nu$. It follows that
 \beq\label{e2.001}
 \trho^{\pm}\kappa u = \mat{\kappa_{\Sigma}}{0}{0}{-\kappa_{\Sigma}}\trho^{\mp}u.
 \eeq

 \begin{proposition}\label{prop2.2}
Assume that $\tD$ satisfies the following reflection property
 \begin{equation}
 \label{e2.2e}
 \tD^{*}= \kappa \tD \kappa.
 \end{equation}
Then
\ben
\item $q\defeq \widetilde{\sigma}\circ\mat{\kappa_{\Sigma}}{0}{0}{-\kappa_{\Sigma}}$ satisfies $q= q^{*}$;
\item one has $\tc^{\pm*}q= q\tc^{\pm}$.
\een
\end{proposition}

In concrete applications, when $\tD$ is obtained by a Wick rotation,  the operator $q$ corresponds to the Lorentzian charge, see \ref{discussion} below.

\medskip

\proof
Let $u,v\in \cinf(\tM; \tV)$. Then using Lemma \ref{lemma1.1} and the fact that $\widetilde{\sigma}= - \widetilde{\sigma}^{*}$ we obtain
\[
\begin{aligned}
&(u|\kappa \tD\kappa v)_{\Omega^{+}}- (\kappa \tD^{*}\kappa u| v)_{\Omega^{+}}=  (\trho^{+}u| \widetilde{\sigma} \trho^{+}v)_{\Sigma}\\
&= (\kappa u| \tD \kappa v)_{\Omega^{-}}- (\tD^{*}\kappa u| \kappa v)_{\Omega^{-}}= - (\trho^{-}\kappa u| \widetilde{\sigma} \trho^{-}\kappa u)_{\Sigma},
\end{aligned}
\]
which implies that $q= q^{*}$ since $\trho^{\pm}= \trho$ on $\cinf(\tM; \tV)$.

To prove 2) we repeat the proof in \cite[Prop~6.2]{Gerard2025}  for the reader's convenience. For $f, g\in \cinf(\Sigma; \tV\otimes\cc^{2})$ we set
 \[
 u= -(\tD^{-1}\trho^{*}\widetilde{\sigma} f)\traa{\Omega^{+}}, \quad v= - (\kappa \tD^{-1} \trho^{*}\widetilde{\sigma} g)\traa{\Omega^{+}}.
 \]
 Since $\tD u= \tD^{*}v= 0$ in $\Omega^{+}$, we deduce from Lemma \ref{lemma1.1} that 
 $ (\trho^{+}v| \widetilde{\sigma} \trho^{+}u)_{\Sigma}=0$.

 Using \eqref{e2.001} we obtain 
 \[
 \trho^{+}u= \tc^{+}f, \quad \trho^{+}v= \mat{\kappa_{\Sigma}}{0}{0}{-\kappa_{\Sigma}}\tc^{-}g.
 \]
 Then $ (\trho^{+}v| \widetilde{\sigma} \trho^{+}u)_{\Sigma}=0$ implies that $c^{-*}q\tc^{+}= 0$ hence $c^{+*} q= q\tc^{+}$ as claimed. \qed

 \subsubsection{\calde projectors in the non-invertible case}\label{sec3.2.3}
Let us assume now that $\Ker \tD\neq \{0\}$. Since $\tD$ is Fredholm,  $\Ran \tD= (\Ker \tD^{*})^{\perp}$, and we can define
\[
\tc^{\pm}f:= \mp\trho^{\pm} \tD^{-1}\trho^{*}\widetilde{\sigma} f
\]
 if $\widetilde{\sigma} f\in(\trho\Ker \tD^{*})^{\perp}$, where   $\tD^{-1}: \Ran \tD\to L^{2}(\tM; \tV)\ominus \Ker \tD$.

 Assume that there exist a reflection $\kappa$ on $\tV$ such that \eqref{e2.2e} holds. Then  it is easy to check that
\beq\label{e2.002}
\trho^{*}\widetilde{\sigma} f\in(\Ker \tD^{*})^{\perp}\hbox{ iff } f\in (\trho\Ker \tD)^{\bf q},
\eeq
where $F^{\bf q}$ denotes the ${\bf q}$-orthogonal of $F$ and the Hermitian form ${\bf q}$ is
\[
\bar{f}\dual{\bf q}f= (u| qu)_{\tV(\Sigma)\otimes \cc^{2}}.
\]
Note that since $\Ker \tD\subset\cinf(\tM; \tV)$ we have $\trho = \trho^{\pm}$ on $\Ker \tD$.

\begin{proposition}\label{prop0.2b} We have:
\ben
\item $\trho\Ker \tD\subset (\trho\Ker \tD)^{\bf q}$ and $[{\bf q}]$ is non-degenerate on $\dfrac{(\trho \Ker \tD)^{\bf q}}{\trho \Ker \tD}$,
\item the maps 
 \[
 [\tc^{\pm}]:\dfrac{(\trho \Ker \tD)^{\bf q}}{\trho \Ker \tD}\to \dfrac{(\trho \Ker \tD)^{\bf q}}{\trho \Ker \tD}
 \]
 are well defined,
 \item $ [\tc^{+}]+ [\tc^{-}]= \one$ and  $[\tc^{\pm}]^{2}= [\tc^{\pm}]$.
 \een
\end{proposition}
\proof Let $u\in \Ker \tD, f= \trho u$. Then $\tD\one_{\Omega^{\pm}}u= \mp \trho^{*}\widetilde{\sigma} f$, hence $\trho^{*}\widetilde{\sigma} f\in (\Ker \tD^{*})^{\perp}$ and thus $f\in(\Ker \tD)^{\bf q}$.  Since $\tD$ is elliptic, $\widetilde{\sigma}$ and hence ${\bf q}$ are non-degenerate. Since $\trho\Ker \tD$ is finite-dimensional we have $( (\trho\Ker \tD)^{\bf q})^{\bf q}= \trho \Ker \tD$, so $[{\bf q}]$ is non-degenerate. 

Moreover $\one_{\Omega^{\pm}}u= \mp \tD^{-1}\trho^{*}\widetilde{\sigma} f+ v^{\pm}$ for $v^{\pm}\in \Ker \tD$ hence  $f= \tc^{\pm}f+ \trho v^{\pm}$, so $\tc^{\pm}$ preserves $\trho \Ker \tD$.
To complete the proof of  (2) we need to show that $\tc^{\pm}$ preserves  $(\trho\Ker \tD)^{\bf q}$. If $f\in (\trho\Ker \tD)^{\bf q}$ and $v= \tD^{-1}\trho^{*}\widetilde{\sigma} f$, $u\in \Ker \tD$
then $\tD v=\tD^{*}\kappa u= 0$ in $\Omega^{\pm}$ so by Green's formula we have $(\trho^{\mp}\kappa u| \widetilde{\sigma} \trho^{\pm}v)_{\Sigma}= \mp (\trho u| q \tc^{\pm}f)_{\Sigma}=0$, so $\tc^{\pm}$ preserves $(\trho \Ker \tD)^{\bf q}$.

Let us now prove (3). We have $\tc^{+}+ \tc^{-}= \one$ on $(\trho\Ker \tD)^{\bf q}$ by the same proof as in the invertible case. This proves the first statement of (3). 
The second statement is proved as before, except at the end: from $\tD\one_{\Omega^{\pm}}v= \trho^{*}\widetilde{\sigma} \tc^{\pm}f$ we obtain $\one_{\Omega^{\pm}}v= \tD^{-1}\trho^{*}\widetilde{\sigma} \tc^{\pm}f$ modulo $\Ker \tD$ hence  $\tc^{\pm}f= (\tc^{\pm})^{2}f$ modulo $\trho \Ker \tD$. \qed

\subsection{Invariance under symmetries}\label{sec3.4}
Let us assume that $(\tM, \trg)$ is Riemannian,  that $\tV$ equals $\cc\otimes\bigotimes_{\s}^{k}T^{*}\tM$ for  $k= 1, 2$ and   that $\tD= \tD_{k, L}- 2\Lambda$ for some $\Lambda\in \rr$,  where $\tD_{k, L}$ are the Lichnerowicz operators for $\trg$ (see Subsect.~\ref{sec1.4}).

We assume that  $\tilde{X}$ is  a Killing vector field of $(\tM, \trg)$ and we denote by $\tilde{\phi}(\cdot)$ the associated flow on $\cinf(\tM; \tV)$. We define its generator $\tilde{L}$ by
\def\tL{\tilde{L}}
\[
\tilde{\phi}(\sigma)\eqdef \e^{\i \sigma\tilde{L}}, \ \ \sigma\in \rr.
\] 
We have $[\tD, \tL]=0$ and $\tL$ is  a first order differential operator, formally selfadjoint for $(\cdot| \cdot)_{\tV(\tM)}$.   If $u\in \bar{\cinf}(\Omega^{\pm}; \tV)$, then $\trho^{\pm} \tL u$ depends only on $\trho^{\pm}u$ and $\tD u\traa{\Sigma}$. We define the operators  $\tL_{\Sigma}$ and  $Z_{\Sigma}$ by
\[
\trho^{\pm} \tL u\eqdef \tL_{\Sigma}\trho^{\pm} u + Z_{\Sigma}\tD u\traa{\Sigma}.
\]
\begin{lemma}\label{killing-lemma}
 Assume that $\tD$ is invertible. Then $[\tL_{\Sigma}, \tc^{\pm}]=0$. 
\end{lemma}

\proof 
Let $u\in \bar{\cinf}(\Omega^{\pm}; \tV)$ with $\tD u= 0$ in $\Omega^{\pm}$ and $f= \trho^{\pm}u$. Then $\tD \tL u=\tL \tD u=0$ in $\Omega^{\pm}$, hence $\tc^{\pm}\tL_{\Sigma}f= \tL_{\Sigma}f= \tL_{\Sigma}\tc^{\pm}f$ since $f= \tc^{\pm}f$. Therefore $\tc^{\mp}\tL_{\Sigma}\tc^{\pm}=0$ which proves the lemma. \qed

\subsection{Wick rotation and Hadamard property}\label{sec3.3}
We recall here some results from \cite{GW2,Gerard2025} in a simple situation, which will be sufficient for the de Sitter spacetime considered in the sequel.

Let $(M, \rg)$ be an {\em analytic} globally hyperbolic spacetime with a {\em compact}  Cauchy surface $\Sigma$. 
We set $D=D_{k}$, $V= V_{k}$, $k= 0, 1, 2$, where $D_{k}, V_{k}$ are introduced in \ref{sec1.5.1b}, \ref{sec1.1.2}.

Using Gaussian normal coordinates to $\Sigma$,  there exists  a neighborhood $U$ of $\Sigma$ in $M$ and an interval $I\ni 0$ such that $(U, \rg)$ is isometric to $(I\times \Sigma, -dt^{2}+ \rh(t, \rx))$.

Clearly  $\rh$ is real analytic in $t$ and by Wick rotation in $t$, we obtain a product $\tilde{I}\times \Sigma$,  $\tilde{I}= \,]-\delta, \delta[$, a complex metric $\trg= ds^{2}+ \widetilde{\rh}(s, \rx)$ on $\tilde{I}\times \Sigma$ and a Wick rotated operator $\tD$ acting on sections of $\tV$. As usual we identify $\Sigma$ with $\{0\}\times \Sigma$.

We will assume that $\trg $ is {\em Riemannian} over $\tilde{I}\times \Sigma$  and   that $\trg$ and $\tD$ extend to  a Riemannian metric $\trg$ and differential operator $\tD$ on a compact analytic manifold $\tM$ with $\tV= \tV_{k}= \cc\otimes\bigotimes^{k}_{\rm s}T^{*}\tM$.  $\tD$ has principal symbol $\xi\dual \trg^{-1}\xi\one_{\tV}$ hence is elliptic. 

Since $\rg$ is real-valued we obtain that
\[
\trg(s, \rx)= \trg(s,\rx)^{*}= \trg(-s, \rx), \ \ s\in I.
\]
By analyticity the reflection $s\mapsto -s$ on $\tilde{I}\times \Sigma$ extends to an involution $\chi$ on $\tM$, which is  isometric for $\trg$. We denote by $\kappa$ the unique lift of $\chi$ to the bundle $\tV$.

Since $D= D^{*}$ for the scalar product $(\cdot| \cdot)_{V(M)}$, we obtain that 
\[
\tD^{*}= \kappa\tD\kappa,
\]
where $\tD^{*}$ is the adjoint of $\tD$ for the scalar product $(\cdot| \cdot)_{\tV(\tM)}$, i.e.~the hypotheses in \ref{sec3.2.2} are satisfied.
\subsubsection{Lorentzian charge}\label{discussion-2}
We define the Lorentzian traces $\varrho_{k}$ as in \ref{sec1.5.2}, using the future directed unit normal $\p_{t}$. We systematically identify Hermitian forms with linear operators using the {\em Hilbertian} scalar product $(\cdot| \cdot)_{\tV_{k}(\Sigma)\otimes \cc^{2}}$. 

Then the Lorentzian charges ${\bf q}_{k}$, see \ref{sec1.5.3} are  given by
\[
\bar{f}\dual {\bf q}_{k}f= (f| q_{k} f)_{_{\tV_{k}(\Sigma)\otimes \cc^{2}}}, \quad q_{k}= \mat{0}{\kappa_{k\Sigma}}{\kappa_{k\Sigma}}{0},
\]
for
\[
\kappa_{0\Sigma}= 1, \quad  \kappa_{1\Sigma}= \mat{-1}{0}{0}{1}, \quad \kappa_{2\Sigma}= \left(\begin{array}{ccc}
1&0&0\\
0&-1&0\\
0&0&1
\end{array}\right),
\]
and we use the decomposition of $(0, k)$-tensors recalled in \ref{sec1.1.4} and \ref{sec1.1.5}.

\subsubsection{Euclidean operators}\label{discussion}
We define the Euclidean traces as in \ref{sec3.1.3}, using the normal vector field $-\p_{s}$. Then  we have
\[
\widetilde{\sigma}_{k}= \mat{0}{-1}{1}{0}, \quad \tq_{k}= \mat{0}{1}{1}{0}, \ \ k= 0, 1, 2.
\]
As claimed in \ref{sec3.2.2} we have  $q_{k}= \widetilde{\sigma}_{k}\circ \mat{\kappa_{k\Sigma}}{0}{0}{- \kappa_{k\Sigma}}$.

\subsubsection{\calde projectors}
Let us finally assume that $\tD: H^{2}(\tM; \tV)\to L^{2}(\tM; \tV)$ is invertible and let $\tc^{\pm}$ be the associated \calde projectors. We  use the coordinates $(t, \rx, \tau, k)$ on $T^{*}(I\times \Sigma)$.

We recall that the operators $\cF_{k\Sigma}$ identifying Cauchy data for $D_{k}$ and traces for $\tD_{k}$ are defined in Subsect.~\ref{sec2.2}.

\begin{theoreme}\label{thm-microloc}
Let us set 
\[
c_{k}^{\pm}\defeq \cF_{k\Sigma}^{-1}\circ \tc_{k}^{\pm}\circ \cF_{k\Sigma}: \cinf(\Sigma; V_{k}\otimes\cc^{2})\to \cinf(\Sigma; V_{k}\otimes\cc^{2}).
\]
Then 
\[
\WF(U_{k}\circ c_{k}^{\pm})'\subset (\cN^{\pm}\cup \{k=0\})\times T^{*}\Sigma,
\]
where $U_{k}$ solves the Cauchy problem for $D_{k}$, see \ref{sec1.5.2}. \end{theoreme}
\proof The theorem is proved in \cite[Sect.~5]{Gerard2025}. In \cite{Gerard2025} a less natural convention is used for the Wick rotation of tensors and differential operators, which consists in replacing $t$ by $\i s$ and $dt$ by $ds$, (except for the background metric $\rg$ which is transformed into $\trg$).  Changing from the convention for operators in \cite{Gerard2025} to the present one is done by conjugation by $\cF_{k}$. Therefore the theorem follows from \cite[Prop~5.4, Prop.~5.21]{Gerard2025}. \qed.

    \section{Wick rotation of  de Sitter space}\init\label{sec4}
    \subsection{Introduction}
  The de Sitter space is $dS^{4}= \rr_{t}\times \bS^{3}$, equipped with the metric
 \beq\label{e1.1}
 \rg= - dt^{2}+ \cosh^{2}(t)\rh, 
 \eeq
   where $\rh$ is the canonical metric on $\bS^{3}= \Sigma$.
   
    If we perform the Wick rotation $t\mapsto \i s$ we obtain
   \beq\label{e1.2}
   \widetilde{\rg}= ds^{2}+ \cos^{2}(s)\rh.
   \eeq
  The manifold $]-\pi/2, \pi/2[_{s}\times \bS^{3}$, equipped with the metric $\widetilde{\rg}$ is the Euclidean sphere $\bS^{4}= \{(x^{0}, \dots, x^{4}): \sum_{i=0}^{4}(x^{i})^{2}=1\}$, with the two poles $x^{0}= \pm 1$ removed. This can easily be seen  using the coordinates $(s, \omega)\in [-\pi/2, \pi/2]\times \bS^{3}$ on $\bS^{4}$ given by
   \[
      x^{0}= \sin s, \quad (x^{1}, \dots x^{4})= \cos s \omega,  \ \  s\in [-\pi/2, \pi/2], \  \omega\in \bS^{3}.
   \]
   
   Therefore the hypotheses in Subsect.~\ref{sec3.3} are satisfied by the de Sitter spacetime $dS^{4}$.
   
    \subsection{Lichnerowicz Laplacians}\label{sec4.1}
   As in Subsect.~\ref{sec2.1} we set
   \[
   \tV_{k}= \cc\otimes\bigotimes^{k}_{\rm s}T^{*}\bS^{4}, \ \ k=0, 1, 2. 
   \]
  Let  $\tD_{k, L}$  be the Lichnerowicz Laplacian on $(\bS^{4}, \trg)$ defined as in Subsect.~\ref{sec1.4} and let $\tD_{k}$ the Wick rotation of $D_{k}$.       
  
  We will denote by $\vec{D}_{k, L}, \vec{d}, \vec{\delta}$ the analogous Lichnerowicz Laplacians, differential and co-differential on the standard sphere $(\bS^{3}, \rh)$, which is Einstein with $\Lambda= 2$.  We denote by $(\trg|$ and $|\trg)$ the operators
     \[
     (\trg|: u_{2}\mapsto (\trg| u_{2})_{\tV_{2}}, \quad |\trg): u_{0}\mapsto  u_{0}\trg
     \]
      and by $(\rh|$,  $|\rh)$ their analogs on $\bS^{3}$. The identities recalled in Subsect.~\ref{sec1.4}  are independent on the signature of the metric. Therefore we have:
   \begin{equation}
   \label{e1.2b}
   \begin{array}{l}
      \tD_{k+1, L}\circ d= d\circ  \tD_{k, L},\quad 
        \delta\circ \tD_{k+1, L}= \tD_{k, L}\circ \delta, \\[2mm]
       (\trg|\circ  \tD_{2, L}=  \tD_{0, L}\circ (\trg|, \quad
        |\trg)\circ  \tD_{0, L}= \tD_{2, L}\circ |\trg),
       \end{array}
   \end{equation}
 and the analogous identities for $\vec{D}_{k, L}, \vec{d}, \vec{\delta}$.  
 
 Since $(\bS^{4}, \trg)$ is Einstein with $\Lambda=3$ we have:
   \[
   \tD_{k}= \tD_{k, L}- 6.
   \]
    Since  on $\bS^{4}$ $\widetilde{{\bf R}}_{abcd}= \trg_{bd}\trg_{ac}- \trg_{ad}\trg_{bc}$,
we have $\widetilde{\Riem}= \one - \12|\trg)(\trg|$ hence 
\beq\label{e0.5b}
\tD_{2}= -\Delta_{2}-|\trg)(\trg|+2\one.
\eeq  

   \subsubsection{Spectrum \texorpdfstring{of $\tD_{k, L}$}{}}\label{sec4.1.1}
   The spectral decompositions of $\tD_{k, L}$ on $\bS^{n}$ for $k=0, 1,2$ have been completely described in \cite{Boucetta1999}. We summarize now the results from \cite{Boucetta1999} that we will need.

We embed $\bS^{3}$ into $\rr^{4}$   so that $\bS^{3}= \{(y_{1}, \dots, y_{4})\in \rr^{4}: \sum_{i=1}^{4}y_{i}^{2}= 1\}$. We denote by $\psi_{i}$, $1\leq i\leq 4$ the function equal to the restriction of $y_{i}$ to $\bS^{3}$  and by $\psi_{jk}$ for $1\leq j<k\leq 4$ the $1$-form equal to the restriction of $y_{j}dy_{k}- y_{k}dy_{j}$ to $\bS^{3}$.

Using the coordinates $(s, \omega)$ on $\bS^{4}$ we set:
\beq\label{killing-1-forms}
\begin{array}{l}
\varphi_{i}= \psi_{i}ds- \sin s\cos s \vd\psi_{i},\  \ 1\leq i\leq 4,\\[2mm]
\varphi_{jk}= \cos^{2}(s)\psi_{jk}, \ \ 1\leq j<k\leq 4.
\end{array}
\eeq 
As is well known $\varphi_{i}, \varphi_{jk}$ span the $10$ dimensional space of Killing $1$-forms  on $\bS^{4}$.
         
   \begin{proposition}\label{prop1.0}
 \ben
 \item $\tD_{2}$ is invertible and $\tD_{2}\geq 2$ on $\Ker (\trg|$.
 \item $ \Ker\tD_{1}= \Ker d= \Ker d\cap \Ker \delta= {\rm Vect}\{\varphi_{i}, \varphi_{jk}: 1\leq i\leq 4, \ 1\leq j<k\leq 4\}$.
  \een
\end{proposition}

\proof We have the orthogonal  decomposition 
\[
\cinf(\bS^{4}; \tV_{2})=\cinf(\bS^{4}; \tV_{0})\trg\oplus\Ker (\trg|,
\] which is preserved by $\tD_{2}$. By \eqref{e0.5b}  $\tD_{2}\geq 2$ on $\Ker (\trg|$ and  $\tD_{2}\trg= -6\trg$, which proves (1).
Let us now prove (2). 
Since $\Ker \tD_{1, L}=\{0\}$ we have the orthogonal  decomposition
\[
\tV_{1}= \Ker \delta_{\rm a}\oplus\Ran d_{\rm a},
\]
where we recall that $d_{\rm a}, \delta_{\rm a}$ are the anti-symmetric differential and co-differential. By  \eqref{e0.5}  this decomposition is  preserved by $ \tD_{1, L}$  hence by $\tD_{1}$.   From the results in \cite[Subsect. 3.6]{Boucetta1999} on the spectrum of $\tD_{1, L}$ on $\bS^{n}$ we obtain that
\[
 \begin{array}{l}
 \sigma(\tD_{1|  \Ran d_{\rm a}})= \{k(k+ 3)-6: \ k\in \nn, \ k \geq 1\},\\[2mm]
\sigma(\tD_{1|\Ker \delta_{\rm a}})= \{k(k+ 3)-4: \ k\in\nn, \ k\geq 1\}.
\end{array}
\]
 The eigenspace of $\tD_{1}$ for the eigenvalue $0$ is included in $\Ker \delta_{\rm a}$ and obtained for $k=1$. Using that   $\delta_{\rm a}= \delta_{\rm s}$ on $\tV_{1}$  and \eqref{e0.6} we obtain that  it is also included in $\Ker d_{\rm s}$.   Moreover following  \cite[Subsect. 3.6]{Boucetta1999}  $\Ker \tD_{1}$ consists of  restrictions to $\bS^{4}$ of $1$-forms on $\rr^{5}$ with linear coefficients. A routine computation then shows that $\Ker\tD_{1}= {\rm Vect}\{\varphi_{i}, \varphi_{jk}\}$.  \qed

\subsubsection{The Lichnerowicz Laplacians on $\bS^{3}$}\label{sec4.1.2}
We now collect various properties of the Lichnerowicz Laplacians $\vec{D}_{i, L}$ on the standard sphere $(\bS^{3}, \rh)$ for $i=0, 1, 2$.
We set
\begin{equation}
\label{def-des-Ij}
\begin{array}{l}
I_{0}\defeq \{k(k+2): k\in \nn\}, \\[2mm]
 I_{1}\defeq  \{k(k+2): k\in \nn,k\geq 1\}\cup\{k(k+2)+1: k\in \nn, k\geq 1\}, \\[2mm]
  I_{2}\defeq\{k(k+2)+4, k\in \nn, k\geq 2\}.
\end{array}
\end{equation}
It is easy to check that 
\beq\label{intersect}
I_{0}\cap I_{2}= I_{1}\cap I_{2}= \emptyset.
\eeq
The next lemma is well-known, see eg  \cite{Boucetta1999}, \cite[Lemma 4.57]{Besse1987}. Recall that  $\bS^{3}$ is also denoted by $\Sigma$.

\begin{lemma}
 \label{dec-sphere}
 We have
 \[
 \begin{array}{rl}
 i)&L^{2}(\Sigma; \tV_{1\Sigma})= \Ker \vec{\delta}\overset{\perp}{\oplus} \Ran \vec{d},\\[2mm]
 ii)&L^{2}(\Sigma; \tV_{2\Sigma})= (\Ker \vec{\delta}\cap \Ker (\rh|)\overset{\perp}{\oplus}(\vec{d}+ \frac{1}{3}|\rh)\circ \vec{\delta})\Ker \one_{\{3, 4\}}(\vD{1})\overset{\perp}{\oplus} \Ran |\rh),
 \end{array}
 \]
 The above decompositions being preserved by $\vD{1}$, $\vD{2}$ respectively.
\end{lemma}
\proof
{\it i)} is the Hodge decomposition.  For {\it ii)} we first uniquely decompose  a section $u_{2}$ of $\tV_{2}$ as $u_{\Sigma\Sigma}= v_{2}\overset{\perp}{\oplus}v_{0}\rh$ with $(\rh| v_{2})=0$. Next, see eg. \cite[Lemma 4.57]{Besse1987} we can write $v_{2}= \bar{v}_{2}\overset{\perp}{\oplus}(dv_{1}+ v_{0}\rh)$ with  $\vec{\delta}\bar{v}_{2}= (\rh| v_{2})=0$. Since $(\rh| v_{2})=0$ we obtain that $v_{0}= \frac{1}{3}\vec{\delta}v_{1}$. Note that $\vec{d}v_{1}+ \frac{1}{3}\vec{\delta}v_{1}\rh=0$ iff $v_{1}$ is a conformal Killing $1$-form on $\bS^{3}$ i.e.~$v_{1}\in \Vect\{\vec{d}\psi_{i}, \psi_{jk}\}= \Ran \one_{\{3, 4\}}(\vD{1})$. This proves {\it ii)}.

The fact that the decompositions are preserved by $\vD_{i}$ follows from the Riemannian versions on $\bS^{3}$ of the identities in Prop. \eqref{prop0.-1}. \qed

\begin{proposition}
 \label{prop1.0b}
 We have:
  \[
 \begin{array}{l}
 1i)\ \sigma(\vec{D}_{0, L})= I_{0},\\[2mm]
1ii)\ \Ker ( \vec{D}_{0, L}-3)= {\rm Vect}\{\psi_{i}: 1\leq i\leq 4\},\\[2mm]
1iii)\ \vec{d}\circ \vec{d}\psi_{i}+ \rh\psi_{i}=0, \ \ 1\leq i\leq 4.\\[2mm]
2i)\ \sigma(\vec{D}_{1, L})= I_{1},\\[2mm]
2ii)\ \Ker ( \vec{D}_{1, L}-3)= {\rm Vect}\{\vec{d}\psi_{i}: 1\leq i\leq 4\},\\[2mm]
2iii)\ \Ker (\vec{D}_{1, L}-4)= {\rm Vect}\{\psi_{j, k}: 1\leq j<k\leq 4\}= \Ker \vec{\delta}\cap \Ker \vec{d},\\[2mm]
3i)\ \sigma(\vD{2})= I_{0}\cup I_{1}\cup I_{2},\\[2mm]
3ii)\ \sigma(\vD{2})_{|\Ran |\rh)}= I_{0},\\[2mm]
3iii)\    \sigma(\vD{2})_{|(\vec{d}+ \frac{1}{3}|\rh)\circ \vec{\delta})\Ker \one_{\{3, 4\}}(\vD{1})}= I_{1},\\[2mm]
3iv)\ \sigma(\vD{2})_{|\Ker \vec{\delta}\cap \Ker (\rh|}= I_{2}.
\end{array}
\]
  \end{proposition}
\proof {\it 1i)}, {\it 1ii)}, {\it 2i)}, {\it 2ii)} and the first statement of {\it 2iii)} are shown in \cite[Subsect.~3.5, 3.6]{Boucetta1999}. {\it 1iii)} follows from \cite[Prop.~3.3]{Boucetta1999}. It is also shown that $\Ker (\vec{D}_{1, L}-4)\subset \Ker \vec{\delta}$, which using that $ \vec{D}_{1, L}-4= \vec{\delta}\circ\vec{d}- \vec{d}\circ\vec{\delta}$ proves the second statement of {\it 2iii)}. {\it 3i)} is shown in \cite[Thm. 3.2]{Boucetta1999}, {\it 3 ii)}, {\it 3iii)} follow from \cite[Prop. 3.18]{Boucetta1999} and {\it 3iv)} from \cite[Prop. 3.19]{Boucetta1999}. \qed

  \subsection{Spatial decomposition \texorpdfstring{of $\tD_{i}$}{}}\label{lalilalou}
  We now explain a useful decomposition of $\tD_{i}$.   We use the identifications in \ref{sec1.1.4}, \ref{sec1.1.5} for $(0,i)$-tensors.

  A lengthy computation using \ref{sec4.0.1} gives for $a(s)= \cos^{2}(s)$: 
  \[
  \begin{aligned}
  (\tD_{1}w)_{s}&=-\p^{2}_{s}w_{s}+ a^{-1}\vec{D}_{0, L}w_{s}- \dot{a}a^{-2}\vec{\delta}w_{\Sigma}+ (\textstyle\frac{3}{4}\dot{a}^{2}a^{-2}- 3)w_{s},\\
  (\tD_{1}w)_{\Sigma}&= -(\p_{s}- \textstyle\12 \dot{a}a^{-1})^{2}w_{\Sigma}+ a^{-1}\vec{D}_{1, L}w_{\Sigma}- \dot{a}a^{-1}\vec{d}w_{s}\\ & \phantom{=} \ + (\textstyle\frac{1}{4}\dot{a}^{2}a^{-2}- 2 a^{-1}-3)w_{\Sigma},
    \end{aligned}
  \]
 \[
  \begin{aligned}
  (\tD_{2}u)_{ss}=&  - \p_{s}^{2}u_{ss}+ a^{-1}\vec{D}_{0, L}u_{ss}- 2 \dot{a}a^{-3}\vec{\delta}u_{s\Sigma}\\
  &+ \textstyle\frac{3}{2}\dot{a}^{2}a^{-2}u_{ss}- (\textstyle\frac{1}{4}\dot{a}^{2}a^{-3}+ 1)(\rh| u_{\Sigma\Sigma}),\\
  (\tD_{2}u)_{s\Sigma}=& - (\p_{s}- \textstyle\12 \dot{a}a^{-1})^{2}u_{s\Sigma}+ a^{-1}\vec{D}_{1,L}u_{s\Sigma}-\textstyle\12 \dot{a}a^{-2}\vec{\delta}u_{\Sigma\Sigma}\\
  &- \dot{a}a^{-1}\vec{d}u_{ss}+ (\textstyle\frac{3}{2}\dot{a}^{2}a^{-2}+ 2 - 2a^{-1})u_{s\Sigma},\\
  (\tD_{2}u)_{\Sigma\Sigma}=& -(\p_{s}- \dot{a}a^{-1})^{2}u_{\Sigma\Sigma}+ a^{-1}\vec{D}_{2, L}u_{\Sigma\Sigma}\\
  &- 2 a^{-1}\dot{a}\vec{d}u_{s\Sigma}+(2+\textstyle \12 a^{-2}\dot{a}^{2}-6a^{-1})u_{\Sigma\Sigma}\\
  &+(a^{-1}-1)\rh(\rh| u_{\Sigma\Sigma})-(2+ \textstyle\12 a^{-1}\dot{a}^{2})\rh u_{ss}.
  \end{aligned}
  \]

We set as operators on $\cinf(\bS^{4}; \tV_{i})$, $0\leq i\leq 2$ with the above identifications (i.e.~$w$ identified with $w_{s}\oplus w_{\Sigma}$, etc.):
  \[\begin{array}{l}
  \hat{D}_{0}= \vec{D}_{0, L},\\[2mm]
  
 \hat{D}_{1}= \vec{D}_{0, L}\oplus \vec{D}_{1, L},\\[2mm]
    \hat{D}_{2}= \vec{D}_{0, L}\oplus \vec{D}_{1, L}\oplus \vec{D}_{2, L}.
\end{array}
  \]
  \subsubsection{Commutation properties}
  $\hat{D}_{i}$ are independent on $s$ and act only in the spatial variables. From the above identities we obtain that
  \[
  [\tD_{i}, \hat{D}_{i}]=0, \hbox{ hence } [\tD_{i}, \one_{\{\lambda\}}(\hat{D}_{i})]=0, \ \ \lambda\in\rr,
  \]
i.e. $\tD_{i}$ commute with the spectral projections of $\hat{D}_{i}$.

Moreover from the expressions \eqref{e.tit2}, \eqref{e.tit3}  in  the Appendix, we obtain that
\beq\label{e.tit4}
\hat{D}_{i}\tK_{ij}= \tK_{ij} \hat{D}_{j},  \quad \hat{D}_{j}\tK_{ij}^{\star}= \tK_{ij}^{\star}\hat{D}_{i}, \quad \hat{D}_{2}\tilde{I}= \tilde{I}\hat{D}_{2}.
\eeq
 
\subsubsection{Spectrum \texorpdfstring{of $\hat{D}_{i}$}{}}
Clearly $\sigma(\hat{D}_{i})= \bigcup_{0\leq j\leq i}\sigma(\vec{D}_{j, L})$. From Prop. \ref{prop1.0b} and \eqref{intersect} we obtain that:
\begin{equation}
\label{ecle.0}
\Ran\one_{I_{2}}(\hat{D}_{2})= \{0\}\oplus \{0\}\oplus (\Ker \vec{\delta}\cap \Ker (\rh|).
\end{equation}
\beq\label{ecle.2}
\begin{array}{l}
\one_{\{3\}}(\hat{D}_{2})= \one_{\{3\}}(\vD{0})\oplus \one_{\{3\}}(\vD{1})\oplus \frac{1}{12\pi^{2}}|\rh)\circ\one_{\{3\}}(\vD{0})\circ (\rh|,\\[2mm]
\one_{\{4\}}(\hat{D}_{2})= 0\oplus \one_{\{4\}}(\vD{1})\oplus 0.
\end{array}
  \eeq

  \subsection{The operators \texorpdfstring{$\tK_{ij\Sigma}$ and $\tK_{ij\Sigma}^{\dag}$}{on Cauchy data}}\label{sec4.2}
    We recall that $\Sigma\defeq\bS^{4}\cap \{s=0\}\sim \bS^{3}$, and we set
    \[
 \Omega^{\pm}\defeq \bS^{4}\cap \{\pm s>0\}.
    \]
Let $u_{1}\in \overline{\cinf}(\Omega^{\pm}; \tV_{1})$ and
  \[
  u_{2}= \tK_{21}u_{1}=  du_{1}- \frac{1}{4}\trg(\trg| d u_{1})_{\tV_{2}}.
  \]
  The trace $\trho_{2}^{\pm}u_{2}$ on $\Sigma$ depends only on $\trho_{1}^{\pm}u_{1}$ and on $\tD_{1}u_{1}\traa{\Sigma}$. 
  
  If $u_{1}\in \overline{\cinf}(\Omega^{\pm}; \tV_{1})$ with $\tD_{1}u_{1}\traa{\Sigma}=0$ one sets:
  \beq\label{e.tit6}
 \trho_{2}^{\pm}\tK_{21} u_{1}\eqdef \tK_{21\Sigma}\trho_{1}^{\pm} u_{1}.
  \eeq
  Similarly if $u_{2}\in \overline{\cinf}(\Omega^{\pm}; \tV_{2})$ with $\tD_{2}u_{2}\traa{\Sigma}=0$ one sets:
       \[
  \trho_{1}^{\pm} \tK_{21}^{\star}u_{2}\eqdef \tK_{21\Sigma}^{\dag} \trho_{2}^{\pm}u_{2}.
  \]
Note that the  notation  for $\tK_{21\Sigma}$, $\tK_{21\Sigma}^{\dag}$ is coherent with the one in \eqref{ew.2}, i.e.~$\tK_{21\Sigma}$, $\tK_{21\Sigma}^{\dag}$ are the Wick rotations of $K_{21\Sigma}$ and $K_{21\Sigma}^{\dag}$.
  In this subsection we will compute the operators $\tK_{21\Sigma}$,  $\tK_{21\Sigma}^{\dag}$. The proofs are given in Appendix \ref{proofs}.
  
\begin{lemma}\label{lemmi}
   If $f= \tK_{21\Sigma}^{\dag}g$ with $f= \col{f_{0}}{f_{1}}\in \cinf(\Sigma; \tV_{1}\otimes \cc^{2})$ and $g= \col{g_{0}}{g_{1}}\in \cinf(\Sigma; \tV_{2}\otimes \cc^{2})$, then:
  \[\begin{cases}
f_{0s}=2g_{1ss}+ 2\vec{\delta}g_{0s\Sigma},\\[2mm]
f_{0\Sigma}= 2 g_{1s\Sigma}+ \vec{\delta}g_{0\Sigma\Sigma},\\[2mm]
 f_{1s}= 2(\vec{D}_{0, L}- 3)g_{0ss}+ 2\vec{\delta} g_{1s\Sigma}-  (\rh|g_{0\Sigma\Sigma}),\\[2mm]
 f_{1\Sigma}= 2(\vec{D}_{1, L}-4)g_{0s\Sigma}+ \vec{\delta}g_{1\Sigma\Sigma}.
  \end{cases}
  \]
 \end{lemma}
     
\begin{lemma}\label{lemmo}
 If $g= \tK_{21\Sigma}f$ with $f= \col{f_{0}}{f_{1}}\in \cinf(\Sigma; \tV_{1}\otimes \cc^{2})$ and $g= \col{g_{0}}{g_{1}}\in \cinf(\Sigma; \tV_{2}\otimes \cc^{2})$, then:
\[\begin{cases}
g_{0ss}= \12(- f_{1s}+ \vec{\delta}f_{0\Sigma}),\\[2mm]
g_{0s\Sigma}= \12 (-f_{1\Sigma}+ \vec{d}f_{0s}),\\[2mm]
g_{0\Sigma\Sigma}= \vec{d}f_{0\Sigma}+ \12 (f_{1s}+ \vec{\delta}f_{0\Sigma})\rh,\\[2mm]
g_{1ss}= \12(-\vec{D}_{0,L}f_{0s}+ \vec{\delta}f_{1\Sigma}),\\[2mm]
g_{1s\Sigma}= \12(-(\vec{D}_{1,L}-4)f_{0\Sigma}+ \vec{d}f_{1s}),\\[2mm]
g_{1\Sigma\Sigma}= \vec{d}f_{1\Sigma}+ \12 \vec{\delta}f_{1\Sigma}\rh+ \12 (\vec{D}_{0, L}- 4)f_{0s}\rh.
\end{cases}
\]
\end{lemma}
We have similar results for  the operators $\tK_{20}$ and $\tK_{20}^{\star}$ defined by
\[
\begin{array}{l}
\tK_{20}u_{0}= \trg u_{0}, \\[2mm]
\tK_{20}^{\star}u_{2}= - (\trg| u_{2})_{\tV_{2}(\bS^{4})}, \ \ u_{k}\in \cinf(\bS^{4}; \tV_{k}).
\end{array}
\]
Note that $\tK_{20}$, $\tK_{20}^{\star}$ are the Wick rotations of $K_{20}$, $K_{20}^{\star}$ defined in Subsect. \ref{sec1.8}.
Moreover if  $u_{2}= \tK_{20}u_{0}$ then $\trho_{2}^{\pm}u_{2}$ depends only on $\trho_{0}^{\pm}u_{0}$ and similarly for $\tK_{20}^{\star}$. We obtain the following expression for  $\tK_{20\Sigma}^{\dag}$.
\begin{lemma}\label{lemmu}
 If $f= \tK_{20\Sigma}^{\dag}g$ then:
 \[
\begin{cases}
f_{0}= -2g_{0ss}- (\rh| g_{0\Sigma\Sigma}),\\[2mm]
f_{1}= -2g_{1ss}- (\rh| g_{1\Sigma\Sigma}).
\end{cases}
 \]
 \end{lemma}
 \subsubsection{Commutation properties \texorpdfstring{with $\hat{D}_{i}$}{}}
 
\begin{lemma}\label{lemmuu}
 We have
 \[
 (\one_{\{\lambda\}}(\hat{D}_{i})\otimes \one_{\cc^{2}}) \circ \tK_{ij\Sigma}= \tK_{ij\Sigma}\circ (\one_{\{\lambda\}}(\hat{D}_{j})\otimes \one_{\cc^{2}})
 \]
 for $(i, j)= (2,1)$ or $(i, j)= (2, 0)$.
\end{lemma}
 \proof We prove the lemma for $(i, j)= (2, 1)$, the proof for $(i, j)= (2, 0)$ being  simpler and left to the reader.  From the expression of $R_{\Sigma}$ in Lemma \ref{lemmo}  we obtain that
 \begin{equation}
 \label{e.tit5}
(\one_{\{\lambda\}}(\hat{D}_{2})\otimes \one_{\cc^{2}})R_{\Sigma}= R_{\Sigma}\one_{\{\lambda\}}(\hat{D}_{1}).
 \end{equation}
Using \eqref{e.tit6}  and the fact that $\hat{D}_{2}\tK_{21}= \tK_{21}\hat{D}_{1}$ we have:
 \[
\begin{aligned}
&\trho_{2}\tK_{21}\one_{\{\lambda\}}(\hat{D}_{1})u_{1}= \trho_{2}\one_{\{\lambda\}}(\hat{D}_{2})\tK_{21}u_{1}= (\one_{\{\lambda\}}(\hat{D}_{2})\otimes \one_{\cc^{2}}) \trho_{2}\tK_{21}u_{1}\\
&=(\one_{\{\lambda\}}(\hat{D}_{2})\otimes \one_{\cc^{2}}) \tK_{21\Sigma}\trho_{1}u_{1}+ (\one_{\{\lambda\}}(\hat{D}_{2})\otimes \one_{\cc^{2}})R_{\Sigma}\tD_{1}u_{1}\traa{\Sigma}.
 \end{aligned} 
\]
From \eqref{e.tit6} and \eqref{e.tit5} we have
  \[
\begin{aligned}
 &\trho_{2}\tK_{21}\one_{\{\lambda\}}(\hat{D}_{1})u_{1}=\tK_{21\Sigma}\trho_{1}\one_{\{\lambda\}}(\hat{D}_{1})u_{1}+ R_{\Sigma}\tD_{1}\one_{\{\lambda\}}(\hat{D}_{1})u_{1}\traa{\Sigma}\\[2mm]
 &=\tK_{21\Sigma}(\one_{\{\lambda\}}(\hat{D}_{1})\otimes \one_{\cc^{2}})\trho_{1}u_{1}+ (\one_{\{\lambda\}}(\hat{D}_{2})\otimes \one_{\cc^{2}})R_{\Sigma}\tD_{1}u_{1}\traa{\Sigma},
\end{aligned} 
\]which proves the lemma. \qed

\subsection{The operators \texorpdfstring{$K_{ij\Sigma}$ and $K_{ij\Sigma}^{\dag}$}{on Cauchy data 2}}\label{sec4.2b}
We obtain the Lorentzian operators  $K_{ij\Sigma}$ and $K_{ij\Sigma}^{\dag}$ by conjugation with $\cF_{i\Sigma}$, $\cF_{j\Sigma}$ as indicated in Subsect.~\ref{sec2.2}, i.e.:
\beq\label{e.add11}
K_{ij\Sigma}^{\dag}= \cF_{i\Sigma}^{-1}\circ \tK_{ij\Sigma}\circ \cF_{j\Sigma}, \quad K_{ij\Sigma}^{\dag}= \cF_{j\Sigma}^{-1}\circ \tK_{ij\Sigma}^{\dag}\cF_{i\Sigma}.
\eeq
The explicit expressions of these operators is not important.  Since $\one_{\{\lambda\}}(\hat{D}_{i})$ commutes with $\cF_{i}$ we obtain from Lemma \ref{lemmuu} that
 \beq\label{e.tit66}
 (\one_{\{\lambda\}}(\hat{D}_{i})\otimes \one_{\cc^{2}}) \circ K_{ij\Sigma}= K_{ij\Sigma}\circ (\one_{\{\lambda\}}(\hat{D}_{j})\otimes \one_{\cc^{2}}),
 \eeq
 for $(i, j)= (2,1)$ or $(i, j)= (2, 0)$.

\section{Properties of the TT  gauge physical space}\init\label{sec4c}
Before studying the Euclidean ``vacuum'' $\omega_{\rm eucl }$ for linearized gravity on $dS^{4}$ it is necessary to investigate the TT  gauge physical phase space $\cE_{\rm TT}$, see Def.~\ref{def-de-phase-spaces}. 

We denote by $\tilde{\cE}_{\rm TT}$, $\tilde{\cF}_{\rm TT}$ the Euclidean versions of $\cE_{\rm TT}$, $\cF_{\rm TT}$ ie:
 \[
\tilde{\cE}_{\rm TT}= \Ker \tK_{21\Sigma}^{\dag}\cap \Ker \tK_{20\Sigma}^{\dag}, \ \tilde{\cF}_{\rm TT}= \Ran \tK_{21\Sigma}\cap \Ker \tK_{20\Sigma}.
\] 
By \eqref{e.add11} we have:
\beq\label{stupidix}
\tilde{\cE}_{\rm TT}=\cF_{2\Sigma}\cE_{\rm TT}, \ \tilde{\cF}_{\rm TT}=\cF_{2\Sigma}\cF_{\rm TT},
\eeq
and by Lemma \ref{lemmuu} $\tilde{\cE}_{\rm TT}$, $\tilde{\cF}_{\rm TT}$ and $\cE_{\rm TT}$, $\cF_{\rm TT}$ are invariant under the spectral projections of $\hat{D}_{2}$.

\subsection{Decomposition \texorpdfstring{of $\tilde{\cE}_{\rm TT}$}{}}\label{sec4c.2}
As a first step we describe a convenient orthogonal decomposition of $\tilde{\cE}_{\rm TT}$. The proof of the next lemma is given in Appendix \ref{sec4.0.3}.

\begin{lemma}\label{monlemmeclean}
 Let $g= \col{g_{0}}{g_{1}}\in \tilde{\cE}_{\rm TT}$. Then there exist unique $u_{i\Sigma\Sigma}\in \cinf(\Sigma; \tV_{2\Sigma})$, $f_{i\Sigma}\in \cinf(\Sigma; \tV_{1\Sigma})$ and $\beta_{s}\in \Ker (\vD{0}-3)$, $\beta_{\Sigma}\in \Ker(\vD{1}-4)$ with
 \[
\vdel u_{i\Sigma\Sigma}= (\rh| u_{i\Sigma\Sigma})=\one_{\{3, 4\}}(\vD{1})f_{i\Sigma}=0
\]
such that:
 \beq\label{e.tit1}
\begin{array}{rl}
1)& g_{0ss}= \vdel f_{0\Sigma}- 3 \beta_{s},\\[2mm]
2)& g_{0s\Sigma}= - \12 (f_{1\Sigma}+ \vd(\vD{0}-6)^{-1}\vdel f_{1\Sigma})+ \beta_{\Sigma},\\[2mm]
3)&g_{0\Sigma\Sigma}= u_{0\Sigma\Sigma}+ \vd f_{0\Sigma}+ \beta_{s}\rh,\\[2mm]
4)& g_{1ss}= (1+ 3(\vD{0}-6)^{-1})\vdel f_{1\Sigma},\\[2mm]
5)&g_{1s\Sigma}= - \12 \vdel\vd f_{0\Sigma}+ \vd \beta_{s},\\[2mm]
6)& g_{1\Sigma\Sigma}= u_{1\Sigma\Sigma}+ \vd f_{1\Sigma}- (\vD{0}-6)^{-1}\vdel f_{1\Sigma}\rh,
\end{array}
\eeq

\end{lemma} 
\begin{definition}\label{defcor.0}
 We set
 \[
 \begin{array}{l}
 \tilde{\cE}_{\rm TT, \rm gauge}\defeq \one_{I_{2}}(\hat{D}_{2}\otimes\one_{\cc^{2}})\tilde{\cE}_{\rm TT},\\[2mm]
  \tilde{\cE}_{{\rm TT}, 4}\defeq \one_{\{4\}}(\hat{D}_{2}\otimes\one_{\cc^{2}})\tilde{\cE}_{\rm TT}
 \end{array}
 \]
where $I_{2}$ is defined in \eqref{def-des-Ij}.
\end{definition}
\begin{lemma}\label{lemmacor.1}
 We have
 \[
 \begin{array}{rl}
 i)&g\in \tilde{\cE}_{\rm TT, \rm gauge}\Leftrightarrow f_{i\Sigma}= \beta_{s}= \beta_{\Sigma}=0,\\[2mm]
 ii)&g\in \tilde{\cE}_{{\rm TT}, 4}\Leftrightarrow u_{i\Sigma\Sigma}= f_{i\Sigma}= \beta_{s}=0.
 \end{array}
\]
 \end{lemma}
\proof {\it i)} follows from \eqref{ecle.0}, {\it ii)}  is a direct computation, using \eqref{ecle.2}. \qed
 \subsection{Decomposition  \texorpdfstring{of $\tilde{\cF}_{\rm TT}$}{} }
 We next perform a similar  decomposition of $\tcF_{\TT}$.
\begin{lemma}\label{lemmacor.2}
 We have
 \[
 g\in \tilde{\cF}_{\rm TT}\Leftrightarrow u_{i\Sigma\Sigma}=0.
 \]
 \end{lemma}
\proof By Lemma \ref{lemmu} we have $\tK_{21\Sigma}f\in \tilde{\cE}_{\rm TT}$ iff $f_{1s}= - \vdel f_{0\Sigma}$ and $f_{0s}= - (\vD{0}-6)^{-1}\vdel f_{1\Sigma}$. From  Lemma \ref{lemmo} we obtain that $g\in \tilde{\cF}_{\rm TT}$ iff 
\[
\begin{array}{rl}
1)& g_{0ss}= \vdel f_{0\Sigma},\\[2mm]
2)& g_{0s\Sigma}= - \12 (f_{1\Sigma}+ \vd(\vD{0}-6)^{-1}\vdel f_{1\Sigma}),\\[2mm]
3)&g_{0\Sigma\Sigma}= \vd f_{0\Sigma},\\[2mm]
4)& g_{1ss}= (1+ 3(\vD{0}-6)^{-1})\vdel f_{1\Sigma},\\[2mm]
5)&g_{1s\Sigma}= - \12 \vdel\vd f_{0\Sigma},\\[2mm]
6)& g_{1\Sigma\Sigma}=  \vd f_{1\Sigma}- (\vD{0}-6)^{-1}\vdel f_{1\Sigma}\rh,
\end{array}
\]
for $f_{i\Sigma}\in \cinf(\Sigma; \tV_{1\Sigma})$.

 We  write $f_{0\Sigma}=\hat{f}_{0\Sigma}- \vd\beta_{s}- \beta'_{\Sigma}$ where $ \beta_{s}\in  \Vect\{\psi_{i}\}, \ \beta'_{\Sigma}\in \Vect\{\psi_{jk}\}$, $\hat{f}_{0\Sigma}\perp \Vect\{\vd \psi_{i}, \psi_{jk}\}$, and similarly $f_{1\Sigma}=\hat{f}_{1\Sigma}- \vd\beta'_{s}-2 \beta_{\Sigma}$.

Using  \eqref{identit} and  Prop. \ref{prop1.0b} {\it 1.)}, {\it 2.)},   we obtain that $g$ is as in the lemma with $f_{i\Sigma}$ replaced by $\hat{f}_{i\Sigma}$, which completes the proof.  \qed
 \subsubsection{Killing \texorpdfstring{$1$}{1}-forms}
 We denote by $\mathcal{K}\subset \Ker_{\rm sc}D_{1}$ the space of {\em Killing} $1$-{\em forms}  on $dS^{4}$.  We have $\mathcal{K}= \Ker d\cap \Ker \delta$ and $\mathcal{K}$ is spanned by the $1$-forms  $- \psi_{i}dt+ \sinh t\cosh t \,\vd \psi_{i}$ and $\cosh^{2}t \,\psi_{jk}$ for $1\leq i\leq 4$, $1\leq j<k\leq 4$. Clearly,  
 \[
 \mathcal{K}= \cF_{1}^{-1}\Ker \tD_{1},
 \]
 since $\Ker \tD_{1}$ is spanned by the Killing $1$-forms on $\bS^{4}$.
\begin{definition}We set
\[
\mathcal{K}_{\Sigma}\defeq \varrho_{1}\mathcal{K},
\]
and denote $\mathcal{K}_{\Sigma}^{{\bf q}_{1}}$ its orthogonal for  the charge ${\bf q}_{1}$ and by $\tilde{\mathcal K}_{\Sigma}^{{\bf q}_{1}}= \cF_{1\Sigma}\mathcal{K}_{\Sigma}^{{\bf q}_{1}}$ its Euclidean version.
\end{definition}
\begin{definition}\label{defcor.1}
 We set
 \[
 \tilde{\cF}_{\rm TT, \rm gauge}\defeq \tK_{21\Sigma}\tilde{\mathcal K}_{\Sigma}^{{\bf q}_{1}}\subset \tcF_{\TT}.
 \]
 \end{definition}
The subspace $\tcF_{\TT, {\rm gauge}}$ will be important when we will discuss the gauge invariance of the Euclidean vacuum in Subsect. \ref{sec5.3b}.

\begin{lemma}\label{lemma5.1}
We have
 \[
 g\in \tilde{\cF}_{\rm TT, \rm gauge}\Leftrightarrow u_{i\Sigma\Sigma}= \beta_{\Sigma}=0.
 \]
 
\end{lemma}
\proof  We have:
\[
\trho_{1}\varphi_{i}= \left(\begin{array}{c}
\psi_{i}\\
0\\
0\\
\vd\psi_{i}
\end{array}\right), \ \trho_{1}\varphi_{jk}= \left(\begin{array}{c}
0\\
\psi_{jk}\\
0\\
0\\
\end{array}\right),
\]
and  $q_{1}= \mat{0}{\kappa_{1}}{\kappa_{1}}{0}$, for $\kappa_{1}= \mat{-1}{0}{0}{\one}$. Therefore
\[
\begin{array}{l}
(\trho_{1}\varphi_{i}| q_{1}f_{1})_{\tV_{1}(\Sigma)\otimes \cc^{2}}= -(\psi_{i}| f_{1s})+ (d\psi_{i}| f_{0\Sigma}), \\[2mm]
(\trho_{1}\varphi_{jk}| q_{1}f_{1})_{\tV_{1}(\Sigma)\otimes \cc^{2}}=(\psi_{jk}| f_{1\Sigma}),
\end{array}
\]
which show that $f\in \tilde{\mathcal K}_{\Sigma}^{{\bf q}_{1}}$ iff $\one_{\{3\}}(\vD{0})(f_{1s}-\vec{\delta}f_{0\Sigma})=0$ and $\one_{\{4\}}(\vD{1})f_{1\Sigma}=0$. If $\tK_{21\Sigma}f\in \tilde{\cE}_{\rm TT}$ then  $f_{1s}= -\vdel f_{0\Sigma}$  so the first condition is automatically satisfied. This completes the proof of the lemma. \qed

\subsection{Direct sum decompositions}
From Subsect. \ref{sec4c.2} and \eqref{stupidix} we obtain the following direct sum decompositions of $\cE_{\rm TT}$.
\begin{proposition}\label{propcor.2}
 
 Let us set
 \[
 \begin{array}{l}
 \cE_{\TT,{\rm gauge}}\defeq \one_{I_{2}}(\hat{D}_{2}\otimes\one_{\cc^{2}})\cE_{\TT}, \quad \cE_{{\rm TT}, 4}\eqdef \one_{\{4\}}(\hat{D}_{2}\otimes\one_{\cc^{2}})\cF_{\rm TT},\\[2mm]
 \cF_{\TT,{\rm gauge}}\defeq K_{21\Sigma}\mathcal{K}_{\Sigma}^{{\bf q}_{1}}.
 \end{array}
 \]
Then 
 \beq\label{decodeco}
 \begin{array}{rl}
i)& \cE_{\rm TT}= \cE_{\rm TT, \rm gauge}\oplus \cF_{\rm TT},\\[2mm]
  ii)&\cE_{\rm TT}= \cE_{\rm TT, \rm gauge}\oplus \cF_{\rm TT, \rm gauge}\oplus \cE_{\rm TT, 4}.
 \end{array}
 \eeq
 \end{proposition}

\subsection{The kernel of the Lorentzian charge}\label{sec4c.1}
From Prop.~\ref{propcor.2} we obtain immediately the following result.
\begin{proposition}\label{propcor.3}
 We have $\Ker {\bf q_{I, 2}}_{|\cE_{\rm TT}}= \cF_{\rm TT}$. Therefore $[{\bf q_{I, 2}}]$ is non-degenerate on $\dfrac{\cE_{\rm TT}}{\cF_{\rm TT}}$.
  \end{proposition}
 \proof Clearly $\cF_{\rm TT}\subset\Ker {\bf q_{I, 2}}_{|\cE_{\rm TT}}$. To prove the converse inclusion it suffices to show that $\cE_{\TT, {\rm gauge}}\cap \Ker {\bf q_{I, 2}}_{|\cE_{\rm TT}}= \{0\}$. If $f\in \cE_{\TT, {\rm gauge}}$, then $f_{i}= (0, 0 ,u_{i\Sigma\Sigma})$ with $u_{i\Sigma\Sigma}\in \Ker \vec{\delta}\cap \Ker (\rh|$ and  if $f\in \Ker {\bf q_{I, 2}}_{|\cE_{\rm TT}}$ then $u_{i\Sigma\Sigma}$ is orthogonal to $ \Ker \vec{\delta}\cap \Ker (\rh|$. Using the orthogonal decomposition in Lemma \ref{dec-sphere} {\it ii)} this implies that $u_{i\Sigma\Sigma}=0$ hence $f= 0$. \qed
 \subsection{Action of symmetries}\label{sec4c.3}
 We now investigate the action of the isometry group $O(1, 4)$ of $dS^{4}$ on $\cE_{\rm TT}$.  
 
It is easier in practice, though  completely equivalent, to  consider the solution space version  of $\cE_{\rm TT}$ equal to $\Ker_{\rm sc}D_{2}\cap \Ker _{\rm sc}K_{21}\cap \Kersc K_{20}$.

We will also consider the subgroup $O^{\uparrow}(1, 4)$ of isometries preserving the time orientation,    the connected component of identity $SO^{\uparrow}(1, 4)$, and the subgroup  $O(4)$ of isometries preserving the hypersurfaces $t= {\rm const}$.  
 
An element $\alpha\in O^{\uparrow}(1, 4)$ acts $\cc$-linearly on tensor fields by 
\[
u_{k}\mapsto \alpha^{*}u_{k}, \ \ u_{k}\in\cinf(dS^{4}; V_{k}).
\]
Clearly $\alpha^{*}$ is unitary for ${\bf q}_{I, 2}$.
\subsubsection{Time reversal}\label{def-de-time-reversal}

The time reversal $\tau: dS^{4}\in (t, \omega)\mapsto (-t, \omega)\in dS^{4}$ acts $\cc$ {\em anti-linearly} on tensor fields by 
\[
Zu_{k}\defeq \overline{\tau^{*}u_{k}}, \ \ u_{k}\in\cinf(dS^{4}; V_{k}).
\]
The map $Z$ is often called the {\em Wigner time reversal}.  Since $Z$ reverses the time orientation we have $Z G_{k}= - G_{k}Z$ hence
$\bar{Z u}_{k}\dual {\bf Q}_{k}Zv_{k}= \bar{v}_{k}\dual {\bf Q}_{k}u_{k}$ so  $Z$ is anti-unitary for ${\bf Q}_{k}$.

Clearly the action of $O(1,4)$ on tensor fields defined as above preserves $\Ran K_{21}$, $\Ker K_{21}^{\star}$ and $\Ker K_{20}^{\star}$, see Subsect.~\ref{sec1.8} for notation. Therefore this  action induces a well-defined action on the TT gauge subspace $\Ker_{\rm sc}D_{2}\cap \Ker _{\rm sc}K_{21}\cap \Kersc K_{20}$ and also on the TT gauge quotient  space 
\[
\frac{\Kersc D_{2}\cap \Kersc K_{21}^{*}\cap \Kersc K_{20}^{\star}}{K_{21}\Kersc D_{1}\cap \Kersc K_{20}^{\star}}.
\]

\section{The Euclidean vacuum on de Sitter space}\init\label{sec4b}
In this section we define the Euclidean ``vacuum'' for linearized gravity on $dS^{4}$ and prove some of its properties.
Our physical phase space is the TT gauge subspace  $\cE_{\rm TT}$ studied in Sect.~\ref{sec4c}.

We will see that the Euclidean vacuum is {\em not a state} on the physical phase space, because  its covariances are not positive. For this reason we will use the words {\em ``pseudo covariances''}  and {\em ``pseudo state''}. Perhaps more importantly, we will see that the Euclidean vacuum is {\em not} strongly  gauge invariant, see Corollary \ref{corr.1} below.

We recall that we  identify Hermitian forms on $\coinf(\Sigma; V_{2}\otimes \cc^{2})$ with linear operators using the Hilbertian scalar product $(\cdot| \cdot)_{\tV_{2}(\Sigma)\otimes \cc^{2}}$, obtained from the Riemannian metric $\trg$, see \ref{discussion-2}.

\subsection{Definition of \texorpdfstring{$\omega_{\rm eucl}$}{Euclidean vacuum} and basic properties}
 We recall that   $\Sigma= \{s=0\}\subset \bS^{4}$, $\Omega^{\pm}= \{\pm s >0\}$.  Since $\tD_{2}$ is invertible, see Prop.~\ref{prop1.0}, we can define the \calde projectors $\tc_{2}^{\pm}$ associated to $\Omega^{\pm}$ and $\tD_{2}$.

\begin{definition}\label{def5b.1}
We set $c_{2}^{\pm}\defeq \cF_{2\Sigma}^{-1}\circ \tc_{2}^{\pm}\circ \cF_{2\Sigma}$ and
 \[
 \bar{f}\dual \boldsymbol\lambda_{2\Sigma}^{\pm}f\defeq \pm \bar{f}\dual {\bf q}_{I, 2}c_{2}^{\pm}f \]
 for  $f\in \coinf(\Sigma; V_{2}\otimes \cc^{2})$. 
   \end{definition}
   It is easy to see that $\boldsymbol\lambda_{2\Sigma}^{\pm}$  are the Cauchy surface covariances of a  pseudo state $\omega_{\rm eucl}$  on $\cinf(\Sigma; V_{2\Sigma}\otimes \cc^{2})$ called the {\em Euclidean vacuum}. The properties of $\omega_{\rm eucl}$ are summarized in the following theorem. 
  \begin{theoreme}\label{thm5.0}
We have
\ben
\item $\boldsymbol\lambda_{2\Sigma}^{\pm}= \boldsymbol\lambda_{2\Sigma}^{\pm*}, \quad  \boldsymbol\lambda_{2\Sigma}^{+}- \boldsymbol\lambda_{2\Sigma}^{-}= {\bf q}_{I, 2}.$
\een
Therefore $\boldsymbol\lambda_{2\Sigma}^{\pm}$  are the Cauchy surface covariances of a  pseudo state $\omega_{\rm eucl}$  on $\cinf(\Sigma; V_{2\Sigma}\otimes \cc^{2})$ called the {\em Euclidean vacuum}.
 \ben\setcounter{smallarabics}{1}
\item $\boldsymbol\lambda_{2\Sigma}^{\pm}\geq 0$ on $\cE_{\rm TT, \rm gauge}$, $\boldsymbol\lambda_{2\Sigma}^{\pm}\leq 0$ on $\cE_{\rm TT, 4}$.
\item $\omega_{\eucl}$ is a Hadamard pseudo state.
\item $\boldsymbol\lambda_{2\Sigma}^{\pm}=0$ on $\cE_{\rm TT}\times \cF_{\rm TT, \rm gauge}$ but $\boldsymbol\lambda_{2\Sigma}^{\pm}\neq 0$ on $\cE_{\rm TT}\times \cF_{\rm TT}$. 
\een
Therefore $\omega_{\eucl}$ is {\em weakly} but {\em not strongly gauge invariant}.
\ben\setcounter{smallarabics}{4}
\item
$\omega_{\rm eucl}$ is invariant under $SO^{\uparrow}(1,4)$ and $O(4)$.
\een
\end{theoreme}
\proof 
(1) follows from  Props.~\ref{prop2.2}  and \ref{prop2.1} {\it i)}. (2) is proved in Props. \ref{prop5.3} and \ref{prop-pos1}. from Thm.~\ref{thm-microloc} and the results recalled in \ref{sec1.7.3}. (3) follows from Thm.~\ref{thm-microloc} and the results recalled in \ref{sec1.7.3}, (4) from Prop. \ref{prop10.1b} and (5) from Prop. \ref{prop-invar}. \qed

\subsubsection{Commutation properties}
From the fact that $[\tD_{2}, \hat{D}_{2}]=0$ and that $\hat{D}_{2}$ acts only in the spatial variables, we obtain immediately that
\begin{equation}
\label{ecle.1}
[\one_{\{\lambda\}}(\hat{D}_{2})\otimes \one_{\cc^{2}}, c_{2}^{\pm}]=0,
\end{equation}
the same property being true for $\tc_{2}^{\pm}$.

\subsection{Gauge invariance}\label{sec5.3b}
In the next proposition we prove that  as a consequence of the existence of Killing $1$-forms, $c_{2}^{\pm}$ do \emph{not} satisfy the strong TT gauge invariance property in the sense that
$$
c_{2}^{\pm}: \cF_{\rm TT}\to \Ran K_{21 \Sigma}+ \Ran K_{20 \Sigma}.
$$
However, $c_{2}^{\pm}$ satisfy the weak TT gauge invariance  explained  in \ref{sss:newTTc}.

Recall that the spaces $\mathcal{K}_{\Sigma}$ and $\mathcal{K}_{\Sigma}^{{\bf q}_{1}}$ are defined in Def. \ref{defcor.1}.
 \begin{proposition}\label{prop10.1b}
One has
 \[
 c_{2}^{\pm}K_{21\Sigma}= K_{21\Sigma}c_{1}^{\pm}\hbox{ on }\mathcal{K}_{\Sigma}^{{\bf q}_{1}},
 \]
 hence
\[\boldsymbol \lambda_{2\Sigma}^{\pm}= 0\hbox{ on }\cE_{\rm TT}\times \cF_{\rm TT, \rm gauge}.
\]
 \end{proposition}
\proof 
We will  work in the Euclidean framework, and   use the results in \ref{sec3.2.3}. If $f_{1}\in \tilde{\mathcal K}_{\Sigma}^{{\bf q}_{1}}$ we can set
\[
u_{1}= \tD_{1}^{-1}\trho_{1}^{*}\tilde{\sigma}_{1}f_{1},
\]
since $\trho_{1}^{*}\tilde{\sigma}_{1}f_{1}$ is orthogonal to $\Ker \tD_{1}$, see \eqref{e2.002}. We have $\tD_{1}u_{1}= 0$ in $\Omega^{\pm}$ and $\trho_{1}^{\pm}u_{1}= \tc_{1}^{\pm}f_{1}$. Setting  $u_{2}= \tK_{21}u_{1}$ we have $\trho_{2}^{\pm}u_{2}= \tK_{21\Sigma}\trho_{1}^{\pm}u_{1}= \tK_{21\Sigma}\tc_{1}^{\pm}f_{1}$ since $\tD_{1}u_{1}\traa{\Sigma}=0$.  Since $\tD_{2}u_{2}=\tK_{21}\tD_{1}u_{1}= 0$ in $\Omega^{\pm}$  we have $\trho_{2}^{\pm}u_{2}= \tc_{2}^{\pm}\trho_{2}^{\pm}u_{2}$ which using that $\tc_{2}^{+}+ \c_{2}^{-}= \one$  and $\tc_{1}^{+}+ \tc_{1}^{-}= \one$ on $\tilde{\mathcal K}_{\Sigma}^{{\bf q}_{1}}$ implies that 
$\tc_{2}^{\pm}\tK_{21}f_{1}= \tK_{21\Sigma}\tc_{1}^{\pm}f_{1}$. \qed

\begin{corollary}\label{corr.1}
The Cauchy surface covariances $\boldsymbol\lambda_{2\Sigma}^{\pm}$  in Def.~\ref{def5b.1} satisfy the weak TT gauge invariance property  but do not satisfy the strong TT gauge invariance property.
\end{corollary}
\proof
We work in the Euclidean framework. We claim that $\Ran \tK_{21\Sigma}^{\dag}\subset \tilde{\mathcal K}_{\Sigma}^{{\bf q}_{1}}$. In fact if $h_{1}\in \tilde{\mathcal K}_{\Sigma}$ then $h_{1}= \trho_{1}u_{1}$ for $u_{1}\in \Ker \tD_{1}= \Ker d$, see Prop. \ref{prop1.0} so $\tK_{21\Sigma}h_{1}= \trho_{2} d u_{1}=0$. Therefore $\bar{h}_{1}\dual {\bf q}_{1}\tK_{21\Sigma}^{\dag}f_{2}= \bar{\tK_{21\Sigma}h_{1}}\dual {\bf q}_{1}f_{2}=0$ so $\tK_{21\Sigma}^{\dag}f_{2}\in \tilde{\mathcal K}_{\Sigma}^{{\bf q}_{1}}$.

By Prop. \ref{prop10.1b} we have $c_{2}^{\pm}: K_{21\Sigma}\mathcal{K}_{\Sigma}^{{\bf q}_{1}}\to \Ran K_{21\Sigma}$ hence 
\[
c_{2}^{\pm}: \Ran K_{21\Sigma}K_{21\Sigma}^{\dag}\to  \Ran K_{21\Sigma},
\] so $\boldsymbol\lambda_{2\Sigma}^{\pm}$  in Def.~\ref{def5b.1} satisfy the weak TT gauge invariance property.

Next from Prop. \ref{prop-pos1} below there exists $f\in \cE_{{\rm TT}, 4}$ such that $\bar{f}\dual\boldsymbol{\lambda}^{\pm}_{2\Sigma}f\neq 0$. By Prop. \ref{propcor.2} we know that $\cE_{{\rm TT}, 4}\in \cF_{\TT}$, so $f= K_{21\Sigma}g$ and $\bar{f}\dual\boldsymbol{\lambda}^{\pm}_{2\Sigma}K_{21\Sigma}g\neq 0$. This contradicts the strong gauge invariance property of $\boldsymbol\lambda_{2\Sigma}^{\pm}$. \qed
\subsection{Positivity}
In this subsection we investigate the positivity of  the Hermitian forms $\boldsymbol\lambda_{2\Sigma}^{\pm}$ on $\cE_{\rm TT}$. 

We use the decomposition \eqref{decodeco} {\it ii)} of $\cE_{\rm TT}$  in Prop. \ref{propcor.2}.  The important observation is that since $\hat{D}_{2}$ is spatial and commutes with $\tD_{2}$ we have
\begin{equation}
\label{commute}
[c_{2}^{\pm}, \one_{\{\lambda\}}(\hat{D}_{2})\otimes \one_{\cc^{2}}]=0. 
\end{equation}

 \subsubsection{Positivity \texorpdfstring{on $\cE_{\rm TT,  gauge}$}{}}\label{sec5.5}
\begin{proposition}
 \label{prop5.3}
 We have
\[
\boldsymbol \lambda_{2\Sigma}^{\pm}\geq 0\hbox{ on }\cE_{\rm TT, \rm gauge}.
\]
\end{proposition}
\proof Let $f\in \cE_{\rm TT, \rm gauge}$ and $\tf= \cF_{2\Sigma}f\in \tcE_{\rm TT, \rm gauge}$. Using that $(c_{2}^{\pm})^{2}= c_{2}^{\pm}$, $[c_{2}^{\pm}, I_{\Sigma}]= 0$ and ${\bf q}_{2}c_{2}^{\pm}= (c_{2}^{\pm})^{*}{\bf q}_{2}$ we obtain that:
\[
\bar{f}\dual\boldsymbol \lambda_{2\Sigma}^{\pm}f=\pm\bar{f}\dual{\bf q}_{2}I_{\Sigma}c_{2}^{\pm}f= \pm\bar{c_{2}^{\pm}f}\dual {\bf q}_{2}c_{2}^{\pm}f= 
 \pm\bar{\tc_{2}^{\pm}\tf}\dual {\bf q}_{2}\tc_{2}^{\pm}\tf.
\]
By \eqref{commute} we know that $\tc_{2}^{\pm}\tf\in \Ran \one_{I_{2}}(\hat{D}_{2}\otimes\one_{\cc^{2}})$, so if $g=\col{g_{0}}{g_{1}}= \tc_{2}^{\pm}\tf$ we have
$g_{iss}= g_{is\Sigma}=0$ by \eqref{ecle.0}. Therefore
\[
\bar{f}\dual\boldsymbol \lambda_{2\Sigma}^{\pm}f=\pm\bar{\tc_{2}^{\pm}\tf}\dual {\bf q}_{2}\tc_{2}^{\pm}\tf= \pm\bar{\tc_{2}^{\pm}\tf}\dual \tilde{\bf q}_{2}\tc_{2}^{\pm}\tf.
\]
We set then $u= \tD_{2}^{-1}\trho_{2}^{*}\tilde{\sigma}_{2}\tf$, so that $\tc_{2}^{\pm}\tf= \mp\trho_{2}^{\pm}u$. By Lemma \ref{lemma1.1} (2), we obtain that
\[
\pm\bar{\tc_{2}^{\pm}\tf}\dual \tilde{\bf q}_{2}\tc_{2}^{\pm}\tf= 2Q(u, u)_{\Omega^{\pm}},
\]
where $Q$ is the sesquilinear form associated to $\tD_{2}$, see \eqref{e2.1}. To complete the proof we use that 
 $\tD_{0}(\trg| u)= (\trg| \tD_{2}u)= 0$ in $\Omega^{\pm}$ and $\trho_{0}^{\pm}(\trg| u)=0$ since $\tf$ and hence $\tc_{2}^{\pm}\tf$ belong to $\Ker \tK_{20}^{\dag}$. Therefore $(\trg| u)=0$ in $\Omega^{\pm}$. By \eqref{e0.5b} we  have:
 \[
 Q(u, u)_{\Omega^{\pm}}= \int_{\Omega^{\pm}}\big(\tnab_{a}\bar{u}\trg^{ab}\tnab_{b}u+ 2\|u\|^{2}\big)d\vol_{\trg}
\geq 0,
 \] 
 which shows that $\bar{f}\dual\boldsymbol\lambda_{2}^{\pm}f\geq 0$. \qed

\subsubsection{Negativity on  $\cE_{\rm TT ,4}$}
\begin{proposition}\label{prop-pos1}
 The covariances $\boldsymbol\lambda_{2\Sigma}^{\pm}$ are {\em negative} on $\cE_{\rm TT ,4}$, and $\bar{f}\cdot \boldsymbol\lambda_{2\Sigma}^{\pm}f=0$  implies that $c_{2}^{\pm}f= 0$  for $f\in \cE_{\rm TT ,4}$. In particular $\boldsymbol\lambda_{2\Sigma}^{+}+ \boldsymbol\lambda_{2\Sigma}^{-}$ is negative definite on $\cE_{\rm TT ,4}$.
\end{proposition}
\proof
Let $f\in \cE_{\TT, 4}$ and $\tf= \cF_{2\Sigma}f\in \tcE_{\TT, 4}$. As in the proof of Prop. \ref{prop5.3} we have
\[
\bar{f}\dual\boldsymbol \lambda_{2\Sigma}^{\pm}f= \pm\bar{\tc_{2}^{\pm}\tf}\dual {\bf q}_{2}\tc_{2}^{\pm}\tf.
\]
Again by \eqref{commute} we know that $\tc_{2}^{\pm}\tf\in \Ran \one_{\{4\}}(\hat{D}_{2}\otimes\one_{\cc^{2}})$ , so if $g=\col{g_{0}}{g_{1}}= \tc_{2}^{\pm}\tf$ we have
$g_{iss}= g_{i\Sigma\Sigma}=0$ by \eqref{ecle.2}. Therefore
\[
\bar{f}\dual\boldsymbol \lambda_{2\Sigma}^{\pm}f= \pm\bar{\tc_{2}^{\pm}\tf}\dual {\bf q}_{2}\tc_{2}^{\pm}\tf= \mp\bar{\tc_{2}^{\pm}\tf}\dual \tilde{\bf q}_{2}\tc_{2}^{\pm}\tf.
\]
Again we set 
 $u=  \tD_{2}^{-1}\trho_{2}^{*}\tilde{\sigma}_{2}\tf$ so that $\tc_{2}^{\pm}\tf= \mp\trho_{2}^{+}u$ and 
 \[
 \mp\bar{\tc_{2}^{\pm}\tf}\dual \tilde{\bf q}_{2}\tc_{2}^{\pm}\tf= -2Q(u, u)_{\Omega^{\pm}}.
 \]
  As before we obtain that $(\trg| u)=0$ in $\Omega^{\pm}$
  and  \[
 Q(u, u)_{\Omega^{\pm}}= \int_{\Omega^{\pm}}\big(\tnab_{a}\bar{u}\trg^{ab}\tnab_{b}u+ 2\|u\|^{2}\big)d\vol_{\trg}
\geq 0,
 \]
 which shows that $\bar{f}\dual\boldsymbol \lambda_{2\Sigma}^{\pm}f\leq 0$. If $\bar{f}\dual\boldsymbol \lambda_{2\Sigma}^{\pm}f=0$ then $u= 0$ in $\Omega^{\pm}$ so $\tc_{2}^{\pm}\tf=0$ as claimed. \qed

\subsection{Invariance under symmetries}
We now discuss the invariance of $\omega_{\rm eucl}$ under the isometry group $O(1, 4)$.
\begin{proposition}\label{prop-invar}
 The pseudo-state $\omega_{\rm eucl}$ is invariant under the action of $SO^{\uparrow}(1, 4)$ and of $O(4)$.
\end{proposition}
  \proof Let $X$ be a generator of ${\rm \textit{so}}(1, 4)$ viewed as a Killing vector field of $dS^{4}$.  We denote by $\phi(\cdot)$ its associated flow on $\cinf(dS^{4}; V_{2})$ and define its generator $L$ by $\phi(\sigma)\eqdef \e^{\i \sigma L}$.  Note that  $L$ commutes with $D_{2}$ and $I$.
  In particular we can define its Cauchy surface version $L_{\Sigma}$ by 
  \[
  L_{\Sigma}= \varrho_{2}L U_{2}.
  \]
  Clearly $[L_{\Sigma}, I_{\Sigma}]=0$ and  $L_{\Sigma}^{*}q_{2\Sigma}= q_{2\Sigma}L_{\Sigma}$. 
  The Wick rotation  of $X$ denoted by $\tilde{X}$ is a Killing vector field of $\bS^{4}$. We use the notation introduced in Subsect.~\ref{sec3.4}. By Lemma \ref{killing-lemma},  $\tilde{L}_{\Sigma}$ commutes with  the \calde projectors $\tc_{2}^{\pm}$.
By analytic continuation, this implies that $L_{\Sigma}$ commutes with the Hadamard projectors $c_{2}^{\pm}$, hence that 
\[
L_{\Sigma}^{*}\boldsymbol\lambda^{\pm}_{2\Sigma}= \boldsymbol\lambda_{2\Sigma}^{\pm}L_{\Sigma},
\]
i.e.~$\boldsymbol\lambda_{2\Sigma}^{\pm}$ and the associated pseudo-state $\omega_{\rm eucl}$ are invariant under the action of $SO^{\uparrow}(1, 4)$.     

  The action of $O(4)$ on tensors commutes with Wick rotation, since it preserves the variable $t$. It is immediate that the \calde projectors $\tc_{2}^{\pm}$ are invariant under the action of $O(4)$ on Cauchy data, which implies as above that $\boldsymbol\lambda_{2\Sigma}^{\pm}$  and $\omega_{\rm eucl}$ are invariant under the action of $O( 4)$.  \qed
  
   \subsubsection{Invariance under time reversal}
   We recall that time reversal  $Z$ is defined in \ref{def-de-time-reversal}.   We denote by $Z_{\Sigma}$ its Cauchy surface version, i.e.~$Z_{\Sigma}= \varrho_{2}ZU_{2}$. 
     \begin{proposition} The Euclidean vacuum pseudo state 
 $\omega_{\rm eucl}$ is invariant under time reversal, i.e.
 \[
 \bar{Z_{\Sigma} f}\dual \boldsymbol\lambda_{2\Sigma}^{\pm}Z_{\Sigma}g= \bar{g}\dual \boldsymbol\lambda_{2\Sigma}^{\pm}f, \quad f, g\in \coinf(\Sigma; V_{2}\otimes \cc^{2}).
 \]
 
\end{proposition}
\proof   
 We have
  \[
  (Z_{\Sigma}f)_{itt}= \bar{f_{itt}},\quad    (Z_{\Sigma}f)_{it\Sigma}= -\bar{f_{it\Sigma}},\quad   (Z_{\Sigma}f)_{i\Sigma\Sigma}= \bar{f_{i\Sigma\Sigma}}, \ \ i= 0, 1.
  \]
   We claim that $Z_{\Sigma}^{-1} c_{2}^{\pm}Z_{\Sigma}= c_{2}^{\pm}$, which is equivalent to $\tilde{Z}_{\Sigma}^{-1}\tc_{2}^{\pm}\tilde{Z}_{\Sigma}= \tc_{2}^{\pm}$ for $\tilde{Z}_{\Sigma}= \cF_{\Sigma}Z_{\Sigma}\cF_{\Sigma}^{-1}$.
 Using the expression of     $\cF_{2}$ in  \eqref{def-de-F_k} we obtain that 
 \beq\label{def-de-T-tilde}
 \tilde{Z}_{\Sigma}f= -\bar{f}.
 \eeq
Since the  Wick rotated operator $\tD_{2}$ is real we obtain that  $\tilde{Z}_{\Sigma}^{-1}\tc_{2}^{\pm}\tilde{Z}_{\Sigma}= \tc_{2}^{\pm}$  Using  then that $Z_{\Sigma}$ is anti-unitary for ${\bf q}_{I, 2}$, this implies the proposition. \qed

\begin{remark}
Let us explain our choice for the time reversal, in connection with real fields. We follow  the exposition given in \cite[Sect.~4.7]{G}. Let $(\cX, \boldsymbol\sigma)$  be a real symplectic space.  A quasi-free state $\omega$ on $\CCR(\cX, \boldsymbol\sigma)$ is determined by its covariance $\boldsymbol\eta\in L_{\rm s}(\cX, \cX')$  defined by $\omega(\phi(x_{1})\phi(x_{2}))= x_{1}\dual \boldsymbol\eta x_{2}+ \frac{\i}{2} x_{1}\dual \boldsymbol\sigma x_{2}$.

Then $(\cc\cX, {\bf q})$ is a Hermitian space, where ${\bf q}= \i \boldsymbol\sigma_{\cc}$ and $\boldsymbol\sigma_{\cc}$ is the sesquilinear extension of $\boldsymbol\sigma$. The extension $\tilde{\omega}$ of $\omega$ to $\CCR(\cc\cX, \i \sigma_{\rm \cc})$ has covariances $\boldsymbol\lambda^{\pm}= \boldsymbol\eta_{\cc}\pm \12 {\bf q}$.  

We apply this to $\cX= \cinf(\Sigma; V_{2, \rr}\otimes \rr^{2})$, the space of Cauchy data of real tensors. We obtain that
\[
\bar{Z_{\Sigma}f}\dual \boldsymbol\eta_{\cc}Z_{\Sigma}g= \overline{\bar{f}\dual \boldsymbol\eta_{\cc}g},
\]
hence if $f, g$ are real tensors we have $Z_{\Sigma}f\dual \boldsymbol \eta Z_{\Sigma}g= f\dual \boldsymbol \eta g$, which is the usual 
meaning of invariance under time reversal. 
\end{remark}
  \subsection{\texorpdfstring{$\alpha$-}{alpha }vacua}
 We now define $\alpha$-{\em vacua}, obtained from $\omega_{\rm eucl}$ by a Bogoliubov transformation, see \cite{Allen1985} for the case of scalar fields.
 Usually $\alpha$-vacua are defined via a mode expansion. It is very easy to define them directly, as we will do in this subsection. We refer the reader to \cite[Sect.~6.4]{G} for a discussion of the mode expansion construction, in connection with the approach used in this paper. 
 
We denote by $S: \coinf(dS^{4}; V_{k})$ the {\em linear} time reversal
\[
Su_{k}= \tau^{*}u_{k}, \ \  u_{k}\in \coinf(dS^{4}; V_{k}).
\]
It is often called the {\em Racah time reversal}.

Again $S$ commutes with $D_{k}$, $K_{ij}, K_{ij}^{\star}$. Denoting by $S_{\Sigma}$ its Cauchy surface version, we see that $S_{\Sigma}$ is well defined on $\cE_{\rm TT}$ and on the quotient space ${\cE_{\rm TT}}/{\cF_{\rm TT}}$.

We also see that  $S_{\Sigma}^{*}{\bf q}_{I, 2}S_{\Sigma}= - {\bf q}_{I, 2}$, which implies that $U_{\alpha}= \e^{\alpha S_{\Sigma}}= \cosh\!\alpha \one + \sinh \!\alpha S_{\Sigma}$  is unitary  for ${\bf q}_{I, 2}$ for $\alpha\in \rr$ .

\begin{definition}
 The $\alpha$-{\em vacuum} pseudo-state $\omega_{\alpha,{\rm eucl}}$ for $\alpha\in \rr$ is defined by the covariances
 \[
 \boldsymbol\lambda_{\alpha, 2\Sigma}^{\pm}\defeq U_{\alpha}^{*}\boldsymbol\lambda_{2\Sigma}^{\pm}U_{\alpha}.
 \]
 \end{definition}
  We now discuss the properties of $\boldsymbol\lambda_{\alpha, 2\Sigma}^{\pm}$.
 
\begin{proposition}\label{prop-alpha-vac}
 $\boldsymbol\lambda_{\alpha, 2\Sigma}^{\pm}$ satisfy the same properties as $\boldsymbol\lambda_{2\Sigma}^{\pm}$ expect for the Hadamard condition. \end{proposition}
\proof The fact that $\boldsymbol\lambda_{\alpha, 2\Sigma}^{+}-\boldsymbol\lambda_{\alpha, 2\Sigma}^{-}= {\bf q}_{I, 2}$ follows from the fact that $U_{\alpha}$ is unitary for $ {\bf q}_{I, 2}$. 

 $U_{\alpha}$ commutes with the spectral projections $\one_{\{\lambda\}}(\hat{D}_{2})\otimes \one_{\cc^{2}}$, hence preserves the subspaces $\cE_{\rm TT, \rm reg}$, $\cE_{\rm TT ,3}$ and $\cE_{\rm TT ,4}$, so the positivity properties of $\boldsymbol\lambda_{\alpha, 2\Sigma}^{\pm}$ are unchanged. 

The invariance properties under the action of $O(1, 4)$ are also unchanged. Finally the Hadamard condition in Prop. \ref{prop:states1} {\it iv)} is not satisfied, because $S$ exchanges $\cN^{+}$ with $\cN^{-}$. \qednoskip

\section{The modified Euclidean vacuum}\init\label{sec5}
In this section we show how to modify the Euclidean pseudo state $\omega_{\rm eucl}$ to obtain a true, positive and gauge invariant state $\omega_{\rm mod}$ on $\cE_{\rm TT}$.  One can view this modification as imposing an additional gauge fixing condition. The price to pay is that $\omega_{\rm mod}$ is no more invariant under the full isometry group $O(1,4)$, but only under  the subgroup $O(4)$.

\begin{definition}\label{def5.1}
 We set 
 \[
 \bar{f}\dual \boldsymbol\lambda_{2\Sigma,{\rm mod}}^{\pm}f\defeq \bar{\pi f}\dual\boldsymbol \lambda_{2\Sigma}^{\pm}\pi f, \  \ f\in \cE_{\rm TT},
 \]
 where 
 \[
 \pi\defeq \one_{\rr\setminus\{4\}}(\hat{D}_{2}\otimes\one_{\cc^{2}}):\cE_{\TT}\to \cE_{\rm TT, \rm gauge}\oplus \cF_{\rm TT, \rm gauge}.
 \]
\end{definition}
\begin{theoreme}\label{thm5.1}
We have:
\ben
\item $\boldsymbol\lambda_{2\Sigma,{\rm mod}}^{\pm}= \boldsymbol\lambda_{2\Sigma,{\rm mod}}^{\pm*}$, $\boldsymbol\lambda_{2\Sigma,{\rm mod}}^{+}- \boldsymbol\lambda_{2\Sigma,{\rm mod}}^{-}= {\bf q}_{I, 2}$ on $\cE_{\TT}$.
\item $\boldsymbol\lambda_{2\Sigma,{\rm mod}}^{\pm}\geq 0$ on $\cE_{\rm TT}$.
\een
Therefore $\boldsymbol\lambda_{2\Sigma,{\rm mod}}^{\pm}$ are the Cauchy surface covariances of a state $\omega_{\rm mod}$ on $\cE_{\rm TT}$. Moreover,
\ben\setcounter{smallarabics}{2}
\item $\omega_{\rm mod}$ is a Hadamard state.
\item $\omega_{\rm mod}$ is {\em fully gauge invariant}, i.e.  
\[
\bar{f}\dual \boldsymbol\lambda_{2\Sigma,{\rm mod}}^{\pm}K_{21\Sigma}g=0 \ \  \forall f\in \cE_{\rm TT}, \ g\in \cinf(\Sigma; V_{1}\otimes \cc^{2}).
\]
\item  $\omega_{\rm mod}$ is invariant under the action of $O(4)$.
\een

\end{theoreme}
\proof  The fact that $\boldsymbol\lambda_{2\Sigma,{\rm mod}}^{\pm}$ are Hermitian is obvious. We have
\[
\boldsymbol\lambda_{2\Sigma,{\rm mod}}^{+}- \boldsymbol\lambda_{2\Sigma,{\rm mod}}^{-}= \pi ^{*}{\bf q}_{I, 2}\pi ,
\]
and since $\pi f- f\in \cF_{\TT}= \Ker {\bf q_{I, 2}}_{|\cE_{\rm TT}}$, see Prop.   \ref{propcor.3}, we have  $\pi ^{*}{\bf q}_{I, 2}\pi ={\bf q}_{I, 2}$ on $\cE_{\TT}$. This completes the proof of (1).

(2) follows from the fact that $\boldsymbol \lambda_{2\Sigma}^{\pm}\geq 0$ on $\cE_{\rm TT, \rm gauge}$ and $\dual\boldsymbol \lambda_{2\Sigma}^{\pm}=0$ on $\cE_{\TT}\times \cF_{\rm TT, \rm gauge}$.

 (3) follows from the fact that the spacetime two-point functions $\Lambda_{2{\rm eucl}}^{\pm}$ and $\Lambda_{2\Sigma,{\rm mod}}^{\pm}$ associated to $\omega_{\rm eucl}$ and $\omega_{\rm mod}$ differ by a finite rank, smoothing operator.

(4) follows from the fact that $\pi:\cF_{\rm TT}\to \cF_{\rm TT, \rm gauge}$ and that $\boldsymbol \lambda_{2\Sigma}^{\pm}=0$ on $\cE_{\TT}\times \cF_{\rm TT, \rm gauge}$.

Finally let $\alpha$ be the  (linear) action  of an element of $O(4)$ on tensors and $\alpha_{\Sigma}$ its Cauchy surface version. Since  the action of $O(4)$ preserves $t= {\rm const}$, we have $\alpha_{\Sigma}= \mat{\alpha}{0}{0}{\alpha}$. Clearly $\alpha$ commutes with $\hat{D}_{2}$ hence $\alpha_{\Sigma}$ commutes with $\pi$. (5) follows then from   Prop.~\ref{prop-invar}. \qed

\begin{remark}
 One can wonder if $\omega_{\rm mod}$ is  invariant under the action of $SO^{\uparrow}(1, 4)$. To check this, one chooses a Killing vector field $X$ on $dS^{4}$ which is not tangent to $\Sigma$. The Wick rotation $\tilde{X}$ of $X$ is then of the form  $\tilde{X}= \trg^{-1}\varphi_{i}$, where $\varphi_{i}$ is one of the Killing $1$-forms on $\bS^{4}$. To compute the associated operator $L_{\Sigma}$ defined in the proof of Prop.~\ref{prop-invar}, one can as well compute its 
 Euclidean version $\tilde{L}_{\Sigma}= \mathcal{F}_{2\Sigma}\circ L_{\Sigma}\circ \mathcal{F}_{\Sigma}^{-1}$  associated to the Killing vector field $\tilde{X}$ on $\bS^{4}$. Using the expression of $\tilde{L}_{\Sigma}$ one can  
 then check that $\cE_{\rm TT, \rm reg}$ is {\em not invariant} under the action of $L_{\Sigma}$, so  the projection $\pi$ does not commute with the action of $\e^{ s L_{\Sigma}}$. Therefore $\omega_{\rm mod}$ is not invariant under the action of $\e^{ s L_{\Sigma}}$.
 \end{remark}

\appendix\section{Auxiliary results}  \label{s:appA}
\subsection{Various proofs}\label{proofs}
    \subsubsection{Covariant derivatives \texorpdfstring{on $\bS^{4}$}{}}\label{sec4.0.1}
    Recall that  we write  $w\in \cinf(\bS^{4}; \tV_{1})$ as $w_{s}ds+ w_{\Sigma}$  and identify $w$ with $(w_{s}, w_{\Sigma})$ . Similarly we write $u\in\cinf(\bS^{4}; \tV_{2})$ as $u_{ss}ds^{2}+ ds\otimes w_{s\Sigma}+ u_{s\Sigma}\otimes ds+ u_{\Sigma\Sigma}$ and identify $u$ with $(u_{ss}, u_{s\Sigma}, u_{\Sigma\Sigma})$. 
     
   We denote  by $\nabla$, resp.~$\vec{\nabla}$, the covariant derivative on $(\bS^{4}, \trg)$, resp.  $(\bS^{3}, \rh)$, and set $\nabla_{0}= \nabla_{\p_{s}}$. Then a straightforward computation shows that if $X\in T\Sigma$:
  \begin{equation}
  \label{e1.10}
  \begin{array}{l}
  (\nabla_{0}w)_{s}= \p_{s}w_{s},\
  (\nabla_{0}w)_{\Sigma}= \p_{s}w_{\Sigma}- \12 a^{-1}\dot{a}w_{\Sigma},\\[2mm]
  (\nabla_{X}w)_{s}= \vec{\nabla}_{X}w_{s}- \12 a^{-1}\dot{a}w_{\Sigma}X,\
  (\nabla_{X}w)_{\Sigma}= \vec{\nabla}_{X}w_{\Sigma}+ \12 \dot{a}\rh X w_{s},
  \end{array}
  \end{equation}
     (where $a(s)= \cos^{2}s$), and
     \begin{equation}
  \label{e1.10b}
  \begin{array}{l}
  (\nabla_{0}u)_{ss}= \p_{s}u_{ss},\
  (\nabla_{0}u)_{s\Sigma}= \p_{s}u_{s\Sigma}- \12 a^{-1}\dot{a}u_{s\Sigma},\
 ( \nabla_{0}u)_{\Sigma\Sigma}= \p_{s}u_{\Sigma\Sigma}- a^{-1}\dot{a}u_{\Sigma\Sigma},\\[2mm]
  (\nabla_{X}u)_{ss}= \vec{\nabla}_{X}u_{ss}- a^{-1}\dot{a}u_{s\Sigma}X,\
  (\nabla_{X}u)_{s\Sigma}= \vec{\nabla}_{X}u_{s\Sigma}-\12 a^{-1}\dot{a}u_{\Sigma\Sigma}X+ \12 \dot{a}\rh X u_{ss},\\[2mm]
  (\nabla_{X}u)_{\Sigma\Sigma}= \vec{\nabla}_{X}u_{\Sigma\Sigma}+ \12 \dot{a}\rh X\otimes u_{s\Sigma}+ \12 \dot{a} u_{s\Sigma}\otimes \rh X.
  \end{array}
  \end{equation}

\subsubsection{Proof of Lemma \ref{lemmi}}\label{sec4.0.2}
  We deduce from \eqref{e1.10b} that
 \beq\label{e.tit2}
  \begin{array}{l}
  (\delta u)_{s}= - 2 \p_{s}u_{ss}+ 2a^{-1}\vec{\delta}u_{s\Sigma}+ \12\dot{a}a^{-2}(\rh|u_{\Sigma\Sigma})- 3 \dot{a}a^{-1}u_{ss},\\[2mm]
  (\delta u)_{\Sigma}= - 2 \p_{s}u_{s\Sigma}+ \vec{\delta}u_{\Sigma\Sigma}-3\dot{a}a^{-1}u_{s\Sigma},
  \end{array}
  \eeq
  for $a(s)= \cos^{2}(s)$  hence:
  \[
  \begin{array}{l}
  (\delta u)_{s}\traa{\Sigma}=(-2\p_{s}u_{ss}+ 2\vec{\delta} u_{s\Sigma})\traa{\Sigma},\\[2mm]
  (\delta u)_{\Sigma}\traa{\Sigma} =(-2\p_{s}u_{s\Sigma}+ \vec{\delta}u_{\Sigma\Sigma})\traa{\Sigma},
  \end{array}
  \]
  and
  \[
  \begin{array}{l}
   -\p_{s}(\delta u)_{s}\traa{\Sigma}= (2\p^{2}_{s}u_{ss} - 2\vec\delta\p_{s}u_{s\Sigma}- 6 u_{ss}+  (\rh| u_{\Sigma\Sigma})\traa{\Sigma},\\[2mm]
   - \p_{s}(\delta u)_{\Sigma}\traa{\Sigma}= (2\p_{s}^{2}u_{s\Sigma}- \vec{\delta}\p_{s}u_{\Sigma\Sigma}- 6 u_{s\Sigma})\traa{\Sigma}.
  \end{array}
  \]
    If  $\tD_{2}u\traa{\Sigma}=0$ then
   \[
    \begin{array}{l}
    -\p_{s}^{2}u_{ss}\traa{\Sigma}= -\vec{D}_{0, L}u_{ss}\traa{\Sigma}+ (\rh| u_{\Sigma\Sigma})\traa{\Sigma},\\[2mm]
    -\p^{2}_{s}u_{s\Sigma}\traa{\Sigma}= - (\vec{D}_{1, L}-1)u_{s\Sigma}\traa{\Sigma},\\[2mm]
    - \p_{s}^{2}u_{\Sigma\Sigma}\traa{\Sigma}= - (\vec{D}_{2, L}- 6)u_{\Sigma\Sigma}- 2\rh u_{ss}\traa{\Sigma},
    \end{array}
    \]
    which proves the lemma. \qed
\subsubsection{Proof of Lemma \ref{lemmo}}\label{zlot}
Again, we deduce from \eqref{e1.10} that
\beq\label{e.tit3}
\begin{array}{l}
(d w)_{ss}= \p_{s}w_{s},\ (dw)_{s\Sigma}
= \12(\p_{s}w_{\Sigma}- \dot{a}a^{-1}w_{\Sigma}+ \vec{d}w_{s}),\\[2mm]
(dw)_{\Sigma\Sigma}= \vec{d}w_{\Sigma}+ \12 \dot{a}\rh w_{s},
\end{array}
\eeq
hence
\[
\begin{array}{l}
(dw)_{ss}\traa{\Sigma}= \p_{s}w_{s}\traa{\Sigma},\ 
(dw)_{s\Sigma}\traa{\Sigma}= \12(\p_{s}w_{\Sigma}+ \vec{d}w_{s})\traa{\Sigma},\ 
(dw)_{\Sigma\Sigma}\traa{\Sigma}= \vec{d}w_{\Sigma}\traa{\Sigma},
\end{array}
\]
and
\[
\begin{array}{l}
-(\p_{s}dw)_{ss}\traa{\Sigma}= -\p^{2}_{s}w_{s}\traa{\Sigma},\\[2mm]
-(\p_{s}dw)_{s\Sigma}\traa{\Sigma}= \12(-\p^{2}_{s}w_{\Sigma}- 2w_{\Sigma}- \vec{d}\p_{s}w_{s})\traa{\Sigma},\\[2mm]
-(\p_{s}dw)_{\Sigma\Sigma}\traa{\Sigma}= -\vec{d}\p_{s}w_{\Sigma}+ \rh w_{s}.
\end{array}
\]
If $\tD_{1}w\traa{\Sigma}=0$
\[
\begin{array}{l}
-\p^{2}_{s}w_{s}\traa{\Sigma}=-(\vec{D}_{0,L}- 3)w_{s}\traa{\Sigma},\ 
-\p^{2}_{s}w_{\Sigma}\traa{\Sigma}=-(\vec{D}_{1,L}- 6)w_{\Sigma}\traa{\Sigma}.
\end{array}
\]
It follows that if $g= \trho_{2}dw$ and $f= \trho_{1}w$ we have
\beq\label{eshit.1}
\begin{array}{l}
g_{0ss}= - f_{1s},\
g_{0s\Sigma}= \12 (-f_{1\Sigma}+ \vec{d}f_{0s}),\ 
g_{0\Sigma\Sigma}= \vec{d}f_{0\Sigma},\\[2mm]
g_{1ss}= -(\vec{D}_{0,L}-3)f_{0s},\ 
g_{1s\Sigma}= \12(-(\vec{D}_{1,L}-4)f_{0\Sigma}+ \vec{d}f_{1s}),\ 
g_{1\Sigma\Sigma}= \vec{d}f_{1\Sigma}+ \rh f_{0s}.
\end{array}
\eeq
To complete the computation of $\tK_{21\Sigma}$ we need to compute $\trho_{2}\tilde{I}dw$, where $\widetilde{I}u = u - \frac{1}{4}(\trg| u)_{\tV_{2}}\trg$, i.e.
\[
\begin{array}{l}
(\widetilde{I}u)_{ss}= \12 u_{ss}- \frac{1}{4} (\rh|u_{\Sigma\Sigma})_{\tV_{2\Sigma}},\
(\widetilde{I}u)_{s\Sigma}= u_{s\Sigma},\\[2mm]
(\widetilde{I}u)_{\Sigma\Sigma}= u_{\Sigma\Sigma}- \frac{1}{4}(2u_{ss}+ (\rh| u_{\Sigma\Sigma})_{\tV_{2\Sigma}})\rh.
\end{array}
\]
If $g= \trho_{2}\widetilde{I}dw$ and $f= \trho_{1}w$, we obtain:
\beq\label{eshit.2}
\begin{array}{l}
g_{0ss}= \12(- f_{1s}+ \vec{\delta}f_{0\Sigma}),\
g_{0s\Sigma}= \12 (-f_{1\Sigma}+ \vec{d}f_{0s}),\\[2mm]
g_{0\Sigma\Sigma}= \vec{d}f_{0\Sigma}+ \12 (f_{1s}+ \vec{\delta}f_{0\Sigma})\rh,\\[2mm]
g_{1ss}= \12(-\vec{D}_{0,L}f_{0s}+ \vec{\delta}f_{1\Sigma}),\
g_{1s\Sigma}= \12(-(\vec{D}_{1,L}-4)f_{0\Sigma}+ \vec{d}f_{1s}),\\[2mm]
g_{1\Sigma\Sigma}= \vec{d}f_{1\Sigma}+ \12 \vec{\delta}f_{1\Sigma}+ \12 (\vec{D}_{0, L}- 4)f_{0s}\rh,
\end{array}
\eeq

which proves the lemma. \qed
\subsubsection{Proof of Lemma \ref{monlemmeclean}}\label{sec4.0.3}
In the sequel we will often use the following identities, see Subsect. \ref{sec4.1} for notation:
\begin{equation}
\label{identit}
\vec{\delta}\circ |\rh)= - 2 \vec{d}, \ (\rh|\circ \vec{d}= - 2\vec{\delta}, \ \vec{d}\circ\vec{d}\circ\vec{d}= 2\vec{d}\circ(\vD{0}-2).
\end{equation}

 If $P$ is selfadjoint with  non-trivial kernel we denote by $P^{-1}: \Ran P\to (\Ker P)^{\perp}$ the generalized inverse of $P$. From Lemmas \ref{lemmi}, \ref{lemmo} we obtain that $g= \col{g_{0}}{g_{1}}\in \Ker \tK^{\dag}_{21\Sigma}\cap\Ker\tK_{20\Sigma}^{\dag}$ iff
 \[
 \begin{array}{rl}
 1)&g_{0ss}+ \12 (\rh| g_{0\Sigma\Sigma})=0,\\[2mm]
2)&g_{1ss}+ \12 (\rh| g_{1\Sigma\Sigma})=0,\\[2mm]
  3)&g_{1ss}+ \vec{\delta}g_{0s\Sigma}=0,\\[2mm]
 4)&2g_{1s\Sigma}+ \vec{\delta}g_{0\Sigma\Sigma}=0,\\[2mm]
5)&2(\vec{D}_{0, L}-3)g_{0ss}+ 2\vec{\delta}g_{1s\Sigma}- (\rh| g_{0\Sigma\Sigma})=0,\\[2mm]
6)&2(\vec{D}_{1, L}-4)g_{0s\Sigma}+ \vec{\delta}g_{1\Sigma\Sigma}=0.
 \end{array}
 \]
An easy computation shows that this is equivalent to
\[
\begin{array}{rl}
1')&g_{0ss}= - \12 (\rh| g_{0\Sigma\Sigma}),\\[2mm]
2')&g_{1ss}= - \12 (\rh| g_{1\Sigma\Sigma}),\\[2mm]
3')&g_{0s\Sigma}= - \12(\vD{1} -4)^{-1}\vec{\delta}g_{1\Sigma\Sigma}+ \beta_{\Sigma},\\[2mm]
4')&g_{1s\Sigma}= - \12 \vec{\delta}g_{0\Sigma\Sigma},\\[2mm]
5')&(\vD{0}- 2)(\rh| g_{0\Sigma\Sigma})= -\vec{\delta}\vec{\delta}g_{0\Sigma\Sigma},\\[2mm]
6')& (\vD{0}-4)(\rh| g_{1\Sigma\Sigma})= -\vec{\delta}\vec{\delta}g_{1\Sigma\Sigma},
\end{array}
\]
for $\beta_{\Sigma}\in \Vect\{\psi_{jk}\}$.  As in the proof of Lemma \ref{dec-sphere} we first write $g_{i\Sigma\Sigma}= \bar{g}_{i\Sigma\Sigma}+ u_{is}$ where $(\rh| g_{i\Sigma\Sigma})= 6u_{is}$ and $\vec{\delta}u_{is}\rh= - 2 \vec{d}u_{is}$, $\vec{\delta}\vec{\delta}u_{is}\rh= -2\vD{0}u_{is}$.

We obtain from 5') and 6)' that
\[
\begin{array}{l}
(\vD{0}- 3) u_{0s}= - \frac{1}{4}\vec{\delta}\vec{\delta} \bar{g}_{0\Sigma\Sigma},\\[2mm]
(\vD{0}- 6) u_{1s}= - \frac{1}{4}\vec{\delta}\vec{\delta} \bar{g}_{1\Sigma\Sigma},
\end{array}
\]
i.e.
\[
\begin{array}{l}
u_{0s}= - \frac{1}{4}(\vD{0}-3)^{-1} \vec{\delta}\vec{\delta}\bar{g}_{0\Sigma\Sigma}+ \beta_{s},\\[2mm]
u_{1s}= - \frac{1}{4}(\vD{0}-6)^{-1} \vec{\delta}\vec{\delta}\bar{g}_{1\Sigma\Sigma},
\end{array}
\]
for $\beta_{s}\in \Vect\{\psi_{i}\}$. This gives:
\[
\begin{array}{l}
\vec{\delta}g_{0\Sigma\Sigma}= (1+ \12 \vec{d}(\vD{0}-3)^{-1}\vdel)\vdel \bar{g}_{0\Sigma\Sigma}- 2 \vd \beta_{s},\\[2mm]
\vdel g_{1\Sigma\Sigma}= (1+ \12 \vd (\vD{0}-6)^{-1}\vdel)\vdel \bar{g}_{1\Sigma\Sigma},
\end{array}
\]
and finally
\beq\label{e.tit77}
\begin{array}{rl}
1)& g_{0ss}= \frac{3}{4}(\vD{0}- 3)^{-1}\vdel\vdel \bar{g}_{0\Sigma\Sigma}- 3 \beta_{s},\\[2mm]
2)& g_{0s\Sigma}= - \12 (\vD{1}-4)^{-1}(1+ \12 \vd (\vD{0}-6)^{-1}\vdel)\vdel \bar{g}_{1\Sigma\Sigma}+ \beta_{\Sigma},\\[2mm]
3)&g_{0\Sigma\Sigma}= - \frac{1}{4}(\vD{0}-3)^{-1} \vec{\delta}\vec{\delta}\bar{g}_{0\Sigma\Sigma}\rh + \bar{g}_{0\Sigma\Sigma}+ \beta_{s}\rh,\\[2mm]
4)& g_{1ss}= \frac{3}{4}(\vD{0}-6)^{-1}\vdel\vdel \bar{g}_{1\Sigma\Sigma},\\[2mm]
5)&g_{1s\Sigma}= - \12  (1+ \12 \vd (\vD{0}-3)^{-1}\vdel)\vdel \bar{g}_{0\Sigma\Sigma}+ \vd \beta_{s},\\[2mm]
6)& g_{1\Sigma\Sigma}= - \frac{1}{4}(\vD{0}-6)^{-1} \vec{\delta}\vec{\delta}\bar{g}_{1\Sigma\Sigma}\rh + \bar{g}_{1\Sigma\Sigma}.
\end{array}
\eeq
Next  as in the proof of Lemma \ref{dec-sphere}  there exist unique couples  $(u_{i\Sigma\Sigma}, f_{i\Sigma})$ with $u_{i\Sigma\Sigma}\in \Ker \vdel\cap \Ker (\rh|$, $f_{i\Sigma}\in \Ker\one_{\{3, 4\}}(\vD{1})$ such that
\beq\label{e.tit7}
\bar{g}_{i\Sigma\Sigma}= u_{i\Sigma\Sigma}\oplus ( \vd f_{i\Sigma}+ \frac{1}{3} \vdel f_{i\Sigma}\rh),\hbox{ with } \vdel u_{i\Sigma\Sigma}= (\rh| u_{i\Sigma\Sigma})=0.
\eeq
Using \eqref{identit} we obtain that
\beq\label{e.tit78}
\begin{array}{l}
\vdel(\vd f_{i\Sigma}+ \frac{1}{3}\vdel f_{i\Sigma} \rh)= (\vD{1}- 4+ \frac{1}{3}\vd\vdel)f_{i\Sigma},\\[2mm]
\vdel\vdel(\vd f_{i\Sigma}+ \frac{1}{3}\vdel f_{i\Sigma} \rh)=\frac{4}{3}(\vD{0}-3)\vdel f_{i\Sigma}.
\end{array}
\eeq

If we plug \eqref{e.tit7} into \eqref{e.tit77} we obtain using \eqref{e.tit78} and the fact that $\vD{1}-4= \vdel\vd- \vd\vdel$ that:
 \[
\begin{array}{rl}
1)& g_{0ss}= \vdel f_{0\Sigma}- 3 \beta_{s},\\[2mm]
2)& g_{0s\Sigma}= - \12 (f_{1\Sigma}+ \vd(\vD{0}-6)^{-1}\vdel f_{1\Sigma})+ \beta_{\Sigma},\\[2mm]
3)&g_{0\Sigma\Sigma}= u_{0\Sigma\Sigma}+ \vd f_{0\Sigma}+ \beta_{s}\rh,\\[2mm]
4)& g_{1ss}= (1+ 3(\vD{0}-6)^{-1})\vdel f_{1\Sigma},\\[2mm]
5)&g_{1s\Sigma}= - \12 \vdel\vd f_{0\Sigma}+ \vd \beta_{s},\\[2mm]
6)& g_{1\Sigma\Sigma}= u_{1\Sigma\Sigma}+ \vd f_{1\Sigma}- (\vD{0}-6)^{-1}\vdel f_{1\Sigma}\rh,
\end{array}
\]
which completes the proof of the lemma. \qednoskip

\section{Maxwell fields} \label{s:maxwell}

\subsection{Maxwell fields on de Sitter space}\label{maxwell}
We briefly treat the analogous case of Maxwell fields on $dS^{4}$. In this case  the bundles $V_{2}, V_{1}$ are replaced by $V_{1}, V_{0}$ respectively, $D_{2}$ is replaced by $D_{1, L}= \delta_{\rm a}\circ d_{\rm a}+ d_{\rm a}\circ \delta_{\rm a}$, $D_{0}$ by $D_{0, L}= - \square_{0}$, $K_{21}$ by $K_{10}= d_{\rm a}$, where the subscript stands for ``anti-symmetric'' and is used to distinguish from the symmetric differential and related operators used in most of  the paper.
The physical charge ${\bf q}_{2, I}$ is replaced by ${\bf q}_{1}$. We set
\[
\cE\defeq \Ker K_{10\Sigma}^{\dag}, \quad \cF\defeq \Ran K_{10\Sigma},
\]
the physical phase space  being then ${\cE}/{\cF}$. 

As before we denote with tildes the Wick rotated objects.

We obtain that
\[
\tD_{0}u=- \p_{s}^{2}u + a^{-1}\vec{D}_{0, L}u+ \textstyle\frac{3}{2}\dot{a}a^{-2}\p_{s}u,
\]
while  the expression for $\tD_{1}w$ is obtained from the one at the beginning of Subsect. \ref{lalilalou} by adding  the constant term $2\Lambda= 6$.

From \cite[Thm.~3.2, Prop.~3.9]{Boucetta1999} we know that the Wick-rotated operator $\tD_{1}$ satisfies $\tD_{1}\geq 4$, hence $\tD_{1}$ is invertible, while $\Ker \tD_{0}= \Ker d= \cc$, the space of constant functions on $\bS^{4}$. 
For coherence of notation with the case of linearized gravity considered earlier, we set
\[
\tilde{\mathcal{K}}= \Ker \tD_{0}\sim \cc, \ \ \tilde{\mathcal{K}}_{\Sigma}= \tilde{\varrho}_{0}\tilde{\mathcal{K}}\sim \cc\oplus \{0\},
\]
and 
\begin{equation}\label{max.e-1}
\mathcal{K}= \cF_{0}^{-1}\tilde{\mathcal{K}}\sim \cc, \  \ \mathcal{K}_{\Sigma}= \cF_{0\Sigma}^{-1}\tilde{\mathcal{K}}_{\Sigma} \sim \cc\oplus \{0\}.
\end{equation}

The \calde projectors $\tc_{1}^{\pm}$ for $\tD_{1}$ are well defined, while those for $\tD_{0}$ exist only on $\tilde{\mathcal{K}}_{\Sigma}^{{\bf q}_{0}}$.  
We set  
\[
I_{0}= \{k(k+2): k\in \nn\}, I_{1}= \{k(k+2)+1:k\in \nn, k\geq 1\}
\]
we obtain from \cite[Sect. 3.6]{Boucetta1999} that
\beq\label{biloute}
\sigma(\vD{1})_{|\Ran \vd_{\rm a}}= I_{0}\setminus\{0\}, \sigma(\vD{1})_{|\Ker \vdel_{\rm a}}= I_{1},
\eeq
and note that $I_{0}\cap I_{1}= \emptyset$.

As before we obtain $[\tc_{1}^{\pm}, \one_{\{\lambda\}}(\hat{D}_{1})\otimes \one_{\cc^{2}}]=0$, and  $(\one_{\{\lambda\}}(\hat{D}_{1})\otimes \one_{\cc^{2}})K_{10\Sigma}= K_{10\Sigma}(\one_{\{\lambda\}}(\hat{D}_{0})\otimes \one_{\cc^{2}})$ so $\one_{\{\lambda\}}(\hat{D}_{1})\otimes \one_{\cc^{2}}$ preserves $\cE$ and $\cF$.

Defining $\tK_{10\Sigma}$ as in Subsect.~\ref{sec4.2} we obtain that if $g= \tK_{10\Sigma}f$ for $f= \col{f_{0}}{f_{1}}\in \cinf(\Sigma; \tV_{0}(\Sigma)\otimes \cc^{2})$ then
\beq\label{max.e1}
\begin{array}{l}
g_{0s}= - f_{1},\
g_{0\Sigma}= \vec{d}f_{0},\\[2mm]
g_{1s}= - \vec{D}_{0, L}f_{0},\
g_{1\Sigma}= \vec{d}f_{1}.
\end{array}
\eeq

Similarly we find that if $f= \tK_{10\Sigma}^{\dag}g$, then 
\beq\label{max.e2}
f_{0}= g_{1s}+ \vec{\delta}g_{0\Sigma},\ f_{1}= \vec{D}_{0, L}g_{0s}+ \vec{\delta}g_{1\Sigma}.
\eeq

The Lorentzian versions of $K_{10\Sigma}, K_{10\Sigma}^{\dag}$ are obtained by conjugation with the $\cF_{i\Sigma}$, as explained in Subsect.~\ref{sec2.2}.
\subsubsection{Properties of \texorpdfstring{$\cE$}{phase space}}
Using the Hodge decomposition  we obtain the following analog of Lemma \ref{monlemmeclean}.
\begin{lemma}\label{blurk}
 Let $g= \col{g_{0}}{g_{1}}\in \tcE$. Then there exist unique  $u_{i\Sigma}\in \cinf(\Sigma; \tV_{1\Sigma})$, $f_{is}\in \cinf(\Sigma; \tV_{0})$, $i=0,1$, and $\beta_{s}\in \Ker \vD{0}$ with 
 \[
 \vdel_{\rm a}u_{i\Sigma}= \one_{\{0\}}(\vD{0})f_{is}=0
 \]
 such that:
 \[
 \begin{array}{l}
 g_{0s}= - f_{1s}+ \beta_{s}, \ \ 
 g_{0\Sigma}= u_{0\Sigma}+ \vd_{\rm a}f_{0s},\\[2mm]
 g_{1s}= - \vD{0}f_{0s},\  \
 g_{1\Sigma}= u_{1\Sigma}+ \vd_{\rm a}f_{1s}.
 \end{array}
 \]
 \end{lemma}
We define the subspaces $\tcE_{\rm gauge}$ and $\tcE_{0}$ by
\[
\tcE_{\rm gauge}\defeq \one_{I_{1}}(\vD{1}\otimes\one_{\cc^{2}})\tcE, \  \tcE_{0}\defeq\one_{\{0\}}(\vD{1}\otimes\one_{\cc^{2}})\tcE,
\]
and obtain that 
\[
g\in \tcE_{\rm gauge}\Leftrightarrow f_{is}= \beta_{s}=0, i=0,1,\quad g\in \tcE_{0}\Leftrightarrow u_{i\Sigma}= f_{is}=0, i=0,1.
\]
Similarly we set 
\[
\tcF_{\rm gauge}\defeq \tK_{10\Sigma}\tilde{\mathcal K}_{\Sigma}^{{\bf q}_{0}},
\]
and obtain that
\[
g\in \tcF_{\rm gauge}\Leftrightarrow u_{i\Sigma}= \beta_{s}=0.
\]
For the Lorentzian versions of the above spaces we have:
\[
\cE= \cE_{\rm gauge}\oplus \cF= \cE_{\rm gauge}\oplus\cF_{\rm gauge}\oplus \cE_{0}.
\]
By the same argument as in Prop. \ref{propcor.3} we obtain that 
\[
{\Ker {\bf q}_{1}}_{| \cE}= \cF
\]
so $[{\bf q}_{1}]$ is non-degenerate on $\dfrac{\cE}{\cF}$.
As noticed in \cite{Sanders2014} (cf.~\cite{WZ} for the analogous statement in the BRST approach), this also follows by a Poincar\'e duality argument and the fact that the Cauchy surface is compact.

\subsubsection{The Euclidean vacuum}
As before we define covariances $\boldsymbol\lambda_{1\Sigma}^{\pm}\defeq \pm {\bf q}_{1}\circ c_{1}^{\pm}$, where $c_{1}^{\pm}= \cF_{1\Sigma}^{-1}\circ \tc_{1}^{\pm}\circ \cF_{1\Sigma}$. The associated pseudo state $\omega_{\rm eucl }$ is again called the {\em Euclidean vacuum}. As in the case of linear gravity we will see that  it is not  positive on the whole of $\cE$.

\subsubsection{Gauge invariance}
The following lemma is proved exactly as Prop. \ref{prop10.1b}.
\begin{lemma}\label{max.lemma1}
One has
\[
c_{1}^{\pm}K_{10\Sigma}= K_{10\Sigma}c_{0}^{\pm}  \hbox{ on }\mathcal{K}_{\Sigma}^{{\bf q}_{0}},
\]
hence $\boldsymbol\lambda_{1\Sigma}^{\pm}=0$ on $\cE\times \cF_{\rm gauge}$.
\end{lemma}
As in the case of linearized gravity, using that $\Ker \tilde{D}_{0}= \Ker d_{\rm a}$ we obtain that $\Ran K_{21\Sigma}^{\dag}\subset \mathcal{K}_{\Sigma}^{{\bf q}_{0}}$ so $\omega_{\rm eucl}$ is weakly gauge invariant. 
 \subsubsection{Positivity}

The following proposition is shown exactly as Props. \ref{prop5.3}, \ref{prop-pos1}, using that by \eqref{biloute} one has ${\bf q}_{1}= \tilde{\bf q}_{1}$ on $\Ran \one_{I_{1}}(\vD{1}\otimes \one_{\cc^{2}})$, ${\bf q}_{1}= -\tilde{\bf q}_{1}$ on $\Ran \one_{\{0\}}(\vD{1}\otimes \one_{\cc^{2}})$, and also the fact that $\tD_{1}\geq 4$.
\begin{proposition}\label{prop-max1}
 The covariances ${\bf \lambda}_{1\Sigma}^{\pm}$ are positive on $\cE_{\rm gauge }$ and negative  on $\cE_{0}$. Moreover $\bar{f}\dual \boldsymbol{\lambda}_{1\Sigma}^{\pm}f=0$ implies that $c_{1}^{\pm}f=0$ for $f\in \cE_{0}$.
\end{proposition}
\subsubsection{The modified Euclidean vacuum}
Let $\pi= \one_{\rr\setminus\{0\}}(\hat{D}_{1})\otimes \cc^{2}$ be the orthogonal projection $\cE\to \cE_{\rm reg}$. For $f\in \cE$ we have $((1-\pi)f)_{0s}= \one_{\{0\}}(\vec{D}_{0, L})f_{0s}$, the other components being equal to $0$.

We define the modified Euclidean vacuum similarly as before by the covariances
\[
{\bf \lambda}_{1\Sigma, {\rm mod}}^{\pm}\defeq \overline{\pi f}\dual {\bf \lambda}_{1\Sigma}^{\pm}\pi f.
\]
where
\[
\pi= \one_{\rr\setminus \{0\}}(\hat{D}_{1}\otimes\one_{\cc^{2}}): \cE\to \cE_{\rm gauge}\oplus \cF_{\rm gauge}.
\]

As in Corollary \ref{corr.1} we have $\Ran (\one - \pi)\subset \Ran K_{10\Sigma}$, $\one- \pi$ is smoothing of rank one, and $\Ran \pi K_{10\Sigma}\subset K_{10\Sigma}\mathcal{K}_{\Sigma}^{{\bf q}_{0}}$. 

The same arguments as in the proof of Thm.~\ref{thm5.1} show that  the associated state $\omega_{\rm mod}$ is a Hadamard, fully gauge invariant state on $\cE$. It is invariant under the action of $O(4)$,  but not under  $SO^{\uparrow}(1, 4)$. 
\section{Link with Bunch--Davies construction} \label{s:bunch-davies}
In this section we explain the relation between the {formal} construction of an Euclidean vacuum by the Bunch-Davies argument and the one we use based on \calde projectors.  We also comment on the  {formal} construction of states for linearized gravity on de Sitter space by mode expansion which is commonly used in the literature, see e.g.~\cite{Allen1986,higuchiweeks,Faizal2012} and we compare it with our construction. 
\subsection{The mode expansion method}
We start with a brief summary of the (formal) construction of states by mode expansion.

For simplicity let us consider a  Klein--Gordon operator $D$ on some spacetime $(M, g)$. We follow our habit of considering  complex fields.  We fix a Cauchy surface $\Sigma$, denote by $\varrho$ the Cauchy trace on $\Sigma$ and by ${\bf q}$ the charge Hermitian form acting on Cauchy data.

A {\em mode expansion} is defined by a family $\{u_{i}^{+}, u_{i}^{-}\}_{i\in I}$ of solutions of $D u=0$, such that for $f_{i}^{\pm}= \varrho u_{i}^{\pm}$ one has
\beq\label{emode.3}
\overline{f^{+}_{i}}\dual {\bf q}f^{+}_{j}= \delta_{ij}, \ \  \overline{f^{+}_{i}}\dual {\bf q}f^{-}_{j}= \overline{f_{i}^{-}}\dual {\bf q}f^{+}_{j}=0,  \ \ \overline{f_{i}^{-}}\dual {\bf q}f^{-}_{j}= - \delta_{ij},
\eeq
and
\beq\label{emode.4}
 \bar{f}\dual {\bf q}f= \sum_{i\in I}\bar{f}\dual {\bf q}f^{+}_{i}\times\overline{f^{+}_{i}}\dual {\bf q}f- \sum_{i\in I} \bar{f}\dual {\bf q}f_{i}^{-}\times\overline{f^{-}_{i}}\dual {\bf q}f, \ f\in \coinf(\Sigma; \cc^{2}).
\eeq
In the physics literature, \eqref{emode.3} resp.~\eqref{emode.4} is the ``orthogonality'', resp. ``completeness'' property of the family $\{u_{i}^{+}, u_{i}^{-}\}_{i\in I}$.

Since the discussion is formal, {we do not} specify the index set $I$, nor the meaning of sums $\sum_{i\in I}a_{i}$, though this is of course a central issue if one seeks to make the construction rigorous. Typically $i\in I$ consists of discrete quantum numbers, for example labelling spherical harmonics, and some real frequency,  so $\sum_{i\in I}$ is a combination of discrete sums and integrals.

One can then define a state by its  covariances $\boldsymbol{\lambda}^{\pm}_{\Sigma}$ defined as
\[
\begin{array}{l}
 \bar{f}\boldsymbol{\lambda}^{+}_{\Sigma}f= \sum_{i\in I}\bar{f}\dual {\bf q}f^{+}_{i}\times\overline{f^{+}_{i}}\dual {\bf q}f,\\[2mm]
\bar{f}\boldsymbol{\lambda}^{-}_{\Sigma}f = \sum_{i\in I} \bar{f}\dual {\bf q}f_{i}^{-}\times\overline{f^{-}_{i}}\dual {\bf q}f.
\end{array}
\]
The associated projections $c^{\pm}$ are
\beq\label{emode.5}
c^{\pm}f= \pm\sum_{i\in I}f^{\pm}_{i}\overline{f^{\pm}_{i}}\dual {\bf q}f,
\eeq
and the associated spacetime covariances are:
\begin{equation}
\label{emode.5b}
\bar{u}\dual \boldsymbol{\Lambda}^{\pm}u= \sum_{i\in I}(u|u_{i}^{\pm})_{M}(u_{i}^{\pm}| u)_{M}, \ u\in \coinf(M).
\end{equation}

\subsection{The Bunch--Davies vacuum and \calde projectors}
 
 Let us assume that $M= \rr_{t}\times \Sigma$ and that we can perform the Wick rotation in $t$. The Bunch--Davies vacuum is obtained by choosing the modes $u_{i}^{\pm}$ so that  $u_{i}^{\pm}$ extend holomorphically to  a strip in $\{\pm \Im\, z >0\}$, where $z= t+ \i s$.

 Their Wick rotations $\tilde{u}_{i}^{\pm}$, obtained by replacing $z$ by $\i s$ are solutions of $\tilde{D}\tilde{u}=0$ in $\Omega^{\pm}$.  Therefore  $f^{\pm}_{i}= \varrho u_{i}^{\pm}$ belong to  $\Ran c^{\pm}$ and the completeness property of the modes formally implies that they span $\Ran c^{\pm}$.  So the projections defined in \eqref{emode.5} are equal to the \calde projectors $c^{\pm}$, hence the Bunch--Davies construction of the  Euclidean vacuum coincide with the construction using \calde projectors. 
 \subsection{Mode expansion in linearized gravity}
 We follow the exposition in \cite[Sect. II]{Faizal2012}.  One considers there {\rm TT} solutions of $D_{2}u=0$, corresponding to working with the space $\cE_{\TT}$ on the level of Cauchy data.
 The modes in \cite{Faizal2012} are denoted by $H^{m, l, \sigma}$  where $m= 0, 1, 2$ and $\sigma$ represents all the other tensor spherical harmonics indices. 
 For $m=0, 2$ one has $l\geq 2$ while for $m= 1$ one has $l\neq 1$.

  Let us write the indices $(m, l, \sigma)$ simply by $i\in I$, with $I= I'\cup I''$, where $I'$ corresponds to $m=2$ and $I''$ to $m= 0, 1$. Similarly we set $u_{i}^{+}= H^{m, l, \sigma}$ and $u_{i}^{-}= \bar{H}^{m, l, \sigma}$. The modes $H^{m, l, \sigma}$ are 'positive frequency solutions', so their Cauchy data should belong to the range of $c_{2}^{+}$. Since the field equation is real, their complex conjugates $\bar{H}^{m, l, \sigma}$ are negative frequency solutions, with Cauchy data in the range of $c_{2}^{-}$.

 The modes $u_{i}^{\pm}$ for $i\in I''$ are pure gauge, i.e.~their Cauchy data belong to $\cF_{\TT}$, while the modes $u_{i}^{\pm}$ for $i\in I'$ have vanishing $u_{ss}$ and $u_{s\Sigma}$ components.
 It is likely that the $u_{i}^{\pm}$ for $i\in I''$ span $\cF_{\TT}$, while the $u_{i}^{\pm}$ for $i\in I'$ span our space $\cE_{\TT, \rm gauge}$, to the extent that this can be made rigorous.

Because the $u_{i}^{\pm}$ for $i\in I''$ are pure gauge, the orthogonality condition \eqref{emode.3} fails, since $\cF_{\TT}= \Ker {\bf q_{I, 2}}_{|\cE_{\rm TT}}$.

In \cite{Faizal2012}, in order to enforce positivity and gauge invariance at the same time, one defines  the Cauchy surface covariances $\boldsymbol \lambda_{2\Sigma\: {\rm FH}}^{\pm}$ by
\[
\bar{f}\dual \boldsymbol\lambda_{2\Sigma\: {\rm FH}}^{\pm}f= \sum_{i\in I'}\bar{f}\dual {\bf q}_{I, 2}u_{i}^{\pm}\times \bar{u}_{i}^{+}\dual{\bf q}_{I, 2}f,
\]
see e.g.~\cite[(3.9)]{Faizal2012} for the corresponding spacetime covariances $\boldsymbol\Lambda_{2\Sigma\:{\rm FH}}^{\pm}$. 
This means that the pure gauge modes $u_{i}^{\pm}$ for $i\in I''$ are  eliminated by hand. 
In other words,  the state constructed in \cite{Faizal2012} is obtained by composing the Euclidean vacuum pseudo state $\omega_{\rm eucl}$ with the orthogonal projection $\pi_{\rm FH}$ on $\cE_{\TT, \rm gauge}$ along  $\cF_{\TT}$, while our state $\omega_{\rm mod}$  is obtained from the projection $\pi$. Note that $\Ker \pi_{\rm FH}= \cF_{\TT}$ is infinite dimensional, while $\Ker \pi= \cE_{\TT, 4}$ is finite dimensional.  

At the level of distributional kernels, we see  that $\Lambda_{2\Sigma\:{\rm FH}}^{+}- \Lambda_{2\Sigma\:{\rm FH}}^{-}$ differs from $\i G_{2}$, 
contrarily to the spacetime covariances $ \Lambda_{2\Sigma}^{\pm}$ of the Euclidean pseudo vacuum. 

On the other hand the restrictions of $\lambda_{2\Sigma\: {\rm FH}}^{\pm}$ and of our modified covariances $\boldsymbol\lambda_{2\Sigma, \rm mod}^{\pm}$ to $\cE_{\TT}$ are expected to coincide.

In conclusion, the two constructions yield the same state on the physical quotient space $\cE_{\TT}/ \cF_{\TT}$ (again, to the extent the former can be made rigorous), the main difference being that we construct a representative satisfying the CCR for the gauge-fixed hyperbolic operator $D_2$ on all $(0,2)$-tensors.

{\small
\subsubsection*{Acknowledgments} The authors would like to thank Vincent Moncrief and Simone Murro for useful discussions.  Support from the grant
ANR-20-CE40-0018 is gratefully acknowledged.  \medskip }
 \bibliographystyle{abbrv}
 \bibliography{desittergravity}

\begin{thebibliography}{10}

\bibitem{Allen1985}
B.~Allen.
\newblock Vacuum states in de sitter space.
\newblock {\em Physical Review D}, 32, 1985.

\bibitem{Allen1986}
B.~Allen.
\newblock Graviton propagator in de sitter space.
\newblock {\em Physical Review D}, 34, 1986.

\bibitem{AFO}
B.~Allen, A.~Folacci, and A.~C. Ottewill.
\newblock {Renormalized graviton stress-energy tensor in curved vacuum
  space-times}.
\newblock {\em Phys. Rev. D}, 38(4):1069--1082, 1988.

\bibitem{BDM}
M.~Benini, C.~Dappiaggi, and S.~Murro.
\newblock {Radiative observables for linearized gravity on asymptotically flat
  spacetimes and their boundary induced states}.
\newblock {\em J. Math. Phys.}, 55(8):082301, 2014.

\bibitem{Besse1987}
A.~L. Besse.
\newblock {\em {Einstein Manifolds}}.
\newblock 1987.

\bibitem{Boucetta1999}
M.~Boucetta.
\newblock Spectre des laplaciens de lichnerowicz sur les sphères et les
  projectifs réels.
\newblock {\em Publicacions Matemàtiques}, 43(2):451--483, 1999.

\bibitem{DS}
C.~Dappiaggi and D.~Siemssen.
\newblock {Hadamard states for the vector potential on asymptotically flat
  spacetimes}.
\newblock {\em Rev. Math. Phys.}, 25(01):1350002, 2013.

\bibitem{Faizal2012}
M.~Faizal and A.~Higuchi.
\newblock Physical equivalence between the covariant and physical graviton
  two-point functions in de sitter spacetime.
\newblock {\em Physical Review D - Particles, Fields, Gravitation and
  Cosmology}, 85, 2012.

\bibitem{FH}
C.~J. Fewster and D.~S. Hunt.
\newblock {Quantization of linearized gravity in cosmological vacuum
  spacetimes}.
\newblock {\em Rev. Math. Phys.}, 25(02):1330003, 2013.

\bibitem{FP}
C.~J. Fewster and M.~J. Pfenning.
\newblock {A quantum weak energy inequality for spin-one fields in curved
  space–time}.
\newblock {\em J. Math. Phys.}, 44(10):4480, 2003.

\bibitem{FS}
F.~Finster and A.~Strohmaier.
\newblock {Gupta–Bleuler quantization of the Maxwell field in globally
  hyperbolic space-times}.
\newblock {\em Ann. Henri Poincar{\'{e}}}, 16(8):1837--1868, 2015.

\bibitem{furlani}
E.~P. Furlani.
\newblock {Quantization of the electromagnetic field on static space–times}.
\newblock {\em J. Math. Phys.}, 36(3):1063--1079, 1995.

\bibitem{Gazeau2023}
J.~P. Gazeau and H.~Pejhan.
\newblock {Covariant quantization of the partially massless graviton field in
  de Sitter spacetime}.
\newblock {\em Phys. Rev. D}, 108(6), 2023.

\bibitem{G}
C.~G{\'{e}}rard.
\newblock {\em {Microlocal Analysis of Quantum Fields on Curved Spacetimes}}.
\newblock European Mathematical Society, arXiv:1901.10175, Z{\"{u}}rich, 2019.

\bibitem{Gerard2024}
C.~G{\'{e}}rard.
\newblock {Hadamard States for Linearized Gravity on Spacetimes with Compact
  Cauchy Surfaces}.
\newblock {\em Commun. Math. Phys.}, 405(6), jun 2024.

\bibitem{Gerard2025}
C.~G{\'{e}}rard, S.~Murro, and M.~Wrochna.
\newblock {Wick Rotation of Linearized Gravity in Gaussian Time and
  Calder{\'{o}}n Projectors}.
\newblock {\em Ann. Henri Poincare}, 26(2):4461--4528, 2025.

\bibitem{GW1}
C.~G{\'{e}}rard and M.~Wrochna.
\newblock {Hadamard states for the linearized Yang–Mills equation on curved
  spacetime}.
\newblock {\em Commun. Math. Phys.}, 337(1):253--320, 2015.

\bibitem{GW2}
C.~G{\'{e}}rard and M.~Wrochna.
\newblock {Analytic Hadamard states, Calder{\'{o}}n projectors and Wick
  rotation near analytic Cauchy surfaces}.
\newblock {\em Commun. Math. Phys.}, 366(1):29--65, 2019.

\bibitem{Glavan2023}
D.~Glavan and T.~Prokopec.
\newblock {Photon propagator in de Sitter space in the general covariant
  gauge}.
\newblock {\em J. High Energy Phys.}, 2023(5), 2023.

\bibitem{HS}
T.-P. Hack and A.~Schenkel.
\newblock {Linear bosonic and fermionic quantum gauge theories on curved
  spacetimes}.
\newblock {\em Gen. Relativ. Gravit.}, 45(5):877--910, 2013.

\bibitem{Higuchi1991a}
A.~Higuchi.
\newblock Quantum linearization instabilities of de sitter spacetime. i.
\newblock {\em Classical and Quantum Gravity}, 8, 1991.

\bibitem{Higuchi1991}
A.~Higuchi.
\newblock Quantum linearization instabilities of de sitter spacetime. ii.
\newblock {\em Classical and Quantum Gravity}, 8, 1991.

\bibitem{Higuchi2001}
A.~Higuchi and S.~S. Kouris.
\newblock {The covariant graviton propagator in de Sitter spacetime}.
\newblock {\em Class. Quantum Gravity}, 18(20), 2001.

\bibitem{dS1}
A.~Higuchi, D.~Marolf, and I.~A. Morrison.
\newblock {de Sitter invariance of the dS graviton vacuum}.
\newblock {\em Class. Quantum Gravity}, 28(24):245012, 2011.

\bibitem{higuchiweeks}
A.~Higuchi and R.~H. Weeks.
\newblock The physical graviton two-point function in de {S}itter spacetime
  with {$\mathbb{S}^3$} spatial sections.
\newblock {\em Classical and Quantum Gravity}, 20(14):3005, jun 2003.

\bibitem{hollands}
S.~Hollands.
\newblock {Renormalized quantum Yang-Mills fields in curved spacetime}.
\newblock {\em Rev. Math. Phys.}, 20(09):1033--1172, 2008.

\bibitem{Iuliano2023}
C.~Iuliano and J.~Zahn.
\newblock {Canonical quantization of Teukolsky fields on Kerr backgrounds}.
\newblock {\em Phys. Rev. D}, 108(12), 2023.

\bibitem{Lichnerowicz1961}
A.~Lichnerowicz.
\newblock Propagateurs et commutateurs en relativité générale.
\newblock {\em Publications Mathématiques de l'Institut des Hautes
  Scientifiques}, 10, 1961.

\bibitem{dS2}
S.~P. Miao, P.~J. Mora, N.~C. Tsamis, and R.~P. Woodard.
\newblock {Perils of analytic continuation}.
\newblock {\em Phys. Rev. D}, 89(10):104004, 2014.

\bibitem{Miao2010}
S.~P. Miao, N.~C. Tsamis, and R.~P. Woodard.
\newblock {De Sitter breaking through infrared divergences}.
\newblock {\em J. Math. Phys.}, 51(7), 2010.

\bibitem{Miao2011}
S.~P. Miao, N.~C. Tsamis, and R.~P. Woodard.
\newblock {The graviton propagator in de Donder gauge on de Sitter background}.
\newblock {\em J. Math. Phys.}, 52(12), 2011.

\bibitem{moncrief}
V.~Moncrief.
\newblock {Decompositions of gravitational perturbations}.
\newblock {\em J. Math. Phys.}, 16(8):1556--1560, 1975.

\bibitem{Moretti2023}
V.~Moretti, S.~Murro, and D.~Volpe.
\newblock {The Quantization of Proca Fields on Globally Hyperbolic Spacetimes:
  Hadamard States and M{\o}ller Operators}.
\newblock {\em Ann. Henri Poincare}, 24(9), 2023.

\bibitem{R}
H.~Ringstr{\"{o}}m.
\newblock {\em {The Cauchy Problem in General Relativity}}.
\newblock European Mathematical Society Publishing House, Z{\"{u}}rich, 2009.

\bibitem{SV}
H.~Sahlmann and R.~Verch.
\newblock {Microlocal spectrum condition and Hadamard form for vector-valued
  quantum fields in curved spacetime}.
\newblock {\em Rev. Math. Phys.}, 13(10):1203--1246, 2001.

\bibitem{Sanders2014}
K.~Sanders, C.~Dappiaggi, and T.~P. Hack.
\newblock Electromagnetism, local covariance, the aharonov-bohm effect and
  gauss' law.
\newblock {\em Communications in Mathematical Physics}, 328, 2014.

\bibitem{WZ}
M.~Wrochna and J.~Zahn.
\newblock {Classical phase space and Hadamard states in the BRST formalism for
  gauge field theories on curved spacetime}.
\newblock {\em Rev. Math. Phys.}, 29(04):1750014, 2017.

\end{thebibliography}
 
\end{document}